\documentclass[a4paper,11pt]{article}
\pdfoutput=1 

\usepackage{jheppub} 
\usepackage[utf8]{inputenc}
\usepackage{setspace}
\usepackage{subfigure}
\usepackage{empheq}
\usepackage{hyperref}
\definecolor{refkey}{gray}{0.45}
\definecolor{labelkey}{RGB}{155,48,48}

\usepackage{mathtools}
\usepackage{physics}
\usepackage{simplewick}
\usepackage[normalem]{ulem}
\usepackage[obeyFinal]{todonotes}

\def\Tr{\text{Tr}}

\def\beq{\begin{eqnarray}}\def\eeq{\end{eqnarray}}
\def\be{\begin{equation}}\def\ee{\end{equation}}

\def\mes[#1]{d^{3}{#1}}

\def\del{\partial}

\newcommand{\half}{\frac{1}{2}}

\def\del{\partial}

\def\order{\ensuremath{\mathcal{O}}}

\reversemarginpar

\definecolor{UI_blue}{RGB}{32, 64, 151}
\definecolor{UI_red}{RGB}{187, 62, 24}
\definecolor{UI_blue2}{RGB}{0, 84, 147}
\definecolor{UI_red2}{RGB}{159, 32, 66}
\definecolor{UI_gray}{RGB}{169, 169, 169}
\definecolor{UI_sepia}{RGB}{112, 66, 20}
\definecolor{UI_bittersweet}{RGB}{254, 111, 94}
\definecolor{UI_emerald}{RGB}{80, 200, 120}
\definecolor{UI_olivegreen}{RGB}{181, 179, 92}
\definecolor{UI_cadetblue}{RGB}{95, 158, 160}
\definecolor{UI_fuchsia}{RGB}{255, 0, 255}
\definecolor{UI_midnightblue}{RGB}{25, 25, 112}
\definecolor{UI_royalblue}{RGB}{0,35, 102}
\definecolor{UI_periwinkle}{RGB}{204, 204, 255}
\definecolor{UI_redorange}{RGB}{255, 83, 73}
\definecolor{UI_brickred}{RGB}{203,65,84}	
\definecolor{UI_forestgreen}{RGB}{34, 139, 34}
\definecolor{UI_tan}{RGB}{210,180,140}	
\definecolor{UI_burlywood}{RGB}{222,184,135}
\definecolor{UI_burlywood}{RGB}{192,64,0}
\definecolor{UI_darkorchid}{RGB}{153,50,204}

\usepackage{titlesec}
\usepackage{verbatim}
\usepackage{fancyhdr}
\usepackage{enumerate}
\hypersetup{colorlinks=true,citecolor=blue,urlcolor=black}
\usepackage{slashed, enumitem}
\usepackage{array}
\newcolumntype{P}[1]{>{\centering\arraybackslash}p{#1}}
\usepackage{anyfontsize}
	\vspace{-11cm}
	\author[a]{Indranil Dey,}
	\author[a,b]{Kanhu Kishore Nanda,}
	\author[a]{Akashdeep Roy,}
	\author[c]{Sunil Kumar Sake,}
	\author[a]{\hspace{0.5cm}Sandip P. Trivedi}
	
	\affiliation[a]{\it Department of Theoretical Physics,
		Tata Institute of Fundamental Research,\\  Colaba, Mumbai, India, 400005\\}
	\affiliation[b]{\it Chennai Mathematical Institute, H1 SIPCOT IT Park, Siruseri, Chennai, India, 603103\\}
	\affiliation[c] {\it Yukawa Institute for Theoretical Physics, Kyoto University, Kyoto, Japan, 606-8502\\}

	\emailAdd{indranil.dey@tifr.res.in}
	\emailAdd{kanhukishore@cmi.res.in}
	\emailAdd{akashdeep.roy@tifr.res.in}
	\emailAdd{sunilsake1@gmail.com}
	\emailAdd{sandip@theory.tifr.res.in}
	
	\vspace{1cm}

	\abstract{We discuss and extend some  aspects pertaining to the canonical quantisation of JT gravity in de Sitter space, including the problem of time and  the construction of a Hilbert space. We then extend this discussion to other two dimensional models obtained by changing the dilaton potential and show that the canonical quantisation procedure can be carried out for a large class of such models. Some discussion leading towards   a  path integral understanding for states, other than  the Hartle Hawking state, is also included here, along with  comments pertaining to Holography and the entropy of de Sitter space.}

	\title{JT Gravity in de Sitter Space and Its Extensions}
	\preprint{\parbox{3cm}{TIFR/TH/24-28 \\
	YITP-24-183}}
	
\begin{document}
		\maketitle
		\flushbottom
		\vskip 10pt
\section{Introduction}
Attempts to apply the laws of Quantum Mechanics to the universe as a whole  are well known to lead to a host of conceptual issues.

Some of these problems  pertain to the meaning and interpretation of the wave function of the universe. Others, relate to the problem of time. 
Solving the problem of time requires one to  find a physical observable in the universe which can act like a  clock,   construct a Hilbert space at  fixed values of this clock,  define operators acting on states, and understand their meaning. This problem is especially acute for universes which are closed, with no spatial boundary. 

The  conceptual questions mentioned above are not directly tied to the UV problems which also beset attempts to quantise gravity. 
Lower dimensional models of gravity are interesting in this context. Cosmologies in these lower dimensional theories have the same conceptual issues   as their higher dimensional counterparts, but their UV behaviour is much better controlled. In particular UV divergences in lower dimensional  models are often tame, 
if not entirely absent,  due to the absence of propagating gravitons.  
This motivates a study of these lower dimensional models with the hope  that lessons learnt from their study would also prove useful in studying higher dimensional, more realistic, theories.

This was the motivation behind the study of  JT gravity in dS space, a two dimensional model of gravity,  carried out in \cite{nanda2023jt}.
In this work the dilaton was taken to be the physical clock; all solutions to the Wheeler DeWitt (WdW) equation were obtained and a bulk Hilbert space was constructed. The number of  classical solutions was found to be  infinite, and it was shown after quantisation, in the single universe sector, that the number of linearly independent states in the quantum theory is also infinite. See also \cite{Maldacena:2019cbz} and \cite{Iliesiu:2020zld} for important related discussion.
Here we further elaborate on the canonical procedure followed in \cite{nanda2023jt}, discussing  in more detail how a conserved norm can be defined on hypersurfaces with a constant value of the dilaton and discussing some of the physical consequences for the resulting cosmologies. 

We then extend our study of canonical quantisation  to a more general class of two dimensional  theories obtained by changing the dilaton potential. 
In  JT theory the dilaton potential $U(\phi)$, defined in  eq.\eqref{jtacta}), is quadratic in $\phi$.
An interesting class of cosmologies, is obtained by  taking different $U(\phi)$ which   asymptotically, for $\phi\rightarrow \infty$, meet the condition, 
\be
\label{asspota}
U(\phi)\rightarrow \phi^2,
\ee
agreeing  with  JT  theory. 

We discuss the  classical solutions  and quantise these theories, following a quantisation procedure similar to the JT case\footnote{There are some important caveats pertaining to which types of classical solutions are included in this quantisation procedure, see below and also section \ref{Jtdsrevsum} for more discussion}, 
\cite{nanda2023jt}.  Once again, one finds an infinite number of states in the single universe sector of the quantum theory.
The qualitative features of the solutions to the WDW equation  are similar to the JT case. In regions which are classically allowed the wave functions are oscillatory and in regions where it is disallowed they are  exponentially damped or growing. 
Not unexpectedly we find that states, in theories where the asymptotic form of $U(\phi)$ satisfies eq.(\ref{asspota}),   can also be interpreted, for large values of the dilaton,  as being states in the JT theory. 

In particular, one might expect that the Hartle Hawking (HH) state in theories with deformed potentials meeting the asymptotic condition, eq.(\ref{asspota}), would lead to physical states in  JT theory, in the asymptotic limit, but that these states  would be different from the HH state of the JT theory. And in this way one can obtain a path integral understanding, to some extent, of states, other than the HH state, in JT theory. 

This expectation is indeed borne out for some cases where 
the corrections to the quadratic term in $U(\phi)$  are exponential in form, 
\be
\label{formexp}
U(\phi)=\phi^2 -2\sum_i \frac{\epsilon_i}{\alpha_i} e^{-\alpha_i\phi}
\ee
and where one can obtain the HH wave function by carrying out the path integral, following \cite{witten2020matrix, eberhardt20232d, lin2023revisiting} explicitly. 
We find that the resulting state at large $\phi$, i.e. at late time,  solves the WDW equation of the JT theory, but is  different from the HH state of the JT theory, as expected. 
The general idea of understanding various states in dS space by changing the underlying theory was discussed in \cite{Chakraborty:2023los, Chakraborty:2023yed}.

It was argued in \cite{Maldacena:2019cbz}, see also \cite{Moitra:2022glw} that the SSS double scaled matrix theory, which describes JT gravity in AdS space,  \cite{Saad:2019lba}, can also be used to describe the HH state for JT gravity in dS space. 
We argue in this paper that other states in the JT dS theory can also be mapped to this matrix model, which can therefore serve as a boundary hologram for the bulk theory. The states in the bulk which are mapped, 
say in the expanding branch, have a   coefficient function $\rho(M)$, defined in eq.\eqref{genex}, with  support only over positive values of $M$. Similarly for 
states in  the contracting branch, the  states which are mapped have a coefficient function,  ${\tilde \rho}$, defined in eq.\eqref{gencont}, having support only over positive values of $M$.  These states therefore correspond to the big bang/big crunch set  of solutions, see section \ref{Jtdsrevsum}. 
The map  allows us to  associate the matrix theory norm and  Hilbert space with such  bulk states in both these branches. This norm and Hilbert space are different from what one obtains after the canonical quantisation procedure on constant dilaton hypersurfaces, described above. 

It is known that  the matrix theory accounts for  topology changing amplitudes, which are interpreted in the JT dS theory  as the creation/ annihilation of one or several universes from/to nothing, or the tunelling of collapsing universes to expanding ones, for the HH state,\cite{Maldacena:2019cbz}. It is perhaps not surprising then that the Hilbert space obtained by the map to matrix theory is different from the one obtained in the  canonical quantisation  procedure, where one imposes  norm preserving boundary conditions in the single universe sector alone. 


A conserved quantity, $M$, defined in eq.\eqref{defMa} plays an important role in our analysis, both for  JT dS  and the  more general cases. We loosely refer to it as the mass in this paper.
A black hole in dS space  corresponds to a geometry with a positive and fixed value of $M$.  An interesting feature of the HH  wave function  in the JT theory is that it can be expressed  as a sum over cosmological spacetimes, each of which is   an  orbifold of a black hole geometry of some mass, $M$. The coefficient function $\rho(M)$, defined as mentioned above in eq.\eqref{genex}, which gives the amplitude to be in such a geometry, is equal to the entropy of the cosmological horizon in the corresponding mass $M$ black hole. 

We find that such a relation persists in several other cases with a deformed potential as well, suggesting an interesting general connection between the HH state and black holes in dS. When the path integral can be done exactly for example for the cases, eq.(\ref{formexp}), this can be explicitly verified\footnote{There are some subtleties with the definition of the exponential operators, $e^{-\alpha_i \phi}$,  since they are composite operators, \cite{witten2020matrix}, as is also discussed in section \ref{condefpot}.}.
In addition, for a wide class of potentials, meeting a few  conditions, as discussed in section \ref{HHwf}, we show that a similar  relation is true at least in the semi-classical limit. 


A parameter $S_0$ in two dimensional gravity, defined in eq.\eqref{sojt},  determines the relative importance of various topologies which need to be summed while calculating amplitudes in the path integral formalism. Restricting to the one universe sector, while canonically quantising gravity,  is justified  from the path integral  perspective when we are in the $S_0\rightarrow \infty$ limit. When JT theory is obtained by dimensional reduction from a near-extremal higher dimensional Nariai black hole, $S_0$ is the entropy of the Nariai black hole ( upto a factor of $2$).
From this higher dimensional perspective it might not seem so surprising  then that we obtained an infinite number of states after canonical quantisation in the single universe sector, since we have  in effect taken $S_0\rightarrow \infty$. 
%
%
%
%
%

In the double scaled limit of the SSS matrix model, referred to above,  the rank of the matrix, $L\rightarrow \infty$. In section \ref{fincom} we make some tentative remarks on the case when the rank is large but finite (or similarly for an SYK model when the number of flavours, $N$, is  large but finite). 
If a modified version of JT theory in the bulk continues to be dual to the finite rank case, the  rank $L$, would equal $e^{S_0}$,  and as we discuss in section \ref{fincom}, this could lead to an understanding of deSitter entropy in terms of the states of the  matrix theory.

%
%
%
%
%

Let us end by drawing attention to two papers which appeared while this manuscript was being completed, \cite{Alonso-Monsalve:2024oii},\cite{Held:2024rmg}. 
These papers pointed out that there was an  additional branch of classical solutions in JT theory which was not included in the discussion in \cite{nanda2023jt}.
The canonical quantisation we discuss here also does not apply to this additional branch, see section \ref{Jtdsrevsum} for additional comments. It would be worth extending the discussion to include these cases as well. 
\cite{Held:2024rmg} also has an illuminating discussion on various aspects of canonical quantisation, including various group invariant norms and other choices for the physical clock, obtained, e.g., by choosing constant extrinsic curvature hypersurfaces. It also discusses a connection of the group invariant norm with the norm that arises from the path integral.  While the discussion is mostly in the context of the mini superspace approximation, the ideas are general and can be adapted for the choice of the Wheeler DeWitt operator made in this paper as well and will surely help further elucidate the properties of the JT theory and other cosmological models. 

{This paper is organised as follows. In the next section \ref{Jtdsrevsum} we review the canonical quantisation procedure followed in \cite{nanda2023jt}. In section \ref{addcom} we discuss some aspects of the quantization of JT gravity which was not explored in \cite{nanda2023jt}. In section \ref{JTMM} we discuss how more general states in JT theory can be mapped to the matrix model. From section \ref{pdp} onwards we will focus on general potentials $U(\phi)$. In section \ref{pdp} we discuss the classical properties and in the following section \ref{canquan} we discuss the canonical quantization. In section \ref{condefpot} we discuss the deformed JT potential with exponential correction and in section \ref{HHwf} we discuss the no-boundary wavefunctions for a class of potentials. In section \ref{fincom} we discuss the possible non-perturbative completion of JT theory using a double scaled matrix model at finite rank. In section \ref{conc} we summarize our discussions and discuss possible further investigations. Some additional related calculations are presented in the appendices.  

Apart from the key references cited above and in various places in the following discussion, here are some additional important references that pertain to dS, general potential and matrix models and so on, \cite{Grumiller:2002nm,Fernandes:2019ige, Betzios:2020nry, Stanford:2020qhm, Anninos:2021ene, Anninos:2022ujl, Witten:2022xxp, Jensen:2023eza, Balasubramanian:2023xyd, Cotler:2024xzz, Kruthoff:2024gxc, Kaushal:2024xob, Anninos:2024iwf,Kolchmeyer:2024fly,Bossi:2024ffa, Blommaert:2024whf,Hertog:2024shf,Buchmuller:2024ksd,Heller:2024ldz,Honda:2024hdr}. While this work was being completed, we were made aware of a study \cite{Iizuka:2025vkl} of the ideas discussed in section \ref{HHwf}, for potentials which are asymptotically AdS with a dS bubble inside. }

\section{JT gravity in dS Space: Discussion and Review }
\label{Jtdsrevsum}
In this section we review the discussion of JT gravity in dS space. The canonical quantisation of this theory was discussed in 
\cite{Hennauxjt,Iliesiu:2020zld,nanda2023jt}, see also \cite{Maldacena:2019cbz} and more recently \cite{Alonso-Monsalve:2024oii} and \cite{Held:2024rmg}. 
Our discussion will mostly be along the lines of \cite{nanda2023jt}. 

The action for JT gravity in dS space is given by 
\begin{equation}
	\label{actionjt}
	S_{\text{JT}}= \frac{-i}{2} \pqty{\int d^2 x\,\sqrt{-g}\, (\phi R- 2\phi)-2\int_{bdy}\sqrt{-\gamma}\phi K }
\end{equation}
where $\phi$ is the dilaton. 

There is an addition term 
\be
\label{sojt}
S_{\text{top}}= \frac{S_0}{4\pi}\left( \int \sqrt{-g} R-2\int  \sqrt{-\gamma}K\right)
\ee
which is a total derivative and which is important in the quantum theory. {When JT gravity in dS is obtained by dimensional reduction from a near extremal black hole in higher dimension, $S_0$ is related to the higher dimensional gravitational entropy.}

Some classical solutions of JT gravity we consider are locally of the form
\be
\label{classjtab}
ds^2=-{dr^2\over r^2-M}+(r^2-M) {dx^2\over A^2}
\ee
with the dilaton being 
\be
\label{valdilab2}
\phi=r
\ee
We will periodically identify $x\sim x+1$. This gives a two parameter worth of solutions, specified by $M,A$. 
We take $A>0$ and $M\in [-\infty,\infty]$.
The parameter $M$ can be expressed as 
\be
\label{defMa}
M=(\nabla \phi)^2+\phi^2
\ee
From the equations of the motion it follows that $M$ is a constant satisfying 
$\partial_r M=\partial_x M=0$.
This constant which we will loosely call the ``mass", will play an important role in the subsequent discussion. 

Note, one only wants to carry out the periodic identification along a space-like direction, this requires 
\be
\label{condaas}
r^2>M. 
\ee

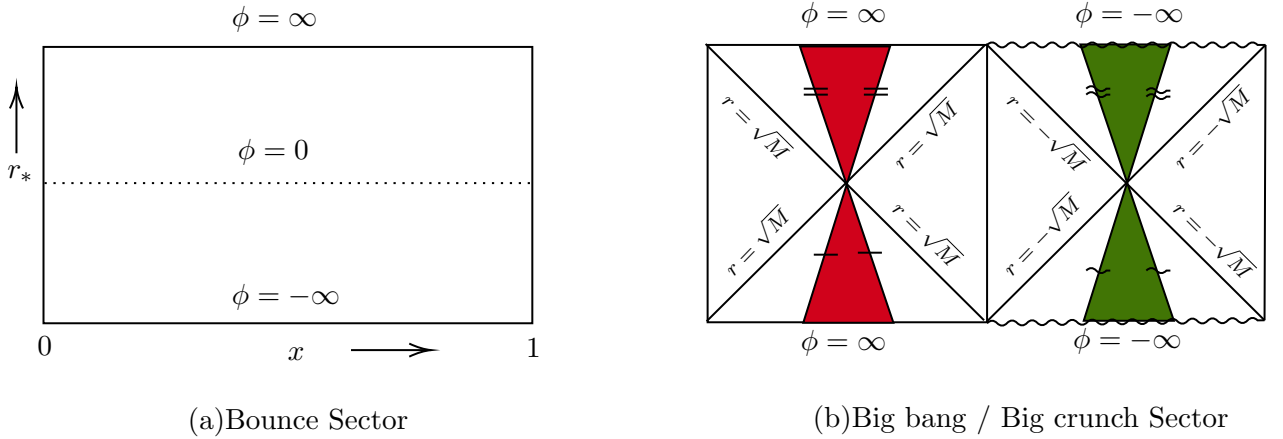
\begin{figure}
	
	\tikzset{every picture/.style={line width=0.75pt}} 
	
	\begin{tikzpicture}[x=0.75pt,y=0.75pt,yscale=-1,xscale=1]
		
		\draw   (49.5,201.5) -- (293.5,201.5) -- (293.5,340.5) -- (49.5,340.5) -- cycle ;
		\draw  [dash pattern={on 0.84pt off 2.51pt}]  (50,270) -- (293,270) ;
		\draw    (36,255) -- (36,223.5) ;
		\draw [shift={(36,221.5)}, rotate = 90] [color={rgb, 255:red, 0; green, 0; blue, 0 }  ][line width=0.75]    (10.93,-3.29) .. controls (6.95,-1.4) and (3.31,-0.3) .. (0,0) .. controls (3.31,0.3) and (6.95,1.4) .. (10.93,3.29)   ;
		\draw    (203,354) -- (241,354) ;
		\draw [shift={(243,354)}, rotate = 180] [color={rgb, 255:red, 0; green, 0; blue, 0 }  ][line width=0.75]    (10.93,-3.29) .. controls (6.95,-1.4) and (3.31,-0.3) .. (0,0) .. controls (3.31,0.3) and (6.95,1.4) .. (10.93,3.29)   ;
		\draw   (381,200.5) -- (520.5,200.5) -- (520.5,340) -- (381,340) -- cycle ;
		\draw    (380,200) -- (519,339) ;
		\draw    (519.5,200.5) -- (380.5,339.5) ;
		\draw    (520,200) -- (659,339) ;
		\draw    (659.5,200.5) -- (520.5,339.5) ;
		\draw    (659.5,200.5) -- (659,339) ;
		\draw  [fill={rgb, 255:red, 208; green, 2; blue, 27 }  ,fill opacity=1 ] (450.42,270.75) -- (473.75,340.25) -- (428.75,340.25) -- cycle ;
		\draw  [fill={rgb, 255:red, 208; green, 2; blue, 27 }  ,fill opacity=1 ] (450.42,270.75) -- (427.08,201.25) -- (472.08,201.25) -- cycle ;
		\draw  [fill={rgb, 255:red, 65; green, 117; blue, 5 }  ,fill opacity=1 ] (590.42,269.75) -- (613.75,339.25) -- (568.75,339.25) -- cycle ;
		\draw  [fill={rgb, 255:red, 65; green, 117; blue, 5 }  ,fill opacity=1 ] (590.42,269.75) -- (567.08,200.25) -- (612.08,200.25) -- cycle ;
		\draw    (520.5,200.5) .. controls (522.17,198.83) and (523.83,198.83) .. (525.5,200.5) .. controls (527.17,202.17) and (528.83,202.17) .. (530.5,200.5) .. controls (532.17,198.83) and (533.83,198.83) .. (535.5,200.5) .. controls (537.17,202.17) and (538.83,202.17) .. (540.5,200.5) .. controls (542.17,198.83) and (543.83,198.83) .. (545.5,200.5) .. controls (547.17,202.17) and (548.83,202.17) .. (550.5,200.5) .. controls (552.17,198.83) and (553.83,198.83) .. (555.5,200.5) .. controls (557.17,202.17) and (558.83,202.17) .. (560.5,200.5) .. controls (562.17,198.83) and (563.83,198.83) .. (565.5,200.5) .. controls (567.17,202.17) and (568.83,202.17) .. (570.5,200.5) .. controls (572.17,198.83) and (573.83,198.83) .. (575.5,200.5) .. controls (577.17,202.17) and (578.83,202.17) .. (580.5,200.5) .. controls (582.17,198.83) and (583.83,198.83) .. (585.5,200.5) .. controls (587.17,202.17) and (588.83,202.17) .. (590.5,200.5) .. controls (592.17,198.83) and (593.83,198.83) .. (595.5,200.5) .. controls (597.17,202.17) and (598.83,202.17) .. (600.5,200.5) .. controls (602.17,198.83) and (603.83,198.83) .. (605.5,200.5) .. controls (607.17,202.17) and (608.83,202.17) .. (610.5,200.5) .. controls (612.17,198.83) and (613.83,198.83) .. (615.5,200.5) .. controls (617.17,202.17) and (618.83,202.17) .. (620.5,200.5) .. controls (622.17,198.83) and (623.83,198.83) .. (625.5,200.5) .. controls (627.17,202.17) and (628.83,202.17) .. (630.5,200.5) .. controls (632.17,198.83) and (633.83,198.83) .. (635.5,200.5) .. controls (637.17,202.17) and (638.83,202.17) .. (640.5,200.5) .. controls (642.17,198.83) and (643.83,198.83) .. (645.5,200.5) .. controls (647.17,202.17) and (648.83,202.17) .. (650.5,200.5) .. controls (652.17,198.83) and (653.83,198.83) .. (655.5,200.5) -- (659.5,200.5) -- (659.5,200.5) ;
		\draw    (520.5,340.5) .. controls (522.15,338.82) and (523.82,338.8) .. (525.5,340.45) .. controls (527.19,342.1) and (528.85,342.08) .. (530.5,340.39) .. controls (532.15,338.71) and (533.82,338.69) .. (535.5,340.34) .. controls (537.19,341.99) and (538.85,341.97) .. (540.5,340.28) .. controls (542.15,338.6) and (543.82,338.58) .. (545.5,340.23) .. controls (547.18,341.88) and (548.85,341.86) .. (550.5,340.18) .. controls (552.15,338.49) and (553.81,338.47) .. (555.5,340.12) .. controls (557.18,341.77) and (558.85,341.75) .. (560.5,340.07) .. controls (562.15,338.38) and (563.81,338.36) .. (565.5,340.01) .. controls (567.18,341.66) and (568.85,341.64) .. (570.5,339.96) .. controls (572.15,338.27) and (573.81,338.25) .. (575.5,339.9) .. controls (577.18,341.55) and (578.85,341.53) .. (580.5,339.85) .. controls (582.15,338.17) and (583.82,338.15) .. (585.5,339.8) .. controls (587.19,341.45) and (588.85,341.43) .. (590.5,339.74) .. controls (592.15,338.06) and (593.82,338.04) .. (595.5,339.69) .. controls (597.19,341.34) and (598.85,341.32) .. (600.5,339.63) .. controls (602.15,337.95) and (603.82,337.93) .. (605.5,339.58) .. controls (607.18,341.23) and (608.84,341.21) .. (610.49,339.53) .. controls (612.14,337.84) and (613.8,337.82) .. (615.49,339.47) .. controls (617.17,341.12) and (618.84,341.1) .. (620.49,339.42) .. controls (622.14,337.73) and (623.8,337.71) .. (625.49,339.36) .. controls (627.17,341.01) and (628.84,340.99) .. (630.49,339.31) .. controls (632.14,337.62) and (633.8,337.6) .. (635.49,339.25) .. controls (637.17,340.9) and (638.84,340.88) .. (640.49,339.2) .. controls (642.14,337.52) and (643.81,337.5) .. (645.49,339.15) .. controls (647.18,340.8) and (648.84,340.78) .. (650.49,339.09) .. controls (652.14,337.41) and (653.81,337.39) .. (655.49,339.04) -- (659,339) -- (659,339) ;
		\draw    (429,222.5) -- (441,222.5)(429,225.5) -- (441,225.5) ;
		\draw    (459,222.5) -- (471,222.5)(459,225.5) -- (471,225.5) ;
		\draw    (570,222.5) .. controls (571.67,220.83) and (573.33,220.83) .. (575,222.5) .. controls (576.67,224.17) and (578.33,224.17) .. (580,222.5) -- (582,222.5) -- (582,222.5)(570,225.5) .. controls (571.67,223.83) and (573.33,223.83) .. (575,225.5) .. controls (576.67,227.17) and (578.33,227.17) .. (580,225.5) -- (582,225.5) -- (582,225.5) ;
		\draw    (600,223.5) .. controls (601.67,221.83) and (603.33,221.83) .. (605,223.5) .. controls (606.67,225.17) and (608.33,225.17) .. (610,223.5) -- (612,223.5) -- (612,223.5)(600,226.5) .. controls (601.67,224.83) and (603.33,224.83) .. (605,226.5) .. controls (606.67,228.17) and (608.33,228.17) .. (610,226.5) -- (612,226.5) -- (612,226.5) ;
		\draw    (434,306) -- (446,306) ;
		\draw    (456,305) -- (468,305) ;
		\draw    (570,314) .. controls (571.67,312.33) and (573.33,312.33) .. (575,314) .. controls (576.67,315.67) and (578.33,315.67) .. (580,314) -- (582,314) -- (582,314) ;
		\draw    (600,314) .. controls (601.67,312.33) and (603.33,312.33) .. (605,314) .. controls (606.67,315.67) and (608.33,315.67) .. (610,314) -- (612,314) -- (612,314) ;
		
		\draw (145,246.4) node [anchor=north west][inner sep=0.75pt]    {$\phi =0$};
		\draw (142,179.4) node [anchor=north west][inner sep=0.75pt]    {$\phi =\infty $};
		\draw (142,319.4) node [anchor=north west][inner sep=0.75pt]    {$\phi =-\infty $};
		\draw (30,259.4) node [anchor=north west][inner sep=0.75pt]    {$r_{\ast }$};
		\draw (170,351.4) node [anchor=north west][inner sep=0.75pt]    {$x$};
		\draw (45,345.4) node [anchor=north west][inner sep=0.75pt]    {$0$};
		\draw (289,346.4) node [anchor=north west][inner sep=0.75pt]    {$1$};
		\draw (120,381) node [anchor=north west][inner sep=0.75pt]   [align=left] {(a)Bounce Sector};
		\draw (426,177.4) node [anchor=north west][inner sep=0.75pt]    {$\phi =\infty $};
		\draw (426,342.4) node [anchor=north west][inner sep=0.75pt]    {$\phi =\infty $};
		\draw (562,341.4) node [anchor=north west][inner sep=0.75pt]    {$\phi =-\infty $};
		\draw (563,177.4) node [anchor=north west][inner sep=0.75pt]    {$\phi =-\infty $};
		\draw (394.72,220.23) node [anchor=north west][inner sep=0.75pt]  [font=\scriptsize,rotate=-45]  {$r=\sqrt{M}$};
		\draw (384.61,310.45) node [anchor=north west][inner sep=0.75pt]  [font=\scriptsize,rotate=-313.89]  {$r=\sqrt{M}$};
		\draw (468.61,256.45) node [anchor=north west][inner sep=0.75pt]  [font=\scriptsize,rotate=-313.89]  {$r=\sqrt{M}$};
		\draw (479.72,274.23) node [anchor=north west][inner sep=0.75pt]  [font=\scriptsize,rotate=-45]  {$r=\sqrt{M}$};
		\draw (536.72,220.23) node [anchor=north west][inner sep=0.75pt]  [font=\scriptsize,rotate=-45]  {$r=-\sqrt{M}$};
		\draw (619.72,274.23) node [anchor=north west][inner sep=0.75pt]  [font=\scriptsize,rotate=-45]  {$r=-\sqrt{M}$};
		\draw (523.61,310.45) node [anchor=north west][inner sep=0.75pt]  [font=\scriptsize,rotate=-313.89]  {$r=-\sqrt{M}$};
		\draw (608.61,256.45) node [anchor=north west][inner sep=0.75pt]  [font=\scriptsize,rotate=-313.89]  {$r=-\sqrt{M}$};
		\draw (432,380) node [anchor=north west][inner sep=0.75pt]   [align=left] {(b)Big bang / Big crunch Sector};

	\end{tikzpicture}
	\caption{Penrose diagram for (a)Bounce Sector and (b) Big Bang/ Big Crunch Sector  }
	\label{dsfig}
\end{figure}

The solutions with $M<0$ and $M>0$ are quite different.  
Global dS lies in the branch $M<0$. In this case  eq.(\ref{condaas}) is automatically met and the radial coordinate can be extended to go from $[-\infty,\infty]$ to obtain a global solution with a Penrose diagram of the form shown in  fig.\ref{dsfig} (a).  The ratio of the length in the horizontal and vertical directions in the Penrose diagram is determined by the ratio $M/A^2$. In these solutions the universe contracts from infinite size in the far past $r<0$,  to a minimum size and then expands back again to infinite size asymptotically. The dilaton goes from being $-\infty$ in the far past to $\infty$ in the far future. Alternatively one can choose to run time backwards by taking $r\rightarrow -r$. Following nomenclature introduced in \cite{Held:2024rmg} we will call this the bounce sector. 

{ For $M>0$  we have to restrict the solution to the region where eq.(\ref{condaas}) is met.  Keeping only the  region $r\ge \sqrt{M}$ one retains  the spacetime outside the cosmological horizons, after a spatial identification. This is shown as the red shaded region in fig.\ref{dsfig}(b). The spacetime  has an orbifold singularity at the bifurcate point where the two cosmological horizons meet. The global solution is obtained by joining two regions described by the metric and dilaton in eq.(\ref{classjtab}) and eq.(\ref{valdilab2}) at the orbifold point $r=\sqrt{M}$. It corresponds to a contracting universe which reaches the orbifold singularity at $r=\sqrt{M}$ and then expands again to asymptotic infinity along the expanding branch. The dilaton is positive everywhere and goes to infinity asymptotically. 
One can also choose to include the region\footnote{Note  in \cite{nanda2023jt} only the region outside the black hole horizon was included.} $r<-\sqrt{M}$, this region lies in the interior of the black hole horizon. Compactifying the spatial coordinate will then leave us with part of spacetime shown as the green shaded region in fig.\ref{dsfig}(b). One then has another orbifold singularity which lies at the bifurcate point of the black hole horizons. In drawing the figure we have taken the minimum value of $\phi$ to be $-\infty$.}

We will refer to this branch as the big bang/big crunch branch solutions. 
The region $ r>\sqrt{M}$ in the future Milne wdge will be called the expanding branch, and for $r>\sqrt{M}$ in the past Milne wedge will be called the contracting branch. 
The region for $r<-\sqrt{M} $ inside the Black hole/White hole region will be called the Black hole (BH) and White hole(WH)  branches respectively. See fig.\ref{dsfig}(b). 

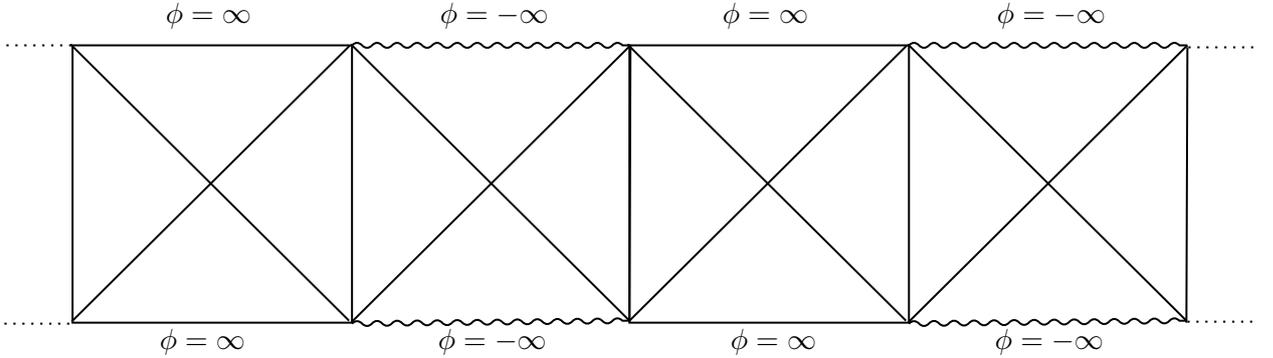
\begin{figure}
	\begin{center}

		\tikzset{every picture/.style={line width=0.75pt}} 
		
		\begin{tikzpicture}[x=0.75pt,y=0.75pt,yscale=-1,xscale=1]
			
			\draw   (46.5,122.5) -- (186,122.5) -- (186,262) -- (46.5,262) -- cycle ;
			\draw    (45.5,122) -- (184.5,261) ;
			\draw    (185,122.5) -- (46,261.5) ;
			\draw    (185.5,122) -- (324.5,261) ;
			\draw    (325,122.5) -- (186,261.5) ;
			\draw    (325,122.5) -- (324.5,261) ;
			\draw    (186,122.5) .. controls (187.67,120.83) and (189.33,120.83) .. (191,122.5) .. controls (192.67,124.17) and (194.33,124.17) .. (196,122.5) .. controls (197.67,120.83) and (199.33,120.83) .. (201,122.5) .. controls (202.67,124.17) and (204.33,124.17) .. (206,122.5) .. controls (207.67,120.83) and (209.33,120.83) .. (211,122.5) .. controls (212.67,124.17) and (214.33,124.17) .. (216,122.5) .. controls (217.67,120.83) and (219.33,120.83) .. (221,122.5) .. controls (222.67,124.17) and (224.33,124.17) .. (226,122.5) .. controls (227.67,120.83) and (229.33,120.83) .. (231,122.5) .. controls (232.67,124.17) and (234.33,124.17) .. (236,122.5) .. controls (237.67,120.83) and (239.33,120.83) .. (241,122.5) .. controls (242.67,124.17) and (244.33,124.17) .. (246,122.5) .. controls (247.67,120.83) and (249.33,120.83) .. (251,122.5) .. controls (252.67,124.17) and (254.33,124.17) .. (256,122.5) .. controls (257.67,120.83) and (259.33,120.83) .. (261,122.5) .. controls (262.67,124.17) and (264.33,124.17) .. (266,122.5) .. controls (267.67,120.83) and (269.33,120.83) .. (271,122.5) .. controls (272.67,124.17) and (274.33,124.17) .. (276,122.5) .. controls (277.67,120.83) and (279.33,120.83) .. (281,122.5) .. controls (282.67,124.17) and (284.33,124.17) .. (286,122.5) .. controls (287.67,120.83) and (289.33,120.83) .. (291,122.5) .. controls (292.67,124.17) and (294.33,124.17) .. (296,122.5) .. controls (297.67,120.83) and (299.33,120.83) .. (301,122.5) .. controls (302.67,124.17) and (304.33,124.17) .. (306,122.5) .. controls (307.67,120.83) and (309.33,120.83) .. (311,122.5) .. controls (312.67,124.17) and (314.33,124.17) .. (316,122.5) .. controls (317.67,120.83) and (319.33,120.83) .. (321,122.5) -- (325,122.5) -- (325,122.5) ;
			\draw    (186,262.5) .. controls (187.65,260.82) and (189.32,260.8) .. (191,262.45) .. controls (192.69,264.1) and (194.35,264.08) .. (196,262.39) .. controls (197.65,260.71) and (199.32,260.69) .. (201,262.34) .. controls (202.69,263.99) and (204.35,263.97) .. (206,262.28) .. controls (207.65,260.6) and (209.32,260.58) .. (211,262.23) .. controls (212.68,263.88) and (214.35,263.86) .. (216,262.18) .. controls (217.65,260.49) and (219.31,260.47) .. (221,262.12) .. controls (222.68,263.77) and (224.35,263.75) .. (226,262.07) .. controls (227.65,260.38) and (229.31,260.36) .. (231,262.01) .. controls (232.68,263.66) and (234.35,263.64) .. (236,261.96) .. controls (237.65,260.27) and (239.31,260.25) .. (241,261.9) .. controls (242.68,263.55) and (244.35,263.53) .. (246,261.85) .. controls (247.65,260.17) and (249.32,260.15) .. (251,261.8) .. controls (252.69,263.45) and (254.35,263.43) .. (256,261.74) .. controls (257.65,260.06) and (259.32,260.04) .. (261,261.69) .. controls (262.69,263.34) and (264.35,263.32) .. (266,261.63) .. controls (267.65,259.95) and (269.32,259.93) .. (271,261.58) .. controls (272.68,263.23) and (274.34,263.21) .. (275.99,261.53) .. controls (277.64,259.84) and (279.3,259.82) .. (280.99,261.47) .. controls (282.67,263.12) and (284.34,263.1) .. (285.99,261.42) .. controls (287.64,259.73) and (289.3,259.71) .. (290.99,261.36) .. controls (292.67,263.01) and (294.34,262.99) .. (295.99,261.31) .. controls (297.64,259.62) and (299.3,259.6) .. (300.99,261.25) .. controls (302.67,262.9) and (304.34,262.88) .. (305.99,261.2) .. controls (307.64,259.52) and (309.31,259.5) .. (310.99,261.15) .. controls (312.68,262.8) and (314.34,262.78) .. (315.99,261.09) .. controls (317.64,259.41) and (319.31,259.39) .. (320.99,261.04) -- (324.5,261) -- (324.5,261) ;
			\draw   (324.5,122.5) -- (464,122.5) -- (464,262) -- (324.5,262) -- cycle ;
			\draw    (323.5,122) -- (462.5,261) ;
			\draw    (463,122.5) -- (324,261.5) ;
			\draw    (463.5,122) -- (602.5,261) ;
			\draw    (603,122.5) -- (464,261.5) ;
			\draw    (464,122.5) .. controls (465.67,120.83) and (467.33,120.83) .. (469,122.5) .. controls (470.67,124.17) and (472.33,124.17) .. (474,122.5) .. controls (475.67,120.83) and (477.33,120.83) .. (479,122.5) .. controls (480.67,124.17) and (482.33,124.17) .. (484,122.5) .. controls (485.67,120.83) and (487.33,120.83) .. (489,122.5) .. controls (490.67,124.17) and (492.33,124.17) .. (494,122.5) .. controls (495.67,120.83) and (497.33,120.83) .. (499,122.5) .. controls (500.67,124.17) and (502.33,124.17) .. (504,122.5) .. controls (505.67,120.83) and (507.33,120.83) .. (509,122.5) .. controls (510.67,124.17) and (512.33,124.17) .. (514,122.5) .. controls (515.67,120.83) and (517.33,120.83) .. (519,122.5) .. controls (520.67,124.17) and (522.33,124.17) .. (524,122.5) .. controls (525.67,120.83) and (527.33,120.83) .. (529,122.5) .. controls (530.67,124.17) and (532.33,124.17) .. (534,122.5) .. controls (535.67,120.83) and (537.33,120.83) .. (539,122.5) .. controls (540.67,124.17) and (542.33,124.17) .. (544,122.5) .. controls (545.67,120.83) and (547.33,120.83) .. (549,122.5) .. controls (550.67,124.17) and (552.33,124.17) .. (554,122.5) .. controls (555.67,120.83) and (557.33,120.83) .. (559,122.5) .. controls (560.67,124.17) and (562.33,124.17) .. (564,122.5) .. controls (565.67,120.83) and (567.33,120.83) .. (569,122.5) .. controls (570.67,124.17) and (572.33,124.17) .. (574,122.5) .. controls (575.67,120.83) and (577.33,120.83) .. (579,122.5) .. controls (580.67,124.17) and (582.33,124.17) .. (584,122.5) .. controls (585.67,120.83) and (587.33,120.83) .. (589,122.5) .. controls (590.67,124.17) and (592.33,124.17) .. (594,122.5) .. controls (595.67,120.83) and (597.33,120.83) .. (599,122.5) -- (603,122.5) -- (603,122.5) ;
			\draw    (464,262.5) .. controls (465.65,260.82) and (467.32,260.8) .. (469,262.45) .. controls (470.69,264.1) and (472.35,264.08) .. (474,262.39) .. controls (475.65,260.71) and (477.32,260.69) .. (479,262.34) .. controls (480.69,263.99) and (482.35,263.97) .. (484,262.28) .. controls (485.65,260.6) and (487.32,260.58) .. (489,262.23) .. controls (490.68,263.88) and (492.35,263.86) .. (494,262.18) .. controls (495.65,260.49) and (497.31,260.47) .. (499,262.12) .. controls (500.68,263.77) and (502.35,263.75) .. (504,262.07) .. controls (505.65,260.38) and (507.31,260.36) .. (509,262.01) .. controls (510.68,263.66) and (512.35,263.64) .. (514,261.96) .. controls (515.65,260.27) and (517.31,260.25) .. (519,261.9) .. controls (520.68,263.55) and (522.35,263.53) .. (524,261.85) .. controls (525.65,260.17) and (527.32,260.15) .. (529,261.8) .. controls (530.69,263.45) and (532.35,263.43) .. (534,261.74) .. controls (535.65,260.06) and (537.32,260.04) .. (539,261.69) .. controls (540.69,263.34) and (542.35,263.32) .. (544,261.63) .. controls (545.65,259.95) and (547.32,259.93) .. (549,261.58) .. controls (550.68,263.23) and (552.34,263.21) .. (553.99,261.53) .. controls (555.64,259.84) and (557.3,259.82) .. (558.99,261.47) .. controls (560.67,263.12) and (562.34,263.1) .. (563.99,261.42) .. controls (565.64,259.73) and (567.3,259.71) .. (568.99,261.36) .. controls (570.67,263.01) and (572.34,262.99) .. (573.99,261.31) .. controls (575.64,259.62) and (577.3,259.6) .. (578.99,261.25) .. controls (580.67,262.9) and (582.34,262.88) .. (583.99,261.2) .. controls (585.64,259.52) and (587.31,259.5) .. (588.99,261.15) .. controls (590.68,262.8) and (592.34,262.78) .. (593.99,261.09) .. controls (595.64,259.41) and (597.31,259.39) .. (598.99,261.04) -- (602.5,261) -- (602.5,261) ;
			\draw    (603,123.5) -- (602.5,262) ;
			\draw  [dash pattern={on 0.84pt off 2.51pt}]  (13.5,122.5) -- (46.5,122.5) ;
			\draw  [dash pattern={on 0.84pt off 2.51pt}]  (12.5,262.5) -- (45.5,262.5) ;
			\draw  [dash pattern={on 0.84pt off 2.51pt}]  (603.5,123.5) -- (636.5,123.5) ;
			\draw  [dash pattern={on 0.84pt off 2.51pt}]  (603,261.5) -- (636,261.5) ;
			
			\draw (91.5,99.4) node [anchor=north west][inner sep=0.75pt]    {$\phi =\infty $};
			\draw (227.5,263.4) node [anchor=north west][inner sep=0.75pt]    {$\phi =-\infty $};
			\draw (228.5,99.4) node [anchor=north west][inner sep=0.75pt]    {$\phi =-\infty $};
			\draw (369.5,99.4) node [anchor=north west][inner sep=0.75pt]    {$\phi =\infty $};
			\draw (505.5,263.4) node [anchor=north west][inner sep=0.75pt]    {$\phi =-\infty $};
			\draw (506.5,99.4) node [anchor=north west][inner sep=0.75pt]    {$\phi =-\infty $};
			\draw (88.5,263.4) node [anchor=north west][inner sep=0.75pt]    {$\phi =\infty $};
			\draw (373.5,263.4) node [anchor=north west][inner sep=0.75pt]    {$\phi =\infty $};

		\end{tikzpicture}
		
	\end{center}
	\caption{Maximally extended JT dS Penrose diagram}
	\label{maxpen}
\end{figure}

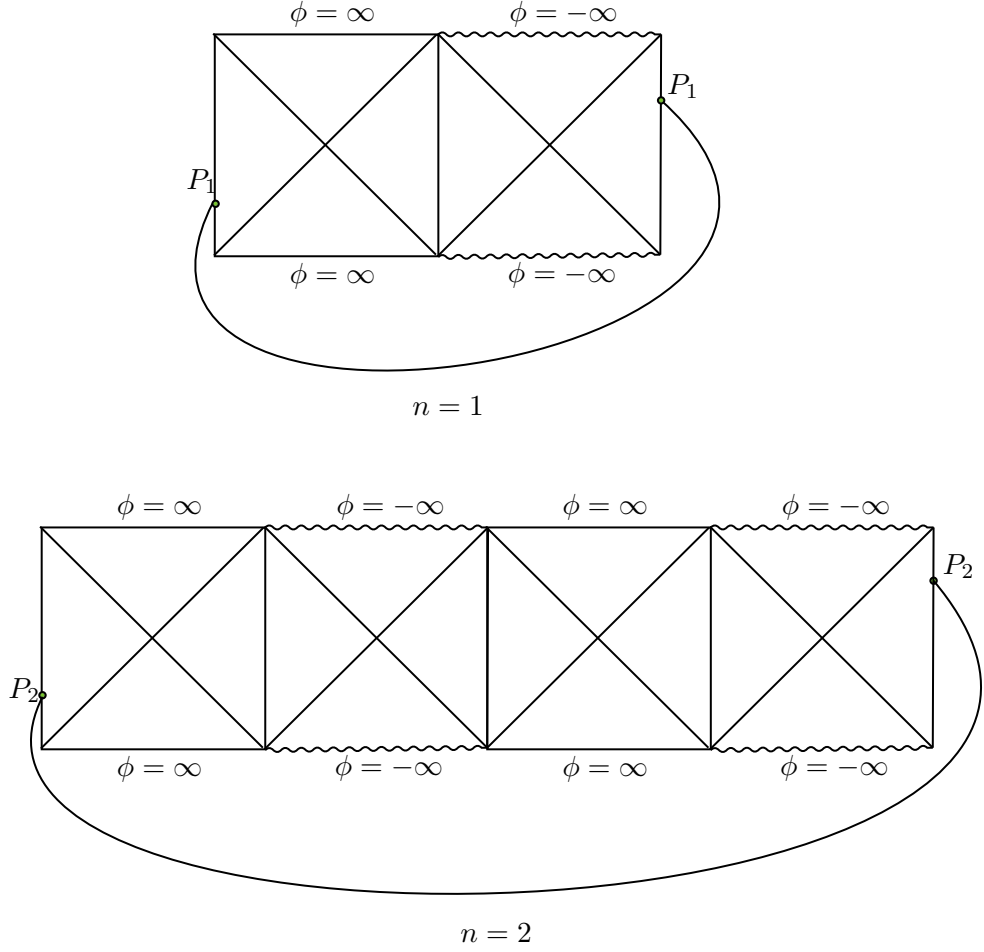
\begin{figure}
	
	\begin{center}
	
	
	\tikzset {_4kyftxquw/.code = {\pgfsetadditionalshadetransform{ \pgftransformshift{\pgfpoint{89.1 bp } { -128.7 bp }  }  \pgftransformscale{1.32 }  }}}
	\pgfdeclareradialshading{_ppss41v7f}{\pgfpoint{-72bp}{104bp}}{rgb(0bp)=(0.49,0.83,0.13);
		rgb(0bp)=(0.49,0.83,0.13);
		rgb(25bp)=(0.4,0.48,0.44);
		rgb(400bp)=(0.4,0.48,0.44)}
	
	
	\tikzset {_pcrl6z1s6/.code = {\pgfsetadditionalshadetransform{ \pgftransformshift{\pgfpoint{89.1 bp } { -128.7 bp }  }  \pgftransformscale{1.32 }  }}}
	\pgfdeclareradialshading{_qbe8aevyc}{\pgfpoint{-72bp}{104bp}}{rgb(0bp)=(0.49,0.83,0.13);
		rgb(0bp)=(0.49,0.83,0.13);
		rgb(25bp)=(0.4,0.48,0.44);
		rgb(400bp)=(0.4,0.48,0.44)}
	
	
	\tikzset {_te5bwuyw7/.code = {\pgfsetadditionalshadetransform{ \pgftransformshift{\pgfpoint{89.1 bp } { -128.7 bp }  }  \pgftransformscale{1.32 }  }}}
	\pgfdeclareradialshading{_7ppad0jbm}{\pgfpoint{-72bp}{104bp}}{rgb(0bp)=(0.49,0.83,0.13);
		rgb(0bp)=(0.49,0.83,0.13);
		rgb(25bp)=(0.4,0.48,0.44);
		rgb(400bp)=(0.4,0.48,0.44)}
	
	
	\tikzset {_z06t8ggo7/.code = {\pgfsetadditionalshadetransform{ \pgftransformshift{\pgfpoint{89.1 bp } { -128.7 bp }  }  \pgftransformscale{1.32 }  }}}
	\pgfdeclareradialshading{_6qu8sik8x}{\pgfpoint{-72bp}{104bp}}{rgb(0bp)=(0.49,0.83,0.13);
		rgb(0bp)=(0.49,0.83,0.13);
		rgb(25bp)=(0.4,0.48,0.44);
		rgb(400bp)=(0.4,0.48,0.44)}
	\tikzset{every picture/.style={line width=0.75pt}} 
	
	\begin{tikzpicture}[x=0.6pt,y=0.6pt,yscale=-1,xscale=1]
		
		\draw   (151.5,50.5) -- (291,50.5) -- (291,190) -- (151.5,190) -- cycle ;
		\draw    (150.5,50) -- (289.5,189) ;
		\draw    (290,50.5) -- (151,189.5) ;
		\draw    (290.5,50) -- (429.5,189) ;
		\draw    (430,50.5) -- (291,189.5) ;
		\draw    (430,50.5) -- (429.5,189) ;
		\draw    (291,50.5) .. controls (292.67,48.83) and (294.33,48.83) .. (296,50.5) .. controls (297.67,52.17) and (299.33,52.17) .. (301,50.5) .. controls (302.67,48.83) and (304.33,48.83) .. (306,50.5) .. controls (307.67,52.17) and (309.33,52.17) .. (311,50.5) .. controls (312.67,48.83) and (314.33,48.83) .. (316,50.5) .. controls (317.67,52.17) and (319.33,52.17) .. (321,50.5) .. controls (322.67,48.83) and (324.33,48.83) .. (326,50.5) .. controls (327.67,52.17) and (329.33,52.17) .. (331,50.5) .. controls (332.67,48.83) and (334.33,48.83) .. (336,50.5) .. controls (337.67,52.17) and (339.33,52.17) .. (341,50.5) .. controls (342.67,48.83) and (344.33,48.83) .. (346,50.5) .. controls (347.67,52.17) and (349.33,52.17) .. (351,50.5) .. controls (352.67,48.83) and (354.33,48.83) .. (356,50.5) .. controls (357.67,52.17) and (359.33,52.17) .. (361,50.5) .. controls (362.67,48.83) and (364.33,48.83) .. (366,50.5) .. controls (367.67,52.17) and (369.33,52.17) .. (371,50.5) .. controls (372.67,48.83) and (374.33,48.83) .. (376,50.5) .. controls (377.67,52.17) and (379.33,52.17) .. (381,50.5) .. controls (382.67,48.83) and (384.33,48.83) .. (386,50.5) .. controls (387.67,52.17) and (389.33,52.17) .. (391,50.5) .. controls (392.67,48.83) and (394.33,48.83) .. (396,50.5) .. controls (397.67,52.17) and (399.33,52.17) .. (401,50.5) .. controls (402.67,48.83) and (404.33,48.83) .. (406,50.5) .. controls (407.67,52.17) and (409.33,52.17) .. (411,50.5) .. controls (412.67,48.83) and (414.33,48.83) .. (416,50.5) .. controls (417.67,52.17) and (419.33,52.17) .. (421,50.5) .. controls (422.67,48.83) and (424.33,48.83) .. (426,50.5) -- (430,50.5) -- (430,50.5) ;
		\draw    (291,190.5) .. controls (292.65,188.82) and (294.32,188.8) .. (296,190.45) .. controls (297.69,192.1) and (299.35,192.08) .. (301,190.39) .. controls (302.65,188.71) and (304.32,188.69) .. (306,190.34) .. controls (307.69,191.99) and (309.35,191.97) .. (311,190.28) .. controls (312.65,188.6) and (314.32,188.58) .. (316,190.23) .. controls (317.68,191.88) and (319.35,191.86) .. (321,190.18) .. controls (322.65,188.49) and (324.31,188.47) .. (326,190.12) .. controls (327.68,191.77) and (329.35,191.75) .. (331,190.07) .. controls (332.65,188.38) and (334.31,188.36) .. (336,190.01) .. controls (337.68,191.66) and (339.35,191.64) .. (341,189.96) .. controls (342.65,188.27) and (344.31,188.25) .. (346,189.9) .. controls (347.68,191.55) and (349.35,191.53) .. (351,189.85) .. controls (352.65,188.17) and (354.32,188.15) .. (356,189.8) .. controls (357.69,191.45) and (359.35,191.43) .. (361,189.74) .. controls (362.65,188.06) and (364.32,188.04) .. (366,189.69) .. controls (367.69,191.34) and (369.35,191.32) .. (371,189.63) .. controls (372.65,187.95) and (374.32,187.93) .. (376,189.58) .. controls (377.68,191.23) and (379.34,191.21) .. (380.99,189.53) .. controls (382.64,187.84) and (384.3,187.82) .. (385.99,189.47) .. controls (387.67,191.12) and (389.34,191.1) .. (390.99,189.42) .. controls (392.64,187.73) and (394.3,187.71) .. (395.99,189.36) .. controls (397.67,191.01) and (399.34,190.99) .. (400.99,189.31) .. controls (402.64,187.62) and (404.3,187.6) .. (405.99,189.25) .. controls (407.67,190.9) and (409.34,190.88) .. (410.99,189.2) .. controls (412.64,187.52) and (414.31,187.5) .. (415.99,189.15) .. controls (417.68,190.8) and (419.34,190.78) .. (420.99,189.09) .. controls (422.64,187.41) and (424.31,187.39) .. (425.99,189.04) -- (429.5,189) -- (429.5,189) ;
		\draw    (150,157) .. controls (57,344) and (610,255) .. (430,92) ;
		\path  [shading=_ppss41v7f,_4kyftxquw] (150,157) .. controls (150,155.9) and (150.9,155) .. (152,155) .. controls (153.1,155) and (154,155.9) .. (154,157) .. controls (154,158.1) and (153.1,159) .. (152,159) .. controls (150.9,159) and (150,158.1) .. (150,157) -- cycle ; 
		\draw   (150,157) .. controls (150,155.9) and (150.9,155) .. (152,155) .. controls (153.1,155) and (154,155.9) .. (154,157) .. controls (154,158.1) and (153.1,159) .. (152,159) .. controls (150.9,159) and (150,158.1) .. (150,157) -- cycle ; 
		
		\path  [shading=_qbe8aevyc,_pcrl6z1s6] (428,92) .. controls (428,90.9) and (428.9,90) .. (430,90) .. controls (431.1,90) and (432,90.9) .. (432,92) .. controls (432,93.1) and (431.1,94) .. (430,94) .. controls (428.9,94) and (428,93.1) .. (428,92) -- cycle ; 
		\draw   (428,92) .. controls (428,90.9) and (428.9,90) .. (430,90) .. controls (431.1,90) and (432,90.9) .. (432,92) .. controls (432,93.1) and (431.1,94) .. (430,94) .. controls (428.9,94) and (428,93.1) .. (428,92) -- cycle ; 
		
		\draw   (43.5,360.5) -- (183,360.5) -- (183,500) -- (43.5,500) -- cycle ;
		\draw    (42.5,360) -- (181.5,499) ;
		\draw    (182,360.5) -- (43,499.5) ;
		\draw    (182.5,360) -- (321.5,499) ;
		\draw    (322,360.5) -- (183,499.5) ;
		\draw    (322,360.5) -- (321.5,499) ;
		\draw    (183,360.5) .. controls (184.67,358.83) and (186.33,358.83) .. (188,360.5) .. controls (189.67,362.17) and (191.33,362.17) .. (193,360.5) .. controls (194.67,358.83) and (196.33,358.83) .. (198,360.5) .. controls (199.67,362.17) and (201.33,362.17) .. (203,360.5) .. controls (204.67,358.83) and (206.33,358.83) .. (208,360.5) .. controls (209.67,362.17) and (211.33,362.17) .. (213,360.5) .. controls (214.67,358.83) and (216.33,358.83) .. (218,360.5) .. controls (219.67,362.17) and (221.33,362.17) .. (223,360.5) .. controls (224.67,358.83) and (226.33,358.83) .. (228,360.5) .. controls (229.67,362.17) and (231.33,362.17) .. (233,360.5) .. controls (234.67,358.83) and (236.33,358.83) .. (238,360.5) .. controls (239.67,362.17) and (241.33,362.17) .. (243,360.5) .. controls (244.67,358.83) and (246.33,358.83) .. (248,360.5) .. controls (249.67,362.17) and (251.33,362.17) .. (253,360.5) .. controls (254.67,358.83) and (256.33,358.83) .. (258,360.5) .. controls (259.67,362.17) and (261.33,362.17) .. (263,360.5) .. controls (264.67,358.83) and (266.33,358.83) .. (268,360.5) .. controls (269.67,362.17) and (271.33,362.17) .. (273,360.5) .. controls (274.67,358.83) and (276.33,358.83) .. (278,360.5) .. controls (279.67,362.17) and (281.33,362.17) .. (283,360.5) .. controls (284.67,358.83) and (286.33,358.83) .. (288,360.5) .. controls (289.67,362.17) and (291.33,362.17) .. (293,360.5) .. controls (294.67,358.83) and (296.33,358.83) .. (298,360.5) .. controls (299.67,362.17) and (301.33,362.17) .. (303,360.5) .. controls (304.67,358.83) and (306.33,358.83) .. (308,360.5) .. controls (309.67,362.17) and (311.33,362.17) .. (313,360.5) .. controls (314.67,358.83) and (316.33,358.83) .. (318,360.5) -- (322,360.5) -- (322,360.5) ;
		\draw    (183,500.5) .. controls (184.65,498.82) and (186.32,498.8) .. (188,500.45) .. controls (189.69,502.1) and (191.35,502.08) .. (193,500.39) .. controls (194.65,498.71) and (196.32,498.69) .. (198,500.34) .. controls (199.69,501.99) and (201.35,501.97) .. (203,500.28) .. controls (204.65,498.6) and (206.32,498.58) .. (208,500.23) .. controls (209.68,501.88) and (211.35,501.86) .. (213,500.18) .. controls (214.65,498.49) and (216.31,498.47) .. (218,500.12) .. controls (219.68,501.77) and (221.35,501.75) .. (223,500.07) .. controls (224.65,498.38) and (226.31,498.36) .. (228,500.01) .. controls (229.68,501.66) and (231.35,501.64) .. (233,499.96) .. controls (234.65,498.27) and (236.31,498.25) .. (238,499.9) .. controls (239.68,501.55) and (241.35,501.53) .. (243,499.85) .. controls (244.65,498.17) and (246.32,498.15) .. (248,499.8) .. controls (249.69,501.45) and (251.35,501.43) .. (253,499.74) .. controls (254.65,498.06) and (256.32,498.04) .. (258,499.69) .. controls (259.69,501.34) and (261.35,501.32) .. (263,499.63) .. controls (264.65,497.95) and (266.32,497.93) .. (268,499.58) .. controls (269.68,501.23) and (271.34,501.21) .. (272.99,499.53) .. controls (274.64,497.84) and (276.3,497.82) .. (277.99,499.47) .. controls (279.67,501.12) and (281.34,501.1) .. (282.99,499.42) .. controls (284.64,497.73) and (286.3,497.71) .. (287.99,499.36) .. controls (289.67,501.01) and (291.34,500.99) .. (292.99,499.31) .. controls (294.64,497.62) and (296.3,497.6) .. (297.99,499.25) .. controls (299.67,500.9) and (301.34,500.88) .. (302.99,499.2) .. controls (304.64,497.52) and (306.31,497.5) .. (307.99,499.15) .. controls (309.68,500.8) and (311.34,500.78) .. (312.99,499.09) .. controls (314.64,497.41) and (316.31,497.39) .. (317.99,499.04) -- (321.5,499) -- (321.5,499) ;
		\path  [shading=_7ppad0jbm,_te5bwuyw7] (42,466) .. controls (42,464.9) and (42.9,464) .. (44,464) .. controls (45.1,464) and (46,464.9) .. (46,466) .. controls (46,467.1) and (45.1,468) .. (44,468) .. controls (42.9,468) and (42,467.1) .. (42,466) -- cycle ; 
		\draw   (42,466) .. controls (42,464.9) and (42.9,464) .. (44,464) .. controls (45.1,464) and (46,464.9) .. (46,466) .. controls (46,467.1) and (45.1,468) .. (44,468) .. controls (42.9,468) and (42,467.1) .. (42,466) -- cycle ; 
		
		\path  [shading=_6qu8sik8x,_z06t8ggo7] (598,394) .. controls (598,392.9) and (598.9,392) .. (600,392) .. controls (601.1,392) and (602,392.9) .. (602,394) .. controls (602,395.1) and (601.1,396) .. (600,396) .. controls (598.9,396) and (598,395.1) .. (598,394) -- cycle ; 
		\draw   (598,394) .. controls (598,392.9) and (598.9,392) .. (600,392) .. controls (601.1,392) and (602,392.9) .. (602,394) .. controls (602,395.1) and (601.1,396) .. (600,396) .. controls (598.9,396) and (598,395.1) .. (598,394) -- cycle ; 
		
		\draw   (321.5,360.5) -- (461,360.5) -- (461,500) -- (321.5,500) -- cycle ;
		\draw    (320.5,360) -- (459.5,499) ;
		\draw    (460,360.5) -- (321,499.5) ;
		\draw    (460.5,360) -- (599.5,499) ;
		\draw    (600,360.5) -- (461,499.5) ;
		\draw    (600,360.5) -- (599.5,499) ;
		\draw    (461,360.5) .. controls (462.67,358.83) and (464.33,358.83) .. (466,360.5) .. controls (467.67,362.17) and (469.33,362.17) .. (471,360.5) .. controls (472.67,358.83) and (474.33,358.83) .. (476,360.5) .. controls (477.67,362.17) and (479.33,362.17) .. (481,360.5) .. controls (482.67,358.83) and (484.33,358.83) .. (486,360.5) .. controls (487.67,362.17) and (489.33,362.17) .. (491,360.5) .. controls (492.67,358.83) and (494.33,358.83) .. (496,360.5) .. controls (497.67,362.17) and (499.33,362.17) .. (501,360.5) .. controls (502.67,358.83) and (504.33,358.83) .. (506,360.5) .. controls (507.67,362.17) and (509.33,362.17) .. (511,360.5) .. controls (512.67,358.83) and (514.33,358.83) .. (516,360.5) .. controls (517.67,362.17) and (519.33,362.17) .. (521,360.5) .. controls (522.67,358.83) and (524.33,358.83) .. (526,360.5) .. controls (527.67,362.17) and (529.33,362.17) .. (531,360.5) .. controls (532.67,358.83) and (534.33,358.83) .. (536,360.5) .. controls (537.67,362.17) and (539.33,362.17) .. (541,360.5) .. controls (542.67,358.83) and (544.33,358.83) .. (546,360.5) .. controls (547.67,362.17) and (549.33,362.17) .. (551,360.5) .. controls (552.67,358.83) and (554.33,358.83) .. (556,360.5) .. controls (557.67,362.17) and (559.33,362.17) .. (561,360.5) .. controls (562.67,358.83) and (564.33,358.83) .. (566,360.5) .. controls (567.67,362.17) and (569.33,362.17) .. (571,360.5) .. controls (572.67,358.83) and (574.33,358.83) .. (576,360.5) .. controls (577.67,362.17) and (579.33,362.17) .. (581,360.5) .. controls (582.67,358.83) and (584.33,358.83) .. (586,360.5) .. controls (587.67,362.17) and (589.33,362.17) .. (591,360.5) .. controls (592.67,358.83) and (594.33,358.83) .. (596,360.5) -- (600,360.5) -- (600,360.5) ;
		\draw    (461,500.5) .. controls (462.65,498.82) and (464.32,498.8) .. (466,500.45) .. controls (467.69,502.1) and (469.35,502.08) .. (471,500.39) .. controls (472.65,498.71) and (474.32,498.69) .. (476,500.34) .. controls (477.69,501.99) and (479.35,501.97) .. (481,500.28) .. controls (482.65,498.6) and (484.32,498.58) .. (486,500.23) .. controls (487.68,501.88) and (489.35,501.86) .. (491,500.18) .. controls (492.65,498.49) and (494.31,498.47) .. (496,500.12) .. controls (497.68,501.77) and (499.35,501.75) .. (501,500.07) .. controls (502.65,498.38) and (504.31,498.36) .. (506,500.01) .. controls (507.68,501.66) and (509.35,501.64) .. (511,499.96) .. controls (512.65,498.27) and (514.31,498.25) .. (516,499.9) .. controls (517.68,501.55) and (519.35,501.53) .. (521,499.85) .. controls (522.65,498.17) and (524.32,498.15) .. (526,499.8) .. controls (527.69,501.45) and (529.35,501.43) .. (531,499.74) .. controls (532.65,498.06) and (534.32,498.04) .. (536,499.69) .. controls (537.69,501.34) and (539.35,501.32) .. (541,499.63) .. controls (542.65,497.95) and (544.32,497.93) .. (546,499.58) .. controls (547.68,501.23) and (549.34,501.21) .. (550.99,499.53) .. controls (552.64,497.84) and (554.3,497.82) .. (555.99,499.47) .. controls (557.67,501.12) and (559.34,501.1) .. (560.99,499.42) .. controls (562.64,497.73) and (564.3,497.71) .. (565.99,499.36) .. controls (567.67,501.01) and (569.34,500.99) .. (570.99,499.31) .. controls (572.64,497.62) and (574.3,497.6) .. (575.99,499.25) .. controls (577.67,500.9) and (579.34,500.88) .. (580.99,499.2) .. controls (582.64,497.52) and (584.31,497.5) .. (585.99,499.15) .. controls (587.68,500.8) and (589.34,500.78) .. (590.99,499.09) .. controls (592.64,497.41) and (594.31,497.39) .. (595.99,499.04) -- (599.5,499) -- (599.5,499) ;
		\draw    (44,467) .. controls (-49,654) and (796,629) .. (600,394) ;
		
		\draw (196.5,27.4) node [anchor=north west][inner sep=0.75pt]    {$\phi =\infty $};
		\draw (196.5,192.4) node [anchor=north west][inner sep=0.75pt]    {$\phi =\infty $};
		\draw (332.5,191.4) node [anchor=north west][inner sep=0.75pt]    {$\phi =-\infty $};
		\draw (333.5,27.4) node [anchor=north west][inner sep=0.75pt]    {$\phi =-\infty $};
		\draw (88.5,337.4) node [anchor=north west][inner sep=0.75pt]    {$\phi =\infty $};
		\draw (88.5,502.4) node [anchor=north west][inner sep=0.75pt]    {$\phi =\infty $};
		\draw (224.5,501.4) node [anchor=north west][inner sep=0.75pt]    {$\phi =-\infty $};
		\draw (225.5,337.4) node [anchor=north west][inner sep=0.75pt]    {$\phi =-\infty $};
		\draw (366.5,337.4) node [anchor=north west][inner sep=0.75pt]    {$\phi =\infty $};
		\draw (366.5,502.4) node [anchor=north west][inner sep=0.75pt]    {$\phi =\infty $};
		\draw (502.5,501.4) node [anchor=north west][inner sep=0.75pt]    {$\phi =-\infty $};
		\draw (503.5,337.4) node [anchor=north west][inner sep=0.75pt]    {$\phi =-\infty $};
		\draw (273,276.4) node [anchor=north west][inner sep=0.75pt]    {$n=1$};
		\draw (303,607.4) node [anchor=north west][inner sep=0.75pt]    {$n=2$};
		\draw (132,133.4) node [anchor=north west][inner sep=0.75pt]    {$P_{1}$};
		\draw (432,73.4) node [anchor=north west][inner sep=0.75pt]    {$P_{1}$};
		\draw (21,453.4) node [anchor=north west][inner sep=0.75pt]    {$P_{2}$};
		\draw (604,375.4) node [anchor=north west][inner sep=0.75pt]    {$P_{2}$};

	\end{tikzpicture}
\end{center}
	\caption{Penrose diagrams for $n=1$ and $n=2$ solutions.}
	\label{ndsfig}
\end{figure}

A careful analysis of the global properties of solutions with  an identification along a spatial direction, of the kind we are considering here,  was carried out in \cite{Alonso-Monsalve:2024oii} and \cite{Held:2024rmg}. 
They emphasized that there was a third   branch of solutions which was not discussed in the earlier work, \cite{nanda2023jt}. This can be described as follows. The maximally extended black hole geometry in dS has the Penrose diagram shown in fig.\ref{maxpen} and consists of alternating cosmological regions - asymptotic to $\cal I^{\pm}$, where the dilaton $\rightarrow \infty$  and black hole/white hole regions- where the dilaton obtains its minimum negative value or asymptotically goes to $-\infty$.  To obtain a compact spatial direction one can include one cosmological and black hole region and identify the end points after a translation along the time direction at the two boundaries as shown in fig.\ref{ndsfig} (a) (this is called the $n=1$ solution in \cite{Held:2024rmg}). Or consider two such geometries, the first one consisting of a cosmological and black hole region  followed by another one containing a second  cosmological and black hole region and identify points along the boundary of the $1st$ cosmological region with that of the $2nd$ black hole region, after a suitable translation along the boundary. This gives the $n=2$ solution in this branch as shown in fig.\ref{ndsfig}(b). And similarly for any $n\ge 1$.

Our discussion of canonical quantisation will only pertain to the first two branches- the bounce and big bang/big crunch branches -  and not the third branch mentioned in the last paragraph. 
These two branches can be foliated by slices of constant values for the dilaton (or its conjugate variable - the extrinsic curvature), unlike the third branch, and we will assume that such a foliation is possible in our discussion of canonical quantisation. 

\subsection{Canonical Quantisation}
\label{canquanti}
Next we turn to a discussion of  canonical quantisation of the the theory. As mentioned above, we will mostly follows the discussion in \cite{nanda2023jt}.
Here we will discuss and clarify some of the salient points which arise in the canonical quantisation procedure.  
The action of JT gravity  was given in eq.\eqref{actionjt} reproduced here for convenience,
\begin{equation}
	S_{\text{JT}}= \frac{-i}{2} \pqty{\int d^2 x\,\sqrt{-g}\, (\phi R- 2\phi)-2\int_{bdy}\sqrt{-\gamma}\phi K }.
\end{equation}

In general the two dimensional metric can be written as 
\be
\label{genfma}
ds^2=-N^2(x,t) dt^2+g_{1}(x,t)(dx+N_{\perp}(x,t)dt)^2
\ee
(with $x\simeq x+1$)
.
The Hamiltonian and Momentum constraints are obtained from the equations of motion of the lapse and shift functions and take the form,
\begin{align}
	\mathcal{H} &= - \frac{\dot{\phi}}{N^2} \left(\frac{N_{\perp}}{2\sqrt{g_1}}{g_1'}+\sqrt{g_1} N_{\perp}'-\frac{\dot{g}_1}{2\sqrt{g_1}}\right)  -\left(\frac{\phi'}{\sqrt{g_1}}\right)' - \sqrt{g_1} \phi \nonumber \\
	& - \frac{\phi'}{N^2} \left(-\sqrt{g_1}N_{\perp}N_{\perp}'+\frac{N_{\perp}}{2\sqrt{g_1}}\dot{g}_1-\frac{N_{\perp}^2}{2\sqrt{g_1}}g_1'\right) \label{hamcons} \\
	\mathcal{P} &= \frac{\dot{\phi}g_1'}{2N\sqrt{g_1}}-\left(\frac{\dot{\phi}\sqrt{g_1}}{N} \right)'+\frac{\phi'\dot{g}_1}{2N\sqrt{g_1}} + \left(\frac{\phi'\sqrt{g_1}}{N} \right)' N_{\perp} - \frac{\phi'g_1' N_{\perp}}{N\sqrt{g_1}} \label{momcons}
\end{align}
In quantising the theory we will work in ADM gauge and set the shift function 
\be
\label{condsh}
N_\perp=0
\ee and the lapse function 
\be
\label{condlapse}
N=1.
\ee
The remaining degrees of freedom are
the metric component $g_1$, dilaton $\phi$, and their conjugate momenta, $\pi_{g_1}$ and $\pi_\phi$ respectively which are given by,. 
\begin{equation}
	\pi_{\phi}=-\frac{\dot{g}_1}{2N\sqrt{g_1}},\quad \pi_{g_1}=-\frac{\dot{\phi}}{2N\sqrt{g_1}}\label{cmominnpz}
\end{equation}
Eq.\eqref{hamcons} and eq.\eqref{momcons} then become
\begin{align}
	& \mathcal{H}=2\pi_\phi \sqrt{g_1}\pi_{g_1}-\left(\frac{\phi'}{\sqrt{g_1}}\right)'- \sqrt{g_1} \phi\label{hamaa} \\
	& \mathcal{P} = 2g_1\pi_{g_1}'+\pi_{g_1}g_1'-\pi_\phi \phi' \label{momaa}
\end{align}
In the quantum theory these become operator constraints which physical states must satisfy. 
These constraints generate residual gauge transformations which preserve the ADM gauge conditions, eq.\eqref{condsh} and eq.\eqref{condlapse}. 
The Hamiltonian constraint, also called the Wheeler DeWitt equation,  ensures that the wave function is invariant under time reparametrisations and the momentum constraint  ensures that it is invariant under spatial reparametrisations. 

An important issue is to deal with   ambiguities which arise in defining these constraints in the quantum theory. These ambiguities are of two types, see \cite{Hennauxjt} and \cite{nanda2023jt}. There are operator ordering ambiguities, present in both $\mathcal{H}$ and $\mathcal{P}$, and an additional $\delta(0)$ divergences present in $\mathcal{H}$, which arises because $\mathcal{H}$ involves terms with  momenta operators, $\pi_\phi, \pi_{g_1}$,  and the metric $g_1$, with  the same spatial arguments.
As discussed in \cite{nanda2023jt} we are guided in defining the quantum constraints  by the fact that the Hartle Hawking (HH) state can be calculated   exactly by evaluating the path integral. Requiring this  state to be physical imposes significant restriction on how the ambiguities are to be resolved, as we will see below.

In general the wave function in ADM gauge is given by $\Psi[g_1(x), \phi(x)]$, and is a functional of $g_1(x), \phi(x)$.
Using the time reparametrisation invariance of the wave function we choose a hypersurface along which the dilaton is constant taking the value $\phi$.  
We note from the discussion in section \ref{Jtdsrevsum} that while this can be done in the classical solutions for the first two branches it cannot be done for the third branch
(see the discussion above subsection \ref{canquanti}). Our quantisation procedure will also therefore only apply to the first two branches. 

From the spatial reparametrisation invariance of the wave function it then follows that for such a hypersurface $\Psi$ will  only be a function  of  the length of spatial 
hypersurface, $l$, besides depending on $\phi$. 
Thus on surfaces with constant $\phi$  the wave function $\Psi(l,\phi)$ simplifies considerably becoming, instead of a  complicated functional, an ordinary  function of $l,\phi$.

We still have to impose one remaining  condition  on  $\Psi(l,\phi)$. This arises from the Hamiltonian constraint 
and can be understood as follows. 
We can use the spatial reparametrisations to set $g_1(x)$ to be $x$ independent, then  the Hamiltonian  as written in eq.(\ref{hamaa})   becomes
\be
\label{hamts}
{\cal H}=2\sqrt{g_1}\pi_{g_1}\pi_\phi-\sqrt{g_1}\phi
\ee
resulting in the WDW equation
\be
\label{hamab}
[2\sqrt{g_1}\pi_{g_1}\pi_\phi-\sqrt{g_1}\phi]\Psi=0
\ee
Substituting, 
\be
\label{defll}
l=\sqrt{g_1}
\ee
\be
\label{ords}
\pi_{g_1}=- {i\over 2 l } \partial_l
\ee
and 
\be
\label{momp}
\pi_\phi=-i \partial_\phi
\ee
eq.(\ref{hamab})  then takes the form
\be
\label{hama}
[\partial_\phi \partial_l+l \phi ]\Psi(l,\phi)=0
\ee
However it turns out that the Hartle Hawking state, whose wave function can be calculated exactly by path integral techniques, does not satisfy eq.(\ref{hama}).

The reason, as was noted above is that the Hamiltonian constraint suffers from normal ordering ambiguities. We have assumed a specific definition, including an ordering prescription when we wrote the constraint in the form given in eq.(\ref{hamts}). We will now instead choose a definition which leads to the HH state satisfying the resulting Wheeler-DeWitt equation. 

The  expanding branch of the HH wave function was calculated in \cite{Maldacena:2019cbz} using path integral techniques and is given at late time, when $\phi\rightarrow \infty, l\rightarrow \infty$, with 
$l/\phi$ held fixed, to be 
\be
\label{hhlt}
\Psi_{HH}={\cal N} e^{S_0} \left({\phi\over l}\right)^{3/2} 
e^{-{i l \phi }+ { 2i \pi^2 \phi \over   l}}  \ee
where ${\cal N}$ is an overall normalisation.

This state does not meet  eq.(\ref{hama}), as noted above, and instead  satisfies the following equation  
\be
\label{wda}
\left(\partial_l\partial_\phi-{1\over l} \partial_\phi +l\phi\right)\Psi(l,\phi)=0
\ee
Accordingly we will adopt eq.(\ref{wda}) to be the correct form of the WDW equation. 

One can check that eq.(\ref{wda})  arises if we take the Hamiltonian constraint   in the zero mode sector - with $\phi$ and $g_1$ being independent of $x$ -  to be of the form 
\be
\label{hamcs2}
{\cal H}=2g_1\pi_{g_1} { ({\sqrt{g_1}})^{-1}} \pi_\phi -\sqrt{g_1}\phi
\ee
instead of eq.(\ref{hamts}).
Classically the two  constraints eq.\eqref{hamts},\eqref{hamcs2}  are  equivalent  but in the quantum theory they are   different. 

Alternatively, we note that requiring, in the zero mode sector, that $\pi_{g_1}$ is conjugate to $g_1$ and satisfies the relation
\be
\label{relac}
[g_1,\pi_{g_1}]=i
\ee
in the quantum theory,
does not uniquely fix the form of the operator $\pi_{g_1}$. Taking it to be 
\be
\label{formpa}
\pi_{g_1}=-i \del_{g_1}+\frac{i}{2g_1}
\ee 
instead of eq.(\ref{ords}) and keeping the Hamiltonian constraint unchanged, and as given in eq.(\ref{hamts}), also gives us the same WdW equation eq.(\ref{hamcs2}).

Let us end this subsection with a few more  comments. 
Note that  it follows from eq.(\ref{wda})  that
\be
\label{foob}
 {\hat \Psi}={\Psi\over l}
 \ee
  satisfies the equation 
\be
\label{wdwh}
-\partial_l\partial_\phi{\hat \Psi}  -l\phi{\hat \Psi}=0
\ee
This can be recast as the Klein Gordon equation in terms of variables $l^2,\phi^2$, which act as null coordinates, for a particle of mass $1/4$, \cite{Maldacena:2019cbz}. 

The asymptotic form of the expanding branch HH wave function was given in eq.\eqref{hhlt} above. More generally since it solves eq.(\ref{wda})  it takes the form 
\be
\label{defhha}
\Psi_{HH}(l,\phi) ={\cal {\hat N} }e ^{S_0}  { l \phi^2\over  (l^2-4\pi^2)}  H_2^{(2)}(\phi\sqrt{l^2-4\pi^2})
\ee
where $H_2^{(2)}$ denotes a Hankel function of index $2$ of the second kind.



We now turn to exploring solutions to eq.(\ref{wda}) in more detail 

\subsection{Solutions}
\label{cansolsinm}
{\bf Eigenstates of ${\hat M}$}

The operator ${\hat M}$ which corresponds to the classically conserved quantity $M$, defined in eq.\eqref{defMa} above is given by 
\be
\label{defmhat}
{\hat M}=\partial^2_{l} + \phi^2
\ee
This operator satisfies the following commutation relation 
\be
\label{coma1}
[\mathcal{H},\hat{M}]=-\frac{2}{l^2} \mathcal{H}
\ee
where
\begin{equation}
	\mathcal{H} = \partial_l\partial_\phi-{1\over l} \partial_\phi +l\phi \label{Hfin} 
\end{equation}
Thus the subspace of physical states with $\mathcal{H}=0$ is left invariant by the action of ${\hat M}$ and we can find solutions to the WdW equation which are also eigenstates of ${\hat M}$. These take the form 
\be
\label{formaa}
\Psi= e^{\mp i l\sqrt{\phi^2-M}}
\ee
The solution varying like $e^{-il\sqrt{\phi^2-M}}$ corresponds to the expanding branch and $e^{il\sqrt{\phi^2-M}}$ to the contracting branch, see \cite{nanda2023jt}.
There are an infinite number of such solutions, for different values of $M$. 

General Solutions in the expanding and contracting branches are then given by 
\be
\label{genex}
\Psi=\int_{-\infty}^\infty dM \rho(M) e^{-il\sqrt{\phi^2-M}}
\ee
and 
\be
\label{gencont}
\Psi=\int_{-\infty}^\infty dM \tilde{\rho}(M) e^{il\sqrt{\phi^2-M}}
\ee
respectively.

More precisely the solutions above for $M>0$ have a branch cut at $\phi=\pm \sqrt{M}$.
A solution given in eq.(\ref{formaa}), for $\phi^2>M$, will become a linear combination of the solutions $e^{\pm l \sqrt{M-\phi^2}}$ in the region   $\phi^2<M$,  
with continuity at $\phi=\pm \sqrt{M}$ relating the solution in the region $\phi^2>M$ to the solution when  $\phi^2<M$. 
E.g., consider  a solution of the form eq.(\ref{formaa}) in the expanding branch, which when $\phi>\sqrt{M}$ is 
\be
\label{begpsi}
\Psi=e^{-il\sqrt{\phi^2-M}}.
\ee Then in the region  $-\sqrt{M}<\phi<\sqrt{M}$, it becomes
\be
\label{solab}
\Psi(\phi,l)=c_1e^{-l\sqrt{M-\phi^2}}+c_2e^{l \sqrt{M-\phi^2}}
\ee
Continuity at $\phi=\sqrt{M}$ gives $
c_1+c_2=1$.

Continuing it further, in the region $\phi<-\sqrt{M}$ this solution becomes 
\be
\label{linca}
\Psi(\phi,l)=d_1 e^{-il\sqrt{\phi^2-M}} + d_2 e^{il\sqrt{\phi^2-M}}
\ee
with $d_1+d_2=c_1+c_2=1$. 
Note from eq.(\ref{classjtab}), eq.(\ref{valdilab2}) we see  that the region $\phi^2<M$ is classically disallowed, which intuitively agrees with the fact that the wave function is exponentially damped or growing in this region.


\subsection{Rindler Basis} 
Instead of diagonalizing the operator ${\hat M}$, eq.\eqref{defmhat} we can work with eigenstates of another operator 
\be
\label{opdefG3}
G_3=-i (l \partial_l- \phi \partial_\phi)
\ee
This operator was discussed in \cite{Held:2024rmg}.
It is easy to see that $G_3$ commutes with $\mathcal{H}$, given in eq.\eqref{Hfin},
\be
\label{condcom}
[G_3,\mathcal{H}]=0
\ee
Eigenstates of $G_3$ with eigenvalue $2i k$, which also have  $\mathcal{H}=0$, are the ``Rindler" modes discussed in \cite{nanda2023jt}. These are of the form 
\be
\label{rform}
J_{\pm i |k|}(\xi) e^{i k\theta}
\ee
where 
\be
\label{defit}
\xi = l \phi, e^\theta = \frac{\phi}{l}
\ee
and $J_\alpha(x)$ are the Bessel functions with index $\alpha$. 

The Rindler modes constitute a complete set and are a useful  basis  for expanding solutions of the WDW equation\footnote{This is discussed in appendix \ref{detrevsum} and follows
	from the fact that the Rindler modes describe the solutions to the WdW equation in the Milne wedges, with  $l,\phi>0$. }.

Note that we chose constant $\phi$ hypersurfaces in our discussion above. We could have instead chosen constant extrinsic curvature hypersurfaces. Denoting the extrinsic curvature by $k$, 
the wave function  in this case,  $\Psi(k,l)$, would be a function of  $l, k$. Since  the extrinsic curvature is conjugate to the dilaton it is given by the operator
$k=-i\partial_\phi$, in the $\phi$ basis. And one can  obtain   $\Psi(l,k)$ by Fourier transforming $\Psi(l,\phi)$,
\be
\label{ftdefa}
\Psi(l,k)=\int d\phi \,e^{ik \phi}\, \Psi(l,\phi)
\ee
However there is one subtlety that we should be aware of. We are assuming in carrying out the Fourier transform  that the dilaton lies in the range $[-\infty,\infty]$; for a more restricted range the relation between the two representations for $\Psi$ would be more complicated. 
In  our discussion using Rindler modes in section \ref{addcom} and appendix \ref{rindcont} we will see that one can extend the range for the dilaton to be  $[-\infty,\infty]$, in this case the Fourier transform in eq.(\ref{ftdefa})  would indeed be valid. We leave a more complete discussion of the system in the $(l,k)$ representation for the future.

%

\subsection{Quantising in the Rindler Basis}
\label{secrindlerquan}
Having understood the solutions to the WdW equation we would now like to define an inner product and obtain a Hilbert space in which states in the quantum theory lie. 
Noting that  $\Psi $ solves the WdW equation, eq.(\ref{wda}) which is closely related to the Klein Gordon  equation motivates us to take  the norm for a state   on a constant $\phi$ slice to be \cite{nanda2023jt}
\be
\label{normaa}
\langle \hat\Psi, \hat\Psi\rangle =\int_{0}^{\infty} dl \,\,i({\hat\Psi }^*\del_l \hat{\Psi} -\hat{\Psi}\del_l {\hat{\Psi}}^*),
\ee
For the norm to be conserved and no ``leakage" of probability flux to occur,  $C_N$, defined by 
\begin{align}
\mathcal{C}_N\equiv	i ( \hat{\Psi}\del_{\phi}\hat{\Psi}^*{}\,-\hat{\Psi}^*{\del_{\phi}}\hat{\Psi} )\label{cnnorm}
\end{align}
must vanish as $l\rightarrow 0, \infty$. 
The condition that $C_N$ vanishes as $l\rightarrow 0$ in particular imposes significant constraints, see \cite{nanda2023jt} and leads to the constraint that $a(k), b(k)$ must vanish for $k<0$. 
This gives
\be
\label{remapsi}
\Psi=\int_{0}^\infty dk \bigl[a(k) e^{i k \theta}J_{-i |k|}( \xi)+ b(k) e^{-i k \theta}J_{i |k|}( \xi)\bigr ]
\ee
The norm \eqref{normaa} then takes the form 
\be
\label{normac}
\langle \hat\Psi, \hat\Psi\rangle =\frac{2}{\pi}\int_{0}^{\infty} dk \, \sinh(k\pi)\left(|a(k)|^2-|b(k)|^2\right)
\ee
and is clearly independent of $\phi$. 

More generally we define the inner product between two states to be 
\be
\label{inpsi}
	\langle \hat{\Psi}_1, \hat{\Psi}_2 \rangle = \frac{i}{2}\int_{0}^{\infty} dl  (\hat{\Psi}_1^* \partial_l \hat{\Psi}_2-\hat{\Psi}_2 \partial_l \hat{\Psi}_1^*) + \frac{i}{2}\int_{0}^{\infty} dl  (\hat{\Psi}_2^* \partial_l \hat{\Psi}_1-\hat{\Psi}_1 \partial_l \hat{\Psi}_2^*)  
\ee
States in the quantum theory then belong to a Hilbert space corresponding to all solutions of the form eq.(\ref{remapsi}), with norm eq.(\ref{normaa}) and inner product eq.(\ref{inpsi}). 

\section{Singularities,  and Time Reversal in JT Theory}
\label{addcom}

\subsection{Singularities and Their Resolution}
The classical solutions in the big bang/big crunch sector have an orbifold singularity, see fig.\ref{dsfig}. 
One would like to understand how the quantum theory behaves  in the presence of such a singularity. 
In the bounce sector on the other hand the dilaton $\phi$ vanishes at the moment of time symmetry, when the universe has minimum size. 
Since $\phi$ is an overall multiplicative factor in the JT action, its vanishing is analogous to $\hbar\rightarrow \infty$, making the region where the dilaton becomes small highly quantum. Here too, one would like to understand better how the quantum theory behaves. Our discussion here will use the Rindler basis in important ways. 
 
\subsection{Orbifold Singularity}
Consider a solution in the expanding branch which takes the form eq.(\ref{genex}). 
As long as the coefficient function $\rho(M)$ satisfies the condition 
\be
\label{condcoeffa}
\int dM \rho(M)=0
\ee
the solution can be expressed in terms of the Rindler basis with $a(k), b(k)$ being determined by $\rho(M)$, see eq.\eqref{aeq1} and \eqref{beq1} of appendix \ref{minrind}. 
By evolving the state in the Rindler basis - ensuring that no leakage of probability occurs at $l=0$ - one can then obtain the wave function in the contracting branch. 
One  finds that the  contracting branch  solution is non-vanishing and in fact has a coefficient function
with 
\be
\label{contaa}
{\tilde \rho}(M)=\rho(M)
\ee
i.e. with the coefficient function of the contracting branch being equal to that of the expanding branch. 

To understand the physical significance of this result,  consider a state where $\rho(M)$ only had support for $M>0$ so that the universe in the far future was in the forward Milne wedge, see fig.\ref{dsfig}(b). of the big bang/big crunch branch. Going back in time one learns that in the quantum theory  it goes through the orbifold singularity at $l=0$ and re-emerges in the past it in the contracting branch. This universe then  extends all the way to the far past, with a  wave function coefficient function given by eq.(\ref{contaa}). Or to turn time around, a contracting universe in the far past goes through the big bang/big crunch singularity and re-emerges as an expanding solution in the far future. The equality of the coefficient functions in the two branches shows that the form of the wave function in the far past and future are closely related, in fact the same upto phase factors. 
A contracting universe,  described by a  Gaussian wave packet which is close to  classical  in the far past, will   passing through the singularity, unscathed, and emerge into    an expanding universe which also becomes classical in the far future.  See appendix \ref{rindcont} for a discussion with a concrete example.

\subsection{ Extending $\Psi$ to  Negative Values of $\phi$:}

We have not been very explicit about the range that the dilaton takes so far. It is true, as was mentioned in section \ref{Jtdsrevsum}, that when the 
JT theory is obtained by dimensional reduction the dilaton should be bigger than a minimum value and cannot go to $-\infty$. 
However here we  examine  the behaviour of JT theory from the two  dimensional  perspective  without worrying about higher dimensional connections. 
We note before proceeding that the region where $\phi\rightarrow 0$ is particularly interesting, because it is analogous to $\hbar\rightarrow \infty$. This follows from that fact that  $\phi$ is an overall multiplicative factor in the JT action,  eq.\eqref{actionjt}, making $\phi\rightarrow 0$    analogous to $\hbar\rightarrow \infty$.
As a result the region where $\phi$ becomes small should be thought of as a very quantum region. 

As discussed in {appendix \ref{detrevsum}} the Rindler basis which has been constructed above to describe  solutions covers the region $l>0,\phi>0$.
In section \ref{rindcont} we also discuss how the Rindler basis modes can be analytically extended  to  the region $l>0,\phi<0$, by imposing continuity  at $\phi=0$.  In fact as we will see below  is quite a natural thing to do in the quantum theory\footnote{Note  that if we do not extend the range of $\phi$  one would have to impose suitable boundary conditions at $\phi=0$, and the wave function can be smoothly extended past $\phi=0$ in the qantum theory.
	We have not studied which boundary conditions would be consistent. Dirichlet boundary conditions at $l=0$, or requiring $\partial_l {\hat \Psi}$ to vanish, set ${\hat \Psi=0} $ identically  and do {not} give any non-trivial solutions to the WdW equation.}. 

Quite interestingly as the analysis below  reveals, the universe evolves  through the quantum region, transiting from large positive values of $\phi$ to large negative values, in a controlled manner. 

\subsubsection{The Big Bang/Big Crunch and Black Hole/White Hole Branches}

For the big bang/big crunch branch with $M>0$, $\phi<0$  corresponds to the region inside
the black hole, see section \ref{Jtdsrevsum} and fig. \ref{dsfig}(b). See appendix \ref{rindcont} for more details. 
As a result of the continuation mentioned above, to $\phi<0$, one finds that  in this region as well the wave function can be expanded in terms of $M$ eigenstates and is given by 
\be
\label{formabp}
\Psi(l,\phi)=\int dM {\hat \rho}(M) e^{-il\sqrt{\phi^2-M}} + \int dM {\tilde {\hat \rho}}(M) e^{il\sqrt{\phi^2-M}}
\ee
By  studying the asymptotic region $l |\phi|\rightarrow \infty$ keeping $l/|\phi|$ fixed, one learns that ${\hat \rho}$ and ${\tilde {\hat \rho}}$ can be expressed in terms of $\rho$ and are given by 
\begin{equation}
\label{relcaa}
	\hat{\rho}(M)=\rho(M)
\end{equation}
and  
\begin{equation}
\label{relcab}
	\tilde{\hat{\rho}}(M) = \rho(M)
\end{equation}
respectively. The components with coefficient functions ${\hat \rho}(M)$
and ${\tilde {\hat \rho}}(M)$ correspond to the wave function in the black hole and white hole regions respectively.  So we learn that the wave function  in the expanding branch, which we saw above on passing through the orbifold singularity gives rise to a component in  the contracting branch, is also related to non-vanishing components components in the black hole/white hole regions. 

In the classical solution the big bang/big crunch branch and the black hole/white hole branch are disconnected; see fig \ref{dsfig} (b) where they are  denoted by the red and green colored regions  respectively.  One might have therefore intuitively expected that if a component arises in  the black hole/white whole branches due to quantum  effects, it would be suppressed. It is quite remarkable then to find that in the quantum theory  the resulting components in  the black hole/white hole branch have   the same coefficient functions,  eq.(\ref{relcaa}), eq.(\ref{relcab}), without any attenuation!

\subsubsection{The Bounce Branch and The Quantum Region:}

For the bounce solutions which correspond to $M<0$, if we start with the expanding branch with coefficient function $\rho(M)$, eq.(\ref{genex}), now having support for $M<0$,  in the region $\phi>0$, there are also three other branches in the wave function. It is worth recalling that in our terminology, in an expanding branch the momentum conjugate to $\phi$, $\pi_{\phi}<0$ (this is because $\pi_{\phi}=-{\dot l}$). Similarly, for contracting branches $\pi_{\phi}>0$. The three other branches can  be described as follows. 
First, there is the contracting branch with $\phi<0$,  in which  the wave function takes the form
\be
\label{bounceaa}
\Psi(l,\phi)=\int_{M<0} {\hat \rho}(M)e^{-il\sqrt{\phi^2-M}}.
\ee
In addition,  there  is one more    contracting  and expanding branch,  with $\phi>0$ and $\phi<0$ respectively. The wave function takes the form 
\be
\label{bounceba}
\Psi(l,\phi)=\int_{M<0} {\tilde \rho}(M)e^{il\sqrt{\phi^2-M}}
\ee
in this contracting branch and 
\be
\label{bouncebb}
\Psi(l,\phi)=\int_{M<0} {\tilde {\hat \rho}}(M)e^{il\sqrt{\phi^2-M}}.
\ee
in this expanding branch.

%

If we consider a classical solution which in the far future goes to $\phi\rightarrow \infty$, after undergoing a bounce, it will in the far past correspond to a solution in the contracting branch with $\phi\rightarrow -\infty$; this contracting branch corresponds to the coefficient function, $ {\hat \rho}$. The remaining two branches might seem surprising at first, but they  in fact correspond to time reversed components as we explain in more detail below. The component with coefficient function ${\tilde \rho}$, eq.(\ref{gencont}), is the time reversed version of $\rho$, and corresponds to a contracting universe where the dilaton asymptotically goes to $\infty$, while the component with a coefficient function ${\tilde {\hat \rho}}$ is the time reversed version of $ {\hat \rho}$ and  corresponds to an expanding universe where $\phi\rightarrow -\infty$. 

As discussed in Appendix \ref{rindcont} one learns by expanding the wave function in Rindler modes that in the bounce  sector also the coefficient functions of all four branches
are equal, meeting the condition, 
\be
\label{secccc}
\rho(M)={\tilde \rho}={\hat \rho}={\tilde {\hat \rho}}.
\ee

%
%

\subsection{Time Reversal:}

The existence of four branches for the bounce sector is related to the action of  the time reversal transformation on the wave function.  Solutions in the bounce sector correspond to $M<0$. If we take a classical solution in this sector, which in the far past is a contracting universe with $\phi\rightarrow -\infty$, it would correspond  to a wave packet made up of solutions of the form 
\be
\label{type1}
\Psi_1= e^{-il\sqrt{\phi^2-M}}, 
\ee
with the far past being  the region where $ \phi\rightarrow -\infty$ (and also $l\rightarrow \infty$), as discussed above.
Classical evolution would take it  to a wave packet made up of expanding branch solutions of the same form 
\be
\label{type2}
\Psi_2=e^{-il\sqrt{\phi^2-M}},
\ee
with the far future corresponding to $ \phi\rightarrow \infty$.

Time reversal exchanges the future with the past. Acting on the wave function, time reversal,   more correctly CPT, transforms it  by complex conjugation, and exchanges an expanding branch with a contracting  branch solution.
As a result in the far past the time reversed solution is a contracting branch solution of the form, 
\be
\label{type3}
\Psi_3=\Psi_2^*=e^{il\sqrt{\phi^2-M}},
\ee
with  $ \phi\rightarrow \infty$.
And in the far future solutions it is an  expanding branch solution, 
\be
\label{type4}
\Psi_4=\Psi_1^*= e^{il\sqrt{\phi^2-M}}, 
\ee
now with $ \phi\rightarrow -\infty$.
These are the four branches we have mentioned above. 

We see from this explicit discussion that under time reversal the sign of the momentum $\pi_{\phi}$ reverses, as one would expect, and therefore expanding and contracting branches are exchanged with each other. On the other hand the dilaton remains unchanged, and is an even function under time reversal, so that a branch describing   the asymptotic region $\phi\rightarrow \infty$  will go into another branch also with the same asymptotic behaviour for $\phi$. These facts lead to the general conclusion that 
the coefficient functions  $\rho^*$ and ${\tilde \rho}$   will be exchanged under time reversal, 
and similarly  ${\hat \rho}^*$ and ${\tilde {\hat \rho}}$  will be exchanged, in accordance with our discussion above.


In the big bang/ big crunch sector also the four branches we found are exchanged under time reversal amongst each other. It follows from our discussion above that the big bang and big crunch branches get exchanged with each other, and the branches in the black hole and white hole regions are also exchanged, under time reversal.

Finally we note that if the wave function is time reversal or CPT invariant then 
\be
\label{condpsia}
\Psi=\Psi^*
\ee
This leads to the conclusion that 
\be
\label{conclua}
\rho(M)={\tilde \rho}(M)^*
\ee
and
\be
\label{conclub}
{\hat \rho}(M)={\tilde {\hat \rho}}^* (M)
\ee

In general this is different from the condition we got which is that all coefficient functions are equal, eq.(\ref{secccc}). Thus the general wave function  will not be time reversal invariant, except in   special cases where the coefficient functions are real (upto an overall phase).   

\section{JT theory and  the Matrix Model}
\label{JTMM}
It was argued previously, \cite{Maldacena:2019cbz,Moitra:2021uiv,nanda2023jt} that the Hartle Hawking state in JT gravity (for dS space) can be mapped to the SSS doubled scaled matrix model. 
In this section we consider how more general states in the JT theory, obtained after canonical quantisation,  can be mapped to this matrix theory. 

We will construct the map in matrix theory for states in the big bang/big crunch sector, with $M>0$. Moreover our map will 
apply only to the expanding and contracting branches of this sector and not the Forward/Backward branches within the Blackhole/White hole region. 

\subsection{Map For the Hartle Hawking State}
To explain the essential idea let us start with the HH state.  For this state it was argued, based on the  AdS space case, \cite{Saad:2019lba},  that the amplitude, in the expanding branch, $\Psi_{HH}^+(l,\phi)$, in the asymptotic limit, when $l,\phi\rightarrow \infty$, with $l\over \phi$ fixed, 
mapped to the matrix theory as follows:
\be
\label{mapmatth}
\Psi_{HH}^+\leftrightarrow e^{-il\phi} \Tr(e^{i l H  \over 2\phi})
\ee
where $H$ on the RHS denotes the matrix in the SSS model. 
By this  equivalence one means, more precisely, that  the Hartle Hawking wave function for the expanding branch is given by 
\be
\label{mapas}
\Psi_{HH}^+(l,\phi)=e^{-il\phi}\langle \Tr(e^{i l H  \over 2\phi})\rangle
\ee
where the expectation value is to be taken in the matrix theory. 
The density of states in the Matrix model is given by 
\be
\label{dosa}
\rho_{MM}(E)=e^{S_0}{1\over 4 \pi^2} \sinh (2\pi \sqrt{E})
\ee
for $E>0$. Here $S_0$ is the topology counting parameter in matrix theory  which we have set equal to the corresponding coefficient which appears in the JT action in eq.(\ref{sojt}).  
The RHS  in eq.(\ref{mapas}) in terms of this density of states  becomes 
\be
\label{rma}
e^{-il\phi}\langle \Tr(e^{i l H  \over 2\phi})\rangle=e^{S_0} e^{-il\phi} \int_{E>0} dE \rho_{MM}(E) e^{ilE\over 2 \phi}
\ee
which can be shown to agree with eq.(\ref{hhlt}). 
(Note that the factor of $e^{S_0}$ in the density of states arises because   we are working with a one boundary correlator which has Euler character $\chi=1$). 

Comparing with the general solution  to the WdW equation, eq.(\ref{genex}) in the asymptotic limit,  we see that the coefficient function for this state is given by 
\be
\label{compa}
\rho_{HH}(M)=\rho_{MM}(M)
\ee
i.e. it is given by  the matrix model density of states, once  we identify the mass $M$ which appears in eq.(\ref{genex}) with the eigenvalue of the random matrix $E$. 

The contracting branch $\Psi_{HH}^-$ is identified in an analogous way as 
\be
\label{contb}
\Psi_{HH}^-= e^{il\phi}\langle \Tr(e^{-{i l H  \over 2\phi}})\rangle
\ee


Let us now summarise the situation regarding multi-boundary amplitudes.
In the path integral description of JT theory  and also in Matrix theory, such amplitudes which involve  topology changing processes can be  calculated and it has been argued that they agree for the HH state. This   agreement follows from the agreement of similar calculations  in the AdS context.

For example,   the amplitude to produce two expanding universes in the HH state with the first, asymptotically, at  (large time)   $\phi_1$ being of  length $l_1$ and the  second asymptotically at 
$\phi_2$ being of length $l_2$, is given by 
\be
\label{mapb}
A(l_1,\phi_1; l_2,\phi_2)=e^{-il_1\phi_1} e^{-il_2 \phi_2} \langle \Tr(e^{i l_1 H  \over 2\phi_1}) \Tr (e^{i l_2 H  \over 2\phi_2}) \rangle.
\ee
An amplitude for an initially contracting universe,  which asymptotically at $\phi_1$ has length $l_1$, to transition in the far future to an expanding universe, with length $l_2$ at $\phi_2$, is given by  
\be
\label{mapc}
T(l_1,\phi_1; l_2,\phi_2)=e^{il_1\phi_1} e^{-il_2 \phi_2} \langle \Tr(e^{-{i l_1 H  \over 2\phi_1}}) \Tr (e^{i l_2 H  \over 2\phi_2}) \rangle
\ee
More general mixed amplitudes involving several contracting universes transitioning to expanding ones can be computed similarly by computing the expectation values
of suitable multiple traces with each universe being mapped to a matrix trace through the correspondence eq.(\ref{mapmatth}), eq.(\ref{contb}),
as described in \cite{Maldacena:2019cbz} and \cite{Moitra:2021uiv}.


Also we note that the  agreement between the path integral results and  matrix theory continue to hold to all orders in the genus expansion for the general multi-boundary case. 

In our discussion of canonical  quantisation in this paper, in contrast to the path integral or matrix theory,  we have worked only in the one-universe sector. In order to allow for processes where the number of universes can change one would need to carry out a ``third quantisation" of this system which we leave for the future. 

\subsection{Map For Other States}

What about states other than the HH state?
Let us first consider  the single universe sector. In canonical quantisation we saw that a solution to WdW equation  in the expanding branch, asymptotically, becomes 
\be
\label{asafa}
\Psi^+(l,\phi)=e^{-il\phi} \int dM \rho(M) e^{ilM\over 2\phi}
\ee
This can be mapped to a ``generalised trace" in the matrix theory as follows. We express $\rho(M)$ in terms of the matrix model density of states $\rho_{MM}$ by defining the function $f(M)$ such that 
\be
\label{deffm}
\rho(M)=f(M) \rho_{MM}(M)
\ee
Note, we are allowing general complex valued functions $f(M)$ here and as a result the coefficient function will be  a general complex function of $M$.
It is then easy to see that  the state eq.(\ref{asafa}) can be expressed in the matrix theory as 
\be
\label{exmat}
\Psi_f^+(l,\phi)=e^{-il\phi} \langle \Tr(f(H) e^{ilH\over 2 \phi})\rangle 
\ee
The RHS is the generalised trace we referred to above, involving both $f(H)$ and $e^{ilH\over 2 \phi}$ now. 
Note, on the LHS the wave function is denoted by $\Psi^+_f$  to indicate schematically that  its depends on $f(M)$, and that it belongs to the expanding branch .

It is natural to suggest this  identification  between the state in canonical quantisation with a coefficient function that is determined by $f(M)$, eq.(\ref{deffm}) and
the generalised trace 
\be
\label{gentrace}
\Psi^+_f(l,\phi)\leftrightarrow e^{-il\phi}\Tr(e^{ilH\over 2 \phi} f(H))
\ee
continues to hold for multi-boundary correlators involving these more general states as well. We do not know how to compute these in canonical quantisation, as was mentioned,  but the claim would be that  can be computed in the matrix theory after this identificaton.

Similarly in the contracting branch, asymptotically for a state in canonically quantisation with a coefficient function of the form 
\be
\label{coeffaa}
{\tilde \rho}(M) =f(M) \rho_{MM}(M),
\ee
we get an identification 
\be
\label{gentracec}
\Psi^-_f(l,\phi)\leftrightarrow e^{il\phi}\Tr(e^{-{ilH\over 2 \phi}} f(H))
\ee

%
Unlike the HH state there is an incomplete  path integral understanding for the more general states (see next two sections for further discussion). 
Thus the correspondence eq.(\ref{gentrace}), eq.(\ref{gentracec}),  for the more general case  is more conjectural and in particular cannot be checked beyond leading order in the genus expansion for the single boundary case.
Nor is there any check for this correspondence that we could carry out   for the  
multi- boundary case. 

If we do accept  the correspondence,, eq.(\ref{gentrace}), eq.(\ref{gentracec}), then as mentioned, one can  compute topology changing processes for more general  states in the Matrix theory.
Analogous to the HH state one would specify the universe asymptotically, for large $\phi,l$, to be in the state $\Psi_f^\pm$. More generally for the multi boundary case one would specify the $i^{th}$  boundary to be asymptotically in the $\Psi_{f_i}^{\pm}$.   Then topology changing amplitude would be given by the expectation value of the corresponding generalised multi -traces in the matrix theory. 

Two more points need to be made. First, we saw in section \ref{Jtdsrevsum}, see discussion around eq.\eqref{condaas},  that even for the JT dS model with $M>0$ there can be  additional branches in  the wave function, corresponding to the BH and WH  regions inside the Black hole/White hole patches, where the wavefunction is oscillatory and these can  admit asymptotic regions where $\phi\rightarrow -\infty$. We will not discuss here whether a matrix theory description can be given for these branches.
Secondly, there is also the bounce sector, corresponding to $M<0$ states, and  we will also not provide a matrix theory description for such states either here. 

%

\subsection{Hilbert Space and Norm}
\label{hsn}

In the discussion above we considered the SSS random matrix theory. However we can instead take  a single realisation of a typical matrix, rather than a  random average theory\footnote{Instead of the SSS Matrix model if we consider the SYK model with $\psi^q$ couplings, then a typical Hamiltonian would correspond to choosing each $\psi^q$ coupling  by  drawing it from a Gaussian distribution of the required variance, and   then holding them fixed.}.
The density of eigenvalues for such a typical realisation  will not differ, in the large N limit, from that of the random average. 

Now  consider the matrix, which can be thought of as a Hamiltonian, $H$,  to be  an operator acting on a vector space, $V$. Unlike the random matrix case, for a single realisation, each eigenvalue of $H$ is associated with a definite state- the corresponding eigenstate in V (degeneracies can be dealt with in the usual manner).
In contrast, in the random average case, the eigenvectors change depending on which element of the ensemble we are considering. 

This means that in  the single realisation  case we can map states in the bulk  to states in $V$ by identifying a bulk state in the expanding branch, which has eigenvalue $M$ for the 
  mass operator ${\hat M}$, eq.\eqref{defmhat}, with the eigenstate of $H$ in $V$ with  the same eigenvalue $M$. 
  Under the map a state with coefficient function eq.(\ref{deffm}) will then be mapped to the state $\int \rho_{MM}(M) \,\, f(M) \,\, |M\rangle \,\,\,  \in   V$.
  The bulk wave function, asymptotically, for large $\phi,l$ can then be expressed in the matrix theory in a manner similar to  eq.(\ref{exmat}), 
  \be
  \label{exmmats}
  \Psi_f^+=e^{-il\phi} \Tr(f(H) e^{ilH\over 2 \phi}) 
  \ee 
  
  In contrast  to the single instantiation case, if we take  the random average model, there is no map between bulk eigenstates  of ${\hat M}$ and eigenstates of the matrix $H$, since  different instantiations of the Hamiltonian  in the random average ensemble  lead to different eigenvectors. 
  However the correspondence between bulk wave functions and generalised traces continues to be true, eq.(\ref{exmat}),  with the trace on the RHS in eq.(\ref{exmmats}) being replaced by an expectation value for the trace in the RMT.

  We should also mention, before we proceed,  that  once we go beyond the single universe sector and consider topology change, the 
single realisation case is very different from  random matrix theory. In particular, there are no connected multi- boundary components in the single realisation case and, as a result, topology change is absent, unlike the random average case. 

  The vector space  $V$ is endowed with a standard inner product. To describe it consider the typical single instantiation,  finite rank,  case first, where the eigenvalues are $M_i, i=1, \cdots, N$ and the eigenvectors are, $|M_i\rangle \,\, ,i=1, \cdots, N$. A state $|f_1\rangle$ can be expanded as 
  \be
  \label{expst}
  |f_1\rangle=\sum_{i}^N f_1(M_i) \,\, |M_i\rangle
  \ee
  Similarly for $|f_2\rangle$. Then the standard  inner product in $V$ between the two states  $|f_1\rangle, |f_2\rangle$   is given by 
    \be
  \label{inpa}
  \langle f_1|f_2\rangle=\sum_i f_1^*(M_i)f_2(M_i)
  \ee
  In the continuum limit where the density of states is given by $\rho_{MM}$ this becomes
  \be
  \label{inpaf}
 \langle f_1|f_2\rangle=\int dM \rho_{MM}(M) f_1(M)^*f_2(M)
 \ee
 Using the map from the bulk states to states in $V$ we can then associate the same inner product with the corresponding bulk states. 
  
The corresponding norm for a state $|f_1\rangle$ then becomes
\be
\label{inpa2}
\langle f_1|f_1\rangle=\int dM \rho_{MM}(M)|f_1(M)|^2
\ee. 

It is also natural to  consider the same inner product and norm for bulk states when the dual is a RMT theory. Strictly speaking the factor $\rho_{MM}$ on the RHS of eq.(\ref{inpaf}), (\ref{inpa2}), should be the density of eigenstates  in the single instantiation case,  and, $\rho_{MM}$, its average value,   in the RMT case,  but they are the same to leading order,    for large rank and typical single realisations.

A similar map to $V$ can also be constructed for states in the contracting branch and an inner product and norm can also be assigned to these states using this map.

Note that the  inner product we have obtained in this way is different from the one discussed earlier in subsection \ref{secrindlerquan}.
As a result the two bulk Hilbert spaces are also quite different. 
For the KG inner product we defined above,  on constant $\phi$ slices, the boundary conditions led to the expanding branch coefficient function being related, in fact being equal, to that of the contracting branch, eq.\eqref{contaa}.
In contrast, here a state in the expanding branch (with no component in the contracting branch), or vice-versa,  is an allowed state in the Hilbert space. 

Note also that though the Hilbert spaces constructed in the manner mentioned above, by mapping bulk states to states in $V$, are similar  for the expanding and contracting branch, these are two  different Hilbert spaces. This is analogous to the situation in a scattering problem. The in- and out- Hilbert spaces, though similar, are two conceptually different Hilbert spaces. In a similar way   the Hilbert space associated to the expanding and contracting branches should be treated differently.

Let us also note that there have been other norms and associated Hilbert spaces that have also been discussed for this theory \cite{Held:2024rmg}. 
For example by considering the group of  time reprametrisations we can construct such a norm. The idea is to take a ``seed wave function" and average it over all time reparametrisations to obtain physical states. An  norm can then be defined which preserves time reparametrisation invariance by following a suitable Faddev Popov procedure. 
Such a procedure, working in the  $l,k$ basis for the wave function, where $k$ is the extrinsic curvature conjugate to $\phi$, was discussed in \cite{Held:2024rmg}, and  gives a different norm compared to the one we defined in subsection \ref{secrindlerquan}  on constant $\phi$ slices and  also a different  one from that  defined here, arising from the map  between the bulk and the matrix theory\footnote{Although the Hamiltonian defined in this paper is different from the minisuperspace Hamiltonian considered in \cite{Held:2024rmg}, similar steps can lead to a norm analogous to eq.(5.13) of \cite{Held:2024rmg} here as well. It is also worth noting that the norm defined via the matrix theory is also time reparametrisation invariant. This is because the  eigenvalues $M$ further classify states which are already solutions to the WdW equation and thus time reparametrisation invariant, thereby also making the norm eq.(\ref{inpa2}) also group invariant. }. 

The conclusion one then  comes to   is that while the solutions to the WdW equation are unambiguously defined, at least once the normal ordering etc issues have been dealt with and a suitable definition of the Hamiltonian constraint has been obtained, the Hilbert space one obtains after canonical quantisation is not uniquely determined. 
Inequivalent Hilbert spaces of course correspond to different quantum theories. 


\section{ General Potentials: Classical Solutions}
\label{pdp}
{Having dealt with the issues of quantization in the JT theory in pure dS spacetime, let us now turn to the analysis of more general potentials. }
\subsection{Basic Setup}
\label{bs}

{Starting in this section  we will consider a generalisation of the JT theory, described by the action eq.\eqref{actionjt}, where  the second term in the action eq.\eqref{actionjt}}, linear in the dilaton, is replaced by a more general dilaton potential resulting in the action, 
\begin{equation}
	S_{\text{U}}= \frac{-i}{2} \pqty{\int d^2 x\,\sqrt{-g}\, (\phi R- U'(\phi))-2\int_{bdy}\sqrt{-\gamma}\phi K }\label{jtacta}
\end{equation}
In this section we will consider the classical behaviour of this  system for various choices of the dilaton potential and  in the next section  turn to canonically quantising it. 

Before proceeding let us note that there is another term which appears in the action of the JT theory, both for $dS_2$ and its generalizations, given by eq.\eqref{sojt}.
This is a topological term and becomes important when one considers the quantum theory but can be ignored in the classical theory where topology change does not occur. 

The equation of motion for the system are,
\begin{align}
	&	R - U''(\phi)=0 \label{Req} \\
	&	\nabla_\mu\nabla_\nu \phi-g_{\mu\nu}\nabla^2\phi- \frac{1}{2}  g_{\mu\nu} U'(\phi)=0 \label{eomJT}
\end{align}
The derivation of the above equations is given in the appendix \ref{varprin}. Taking the trace of the second equation above and using it to re-express $\nabla^2\phi$ in terms of the potential, leads to the following simplified equation
\begin{equation}
	\nabla_\mu\nabla_\nu \phi + \frac{1}{2}  g_{\mu\nu} U'(\phi) =0 \label{phieom}
\end{equation}
We will analyze the solutions for  some interesting choice of potentials $U(\phi)$ in what follows. However, even without the knowledge of the explicit form of the potential, a few important properties about the classical solutions can be deduced from the above equations of motion \footnote{See \cite{Grumiller:2002nm} for a more detailed discussion of the solutions.}. In solving the above system of equations, a central role is played by the following conserved scalar quantity, $M$, defined as below,
\begin{equation}
	\label{Meq}
	M = \nabla_{\mu} \phi \nabla^{\mu} \phi + U(\phi)
\end{equation}
Note that this matches with the analogous quantity defined in eq.\eqref{defMa} for the dS spacetime. 
The conservation of this quantity is easy to verify by considering its variation,
\begin{equation}
	\nabla_{\alpha} M= 2 \nabla ^{\mu} \phi (\nabla_{\mu} \nabla_{\alpha} \phi  + \frac{1}{2} g_{\mu \alpha} U'(\phi)) =0
\end{equation}
where we used eq.\eqref{phieom}. 
Also, the equations of motion guarantee the existence of a Killing vector defined as, 
\begin{equation}
	\xi^\mu = \epsilon^{\mu \nu} \nabla_\nu \phi \label{defxi}
\end{equation}
 Taking a derivative of both sides we get,
\begin{equation*}
	\nabla^{\alpha} \xi^\mu = \epsilon^{\mu \nu} \nabla^{\alpha} \nabla_\nu \phi = - \frac{1}{2} \epsilon^{\mu \nu} \delta^{\alpha}_{\nu} U'(\phi) = -\frac{1}{2} \epsilon^{\mu \alpha} U'(\phi)
\end{equation*}
where we have used eq.\eqref{phieom}. As can be seen from above $\nabla^{\alpha} \xi^\mu$ is an anti-symmetric tensor and hence,
\begin{equation}
	\nabla^{\alpha} \xi^\mu + \nabla^{\mu} \xi^\alpha=0.
\end{equation}
Thus $\xi$ is a killing vector. The norm of this vector is given by 
\begin{equation}
	\xi^\mu \xi_{\mu} = \epsilon^{\mu \nu} \nabla_\nu \phi \epsilon_{\mu \alpha} \nabla^\alpha \phi = -  \nabla_\nu \phi  \nabla^\nu \phi = U(\phi)-M
\end{equation}
Here we have used $\epsilon^{\mu \nu} \epsilon_{\mu \alpha}= - \delta^{\nu}_{\alpha}$ in getting the third equality and eq.\eqref{Meq} in the last equality. From the above it can be deduced that  $\xi$ is spacelike for $U(\phi) >M$ and timelike when $U(\phi) < M$. 
\subsection{General solutions}
\label{solpdp}
It is convenient to  consider a coordinate system in which the Killing symmetry is manifest. This can always be done. Let us consider the situation where the Killing vector is spacelike, as happens when  $U(\phi) > M$. Then we can write the metric in the form, 
\begin{align}
	ds^2 &= - N(t)^2 dt^2 + a(t)^2 dx^2, \,\nonumber\\
	 \phi&= \phi(t) \label{met}
\end{align}
where the killing vector is given by  ${\partial_x}$. The metric can always be put in a diagonal form as above by a suitable coordinate transformation. Since the killing vector has only $x$ component, its $t$ component vanishes which in turn means the $x$ derivative of $\phi$ is zero. Hence $\phi$ is a function of $t$ only.
Inserting this form of the fields in the  action eq.\eqref{jtacta} we get,
\begin{equation}
	S_{\text{U}}= (-i)\left(\int d^2 x \left(\phi(t) \left(\frac{\ddot{a}(t)}{N(t)}- \frac{\dot{a}(t) \dot{N}(t)}{N(t)^2}\right) - \frac{N(t)}{2} a(t)U'(\phi)\right) - \int_{bdy}\sqrt{-\gamma}\phi K\right) \label{I1}
\end{equation}
Doing an integration by parts gives,
\begin{equation}
	S_{\text{U}}=(-i) \int d^2 x \left(- \frac{\dot{\phi}\,\dot{a}}{N} - \frac{1}{2}a NU'(\phi)\right) \label{I2}
\end{equation}
where the boundary term  has been canceled by a  boundary term in the action involving the extrinsic curvature $K$. The Hamiltonian constraint is given by,
\begin{equation}
	\mathcal{H} = \frac{\delta S_{U}}{\delta N}\bigg\vert_{N(t)=1} = \dot{\phi} \,\dot{a}- \frac{1}{2} a\,U'(\phi) =0 \label{HCons}
\end{equation}
where we have considered the gauge choice $N(t)=1$, as the function $N(t)$ can always be set to unity by a redefinition of the coordinate $t$ in eq.\eqref{met}.
The conserved quantity $M$, defined by eq.\eqref{Meq}, then takes the form,
\begin{equation}
    M = - \dot{\phi}^2 + U(\phi) \label{mphidd}
\end{equation}
This can be rearranged to give,
\begin{equation}
\dot{\phi} = \sqrt{U(\phi)-M} \implies dt = \frac{d \phi}{\sqrt{U(\phi)-M}} \label{phidoteq}
\end{equation}
From eq.\eqref{HCons} and \eqref{phidoteq} we get,
\begin{equation}
	2 \frac{\dot{a}}{a} dt = \frac{U'(\phi)}{U(\phi)-M} d\phi\label{adupphi}
\end{equation}
Integrating this we find the following relation for the scale factor $a$ as a function of the dilaton, 
\begin{equation}
	a(\phi) = \sqrt{U(\phi)-M} \label{aeq}
\end{equation}
This expression for scale factor, combined with eq.\eqref{phidoteq}, can be used to express the metric with $\phi$ as one of the coordinates,
\begin{equation}
ds^2 = -\frac{d\phi^2}{U(\phi)-M} + (U(\phi)-M) dx^2\label{solds}
\end{equation}
In the subsequent discussion we will sometime refer to $M$ as the mass.

It can be seen from eq.(\ref{solds})  that at $U(\phi)=M$, the metric becomes singular. However, the Ricci scalar remains finite as long as $U''(\phi)$ is finite at this location, see eq.\eqref{Req}. In such cases we will somewhat loosely refer to   the locus $U(\phi)=M$ as a horizon. 
The solution can be extended to the region  $U(\phi) < M$, leading to the same metric as above but now with $x$ being  a time-like coordinate, and $\phi$ being a space-like direction.  

\begin{figure}
	
	\begin{center}
	
	\tikzset{every picture/.style={line width=0.75pt}} 
	
	\begin{tikzpicture}[x=0.75pt,y=0.75pt,yscale=-1,xscale=1]
		
		\draw    (148,388) -- (148,84) -- (148,53) ;
		\draw [shift={(148,51)}, rotate = 90] [color={rgb, 255:red, 0; green, 0; blue, 0 }  ][line width=0.75]    (10.93,-3.29) .. controls (6.95,-1.4) and (3.31,-0.3) .. (0,0) .. controls (3.31,0.3) and (6.95,1.4) .. (10.93,3.29)   ;
		\draw    (148,219.5) -- (553,219.5) ;
		\draw [shift={(555,219.5)}, rotate = 180] [color={rgb, 255:red, 0; green, 0; blue, 0 }  ][line width=0.75]    (10.93,-3.29) .. controls (6.95,-1.4) and (3.31,-0.3) .. (0,0) .. controls (3.31,0.3) and (6.95,1.4) .. (10.93,3.29)   ;
		\draw    (148,219.5) .. controls (254,523) and (335,-75) .. (419,149) .. controls (503,373) and (534.62,294.1) .. (537,305) ;
		\draw  [fill={rgb, 255:red, 0; green, 0; blue, 0 }  ,fill opacity=1 ] (160,258) .. controls (160,255.79) and (161.79,254) .. (164,254) .. controls (166.21,254) and (168,255.79) .. (168,258) .. controls (168,260.21) and (166.21,262) .. (164,262) .. controls (161.79,262) and (160,260.21) .. (160,258) -- cycle ;
		\draw  [dash pattern={on 4.5pt off 4.5pt}]  (539,258) -- (164,258) -- (146.5,258) ;
		\draw  [fill={rgb, 255:red, 0; green, 0; blue, 0 }  ,fill opacity=1 ] (263,258) .. controls (263,255.79) and (264.79,254) .. (267,254) .. controls (269.21,254) and (271,255.79) .. (271,258) .. controls (271,260.21) and (269.21,262) .. (267,262) .. controls (264.79,262) and (263,260.21) .. (263,258) -- cycle ;
		\draw  [fill={rgb, 255:red, 0; green, 0; blue, 0 }  ,fill opacity=1 ] (464,258) .. controls (464,255.79) and (465.79,254) .. (468,254) .. controls (470.21,254) and (472,255.79) .. (472,258) .. controls (472,260.21) and (470.21,262) .. (468,262) .. controls (465.79,262) and (464,260.21) .. (464,258) -- cycle ;
		
		\draw (117,250.4) node [anchor=north west][inner sep=0.75pt]    {$-M$};
		\draw (562,208.4) node [anchor=north west][inner sep=0.75pt]    {$\phi $};
		\draw (124,28.4) node [anchor=north west][inner sep=0.75pt]    {$-U(\phi)$};
		\draw (167,238.4) node [anchor=north west][inner sep=0.75pt]    {$A$};
		\draw (276,238.4) node [anchor=north west][inner sep=0.75pt]    {$B$};
		\draw (472,237.4) node [anchor=north west][inner sep=0.75pt]    {$C$};
	\end{tikzpicture}
	\end{center}
	\caption{This plot describes a  general potential with $M>0$ having multiple horizons.}
	\label{plotpot3}
\end{figure}
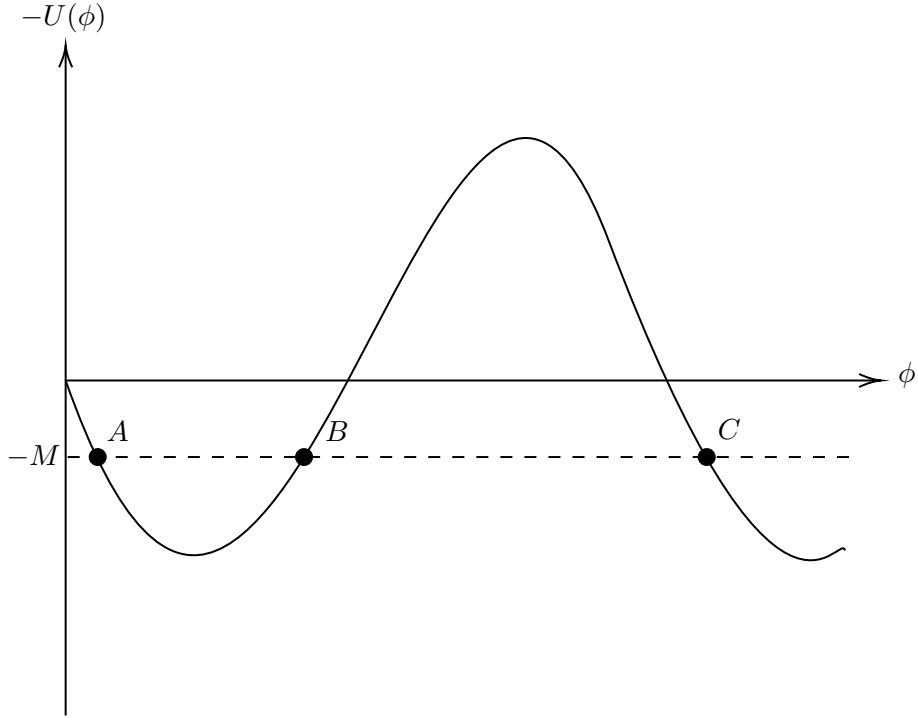

An intuitive way to understand the qualitative nature of the spacetimes for a general potential is as follows. 
Rewriting the conserved quantity in eq.\eqref{mphidd} as
\begin{equation}
 \dot{\phi}^2 - U(\phi)= - M\label{mpar}
\end{equation}
we see that the above equation is akin to a particle moving in a potential $-U(\phi)$ with  energy $-M$. Let us take $\phi$ to be non-negative such that $U(\phi)$ is also  non-negative and a monotonically increasing function of $\phi$, see fig. \ref{plotpot1}. Then for $M>0$,  starting from the point A which corresponds to a  horizon   where $U(\phi)=M$ and ${\dot \phi}=0$, 
 the particle will keep moving down the potential ``hill". In contrast if $M<0$ then there is no horizon. In this case starting at some point A, with  initial velocity ${\dot \phi}>0$, the particle will keep 
 moving to larger value of $\phi$. While  for initial velocity ${\dot \phi}<0$, the particle will evolve to smaller values of $\phi$ till it reaches  $\phi=0$, with its subsequent evolution depending on the behaviour of $U(\phi)$ for $\phi<0$. 
 If $U(\phi)$ is non-monotonic as shown in fig.\ref{plotpot3} the behaviour can be more interesting. For example consider the potential shown in fig.\ref{plotpot3}. In this case   for $M>0$, starting at point A which is a horizon, $\phi$ will evolve to larger values till it reaches point B. At that stage one encounters another horizon and between $\phi\in [B,C]$ the $\phi $ direction will be space like. When $\phi>\phi_C$ the $\phi$ direction  will again be a time- like direction.

 As these examples illustrate, for a general potential there can be multiple horizons or even none depending on the nature of the potential and the value of $M$ chosen, see fig.\ref{plotpot3}.

\begin{figure}[h!]
	\begin{center}
		
	\tikzset{every picture/.style={line width=0.75pt}} 

\begin{tikzpicture}[x=0.75pt,y=0.75pt,yscale=-1,xscale=1]
	
	\draw    (148,388) -- (148,105) ;
	\draw [shift={(148,103)}, rotate = 90] [color={rgb, 255:red, 0; green, 0; blue, 0 }  ][line width=0.75]    (10.93,-3.29) .. controls (6.95,-1.4) and (3.31,-0.3) .. (0,0) .. controls (3.31,0.3) and (6.95,1.4) .. (10.93,3.29)   ;
	\draw    (148,219.5) -- (459,219.5) ;
	\draw [shift={(461,219.5)}, rotate = 180] [color={rgb, 255:red, 0; green, 0; blue, 0 }  ][line width=0.75]    (10.93,-3.29) .. controls (6.95,-1.4) and (3.31,-0.3) .. (0,0) .. controls (3.31,0.3) and (6.95,1.4) .. (10.93,3.29)   ;
	\draw  [fill={rgb, 255:red, 0; green, 0; blue, 0 }  ,fill opacity=1 ] (285,301) .. controls (285,298.79) and (286.79,297) .. (289,297) .. controls (291.21,297) and (293,298.79) .. (293,301) .. controls (293,303.21) and (291.21,305) .. (289,305) .. controls (286.79,305) and (285,303.21) .. (285,301) -- cycle ;
	\draw  [dash pattern={on 4.5pt off 4.5pt}]  (441,301) -- (296,301) -- (164,301) -- (146.5,301) ;
	\draw    (148,219.5) .. controls (221,221) and (320,317) .. (331,366) ;
	
	\draw (117,291.4) node [anchor=north west][inner sep=0.75pt]    {$-M$};
	\draw (466,208.4) node [anchor=north west][inner sep=0.75pt]    {$\phi $};
	\draw (124,80.4) node [anchor=north west][inner sep=0.75pt]    {$-U(\phi)$};
	\draw (292,278.4) node [anchor=north west][inner sep=0.75pt]    {$A$};

\end{tikzpicture}

	\end{center}
	\caption{The figure describes the case of a potential that increases monotonically with $\phi$. In this case $M>0$. }
	\label{plotpot1}
\end{figure}
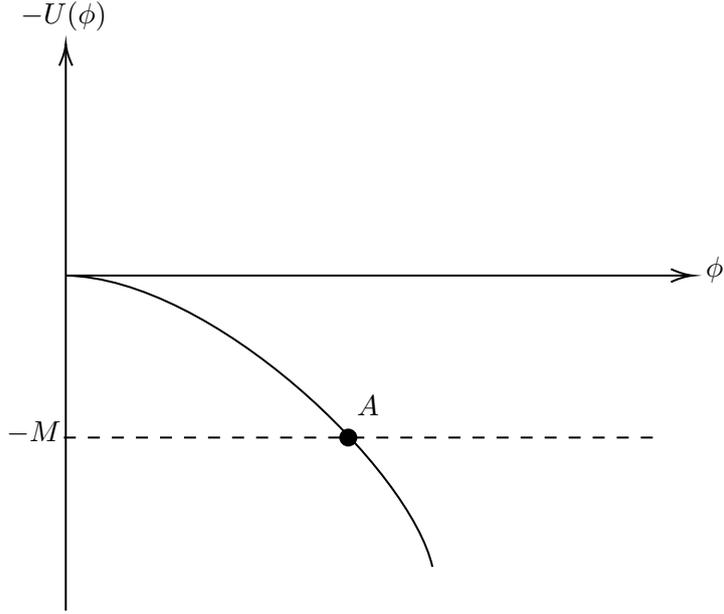

A few more comments are in order. In the two dimensional theory depending on the nature of $U(\phi)$ or  the sign of $M$, one might get a naked time-like curvature singularity in a static patch (where the killing vector ${ \partial_x}$ is time-like) of the cosmology. The singularity could be either a locus where the curvature blows up, or, in cases of dilaton having a bounded range, where  the dilaton attains its minimum value. To avoid such naked singularities one might want to exclude these choices of the parameter $M$, or of the dilaton potential $U(\phi)$.

Finally, we can take the $x$ coordinate, corresponding to the Killing symmetry direction, eq.(\ref{solds}),  to be compact in regions where it is space-like, by making an identification $x\sim x+a$. In fact we will do this when we turn to quantising the system in section \ref{canquan}. After  such an identification only regions where the Killing vector is space-like can be retained - these are free from the naked singularities mentioned above.
\subsection{Examples}
\label{examples}
We now turn to discussing a few examples more explicitly. 
\subsubsection{$U(\phi)=\phi^2+ a e^{-\alpha \phi}$}
We will be especially interested in our discussion of the quantum theory in  what happens for potentials which  asymptote, for large $\phi$, to the deSitter form, i.e. where 
$U(\phi)\rightarrow \phi^2$ when   $\phi\rightarrow \infty$. As an example consider the case where 
 \be
 \label{altu}
 U(\phi)=\phi^2+a e^{-\alpha \phi}
 \ee
 where $a, \alpha>0$. 
 
{\begin{figure}[h!]
		\centering
		{\includegraphics[width=0.6\textwidth]{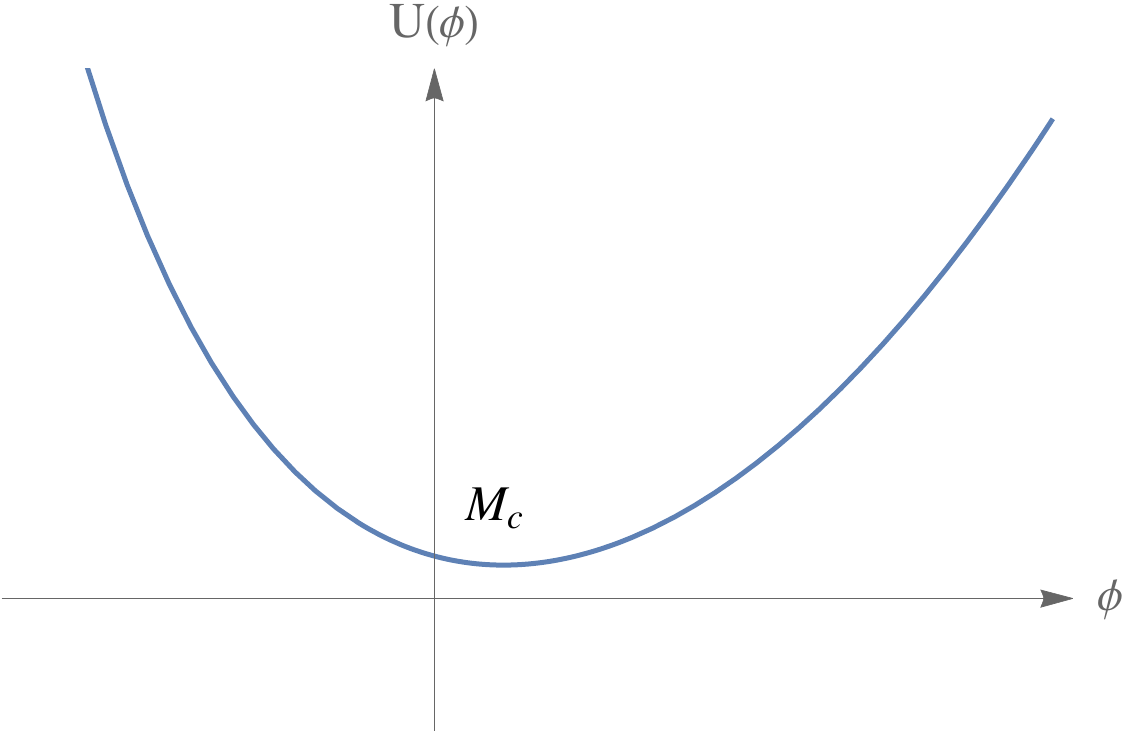}}
		\caption{$M_c$ is the critical mass determined by $U(0)=M_C$.}
		\label{exppot}
\end{figure}}

 In this case two  horizons- a cosmological and black hole horizon- are   present as long as $M$ is bigger than a critical value, $M_c$, see fig.\ref{exppot}. 
 For $M_c$ we have an extremal limit, akin to a Naria black hole. The Penrose diagram for $M>M_c$ is similar to that of a black hole in $dS_2$.

\subsubsection{Polynomial $U(\phi)$}
\label{exsub}
Next consider a general power 
\begin{equation}
	U(\phi) = \phi^n. \label{nphipot}
\end{equation}
From eq.\eqref{Req}, it can be seen that 
\begin{equation}
	R = n(n-1) \phi^{n-2}\label{ricmon}
\end{equation}
The case $n=1$ gives flat space. 
For $n>2$ we get a spacetime whose curvature increases as $\phi$ increases and asymptotically for $\phi\rightarrow \infty$ becomes infinty. 
The affine time along a null geodesic to reach  $\phi\rightarrow \infty$ can be easily seen to be infinite. 
Whereas the proper time along a time-like geodesic to reach the curvature singularity at $\phi\rightarrow \infty$ is finite. For $M>0$ a cosmological horizon is present at $\phi=M^{1/n}$. 
In addition if $n$ is even there is a black hole horizon at $\phi=-M^{1/n}$. {We describe a few properties of the geodesics in the spacetime described by eq.\eqref{nphipot} in the appendix \ref{asymdiffds}}. 

If $U(\phi)$ is a a more general polynomial
 \begin{equation}
 	U(\phi) = \sum_{n} c_{n} \phi^n \label{sumpot}
 \end{equation}   the resulting spacetime can be more interesting. 
The behaviour at large $\phi$ would  of course dominated by the largest power in the polynomial but there can be multiple horizons. 

Finally consider an exponential  potential 
$U(\phi)=e^\phi$. This cosmology is similar to the $n>2$  case in its  behaviour at large $\phi$- the curvature blows up as $\phi\rightarrow \infty$, with the affine time to reach the singularity being infinite along null geodesic and the 
proper time along a time- like geodesic being finite. 
For $M>0$  in this case there is one cosmological horizon.

One can also consider non-polynomial power law potentials. We will analyse one example when we consider the case obtained by dimensional reduction. 
{The potential that one obtains  in such a dimensional reduction from a 4D black hole is of the form 
\begin{equation}
	U(\phi) = 4 \left(\sqrt{\phi^3}- \sqrt{\phi}\right) \label{pot22}
\end{equation}
This will be an interesting example to study in the quantum theory, which we do in section \ref{exam}. The details of the dimensional reduction can be found in the appendix \ref{dimred}.} 

More generally, one can also consider potentials where the Ricci scalar changes sign in different regions of spacetime. For example, one such potential could correspond to a situation of asymptotic dS region with an AdS bubble inside or vice-versa. An example of such a potential is discussed in appendix \ref{rchanpot}.

\section{General Potentials: Canonical Quantization}
\label{canquan}
In this section we canonically quantize the system given in eq.\eqref{jtacta}. Our discussion will closely follow earlier work, \cite{nanda2023jt} and also \cite{Hennauxjt,Iliesiu:2019xuh}, . 
The metric for a general solution in   the classical theory was given in eq.\eqref{solds}. Here it will be convenient to express this metric in the form
\be
\label{metcanquant}
ds^2=-{dr^2\over (U(r)-M)}+{(U(r)-M)} {dx^2 \over A^2}
\ee
with dilaton taking value 
\be
\label{valdil}
\phi= r
\ee
We will be interested in a region of spacetime where $U(r)\ge M$. In such a region $\phi$ will vary along a time-like direction. 

The spacetime we will be actually interested in is obtained by making the identification $x\sim x+1$.  After this identification the constant $A$ which appears in eq.(\ref{metcanquant}) is physically meaningful and cannot be rescaled out of the metric. 
The two constants $M,A$ which appear in eq.(\ref{metcanquant})  parameterize the resulting set of   solutions, upto gauge transformations, showing that the covariant phase space is two dimensional.  We note that this identification  results in an  orbifold singularity  at  $U(r)=M$.

Let us mention that there are other branches, analogous to the discussion in the JT case which are not included in the set being considered here. 

Our subsequent discussion  below will  follow that of the JT theory  closely. Starting with the metric
\begin{align}
	ds^2=g_{\mu\nu}dx^\mu dx^\nu=-N^2(x,t) dt^2+g_{1}(x,t)(dx+N_{\perp}(x,t)dt)^2\label{adm}
\end{align}
we  go to ADM gauge setting $N_\perp=0$ and $N=1$.

The Hamiltonian and Momentum constraints are then given by 
\begin{align}
\mathcal{H} & =  2\pi_\phi \pi_{g_1}\sqrt{g_1}-\left(\frac{\phi'}{\sqrt{g_1}}\right)'-\frac{\sqrt{g_1} U'(\phi)}{2} \nonumber \\
\mathcal{P} & = 2g_1\pi_{g_1}'+\pi_{g_1}g_1'-\pi_\phi \phi' \label{hammomcon}
\end{align}
One can show that the Poisson brackets of Hamiltonian and Momentum  constraints  in the classical theory close, as is discussed in appendix \ref{conana}. 
Also, the Poisson bracket of $M,A$, is given by 
\be
\label{pbma}
\left\{M,\frac{1}{A}\right\}= 2
\ee

We will solve the quantum theory by obtaining all states which solve the constraints. 
As a first step one needs to make the constraints in the quantum theory well defined since they suffer from ordering ambiguities and also (for the Hamiltonian constraint) ``$\delta(0)$" singularities (see \cite{nanda2023jt} for more details). We will use the same arguments laid out in the section \ref{canquanti} to define the constraints in the case of a general potential $U(\phi)$. Following our discussion of the JT case we do so as follows. First we choose spatial slices along which the dilaton is a constant. 
Then on each slice we use the spatial reparametrisation invariance to set $g_1(x)$ to be a constant. 
The resulting wavefunction can then only be a function of $\phi$ and $l$, the length of the universe. 

We take the WdW equation  which acts on $\Psi(\phi,l)$ to be of the form 
\begin{equation}
	\mathcal{H}\Psi(l,\phi) = \left(\partial_l \partial_{\phi}- \frac{1}{l} \partial_\phi + \frac{1}{2} l U'(\phi)\right)\Psi(l,\phi)=0\label{hcongenph}
\end{equation}
The second term ${1\over l} \partial_\phi \Psi$ is non-trivial. Our motivation to include it is twofold. First, it is easy to see that for the JT dS potential the above equation reduces to eq.\eqref{wda}. Second,  we will see in the section \ref{condefpot} and appendix \ref{moredetpotcd} that for the JT potential with an exponential correction, the path integral answer for the wavefunction satisfies the above WdW equation.

%
An   operator $\hat{M}$, similar to the JT case can also be defined and  now takes the form,
\begin{equation}
	\hat{M} = \del_l^2 + U(\phi) \label{hatm}
\end{equation}
This is the operator in the quantum theory corresponding to the classically conserved mass $M$ defined in eq.\eqref{Meq}.

Eigenvectors of ${\hat M}$ with eigenvalue $M$ take the form 
\be
\label{evec}
\Psi=e^{\pm i l \sqrt{U(\phi)-M}}
\ee
where $l$ is the size of the universe on a spatial slice where the dilaton takes the constant value $\phi$. 
This is the generalisation of the result which was obtained in \cite{nanda2023jt} with now the factor $\sqrt{U(\phi)-M} $ replacing $\sqrt{\phi^2-M}$ in the exponent. 
The component with wave function $e^{-i l \sqrt{U-M}}$, $e^{i l \sqrt{ U-M}}$ correspond to the expanding and contracting branches respectively of the wavefunction. 
See appendix \ref{moredetquan} for more details. 

A general gauge invariant state can be obtained by summing over states with  different values of $M$:
\begin{equation}
	\Psi = \int dM \rho(M) e^{-il\sqrt{U(\phi)-M}} + \int dM \tilde{\rho} (M) e^{ il\sqrt{U(\phi)-M}}  \label{genpsimod}
\end{equation}
The integral over $M$  goes from $[-\infty,\infty]$. 

The above wavefunction is the solution to the Wheeler deWitt equation given by,
\begin{equation}
	\left(\partial_l \partial_{\phi}- \frac{1}{l} \partial_\phi + \frac{1}{2} l U'(\phi)\right) \Psi=0           \label{WdW2}
\end{equation}

In fact for specifying the wave function we need to be more precise in defining the $\sqrt{U(\phi)-M}$ function which appears in eq.(\ref{genpsimod}). 
Consider a solution  $\Psi_M(l,\phi)$ to the WDW equation corresponding to an eigenvalue $M$.
For values of $\phi$ where $U(\phi)-M>0$ it takes the form 
\be
\label{forma}
\Psi_M(l,\phi)= \rho(M) e^{-il\sqrt{U-M}}+
{\tilde \rho}(M) e^{il\sqrt{U-M}}
\ee
While for  $U(\phi)-M<0$ we take it to have  the form 
\be
\label{formb}
\Psi_M(l,\phi)=\rho_1 e^{-l\sqrt{M-U}}+\rho_2 e^{l\sqrt{M-U}}
\ee
Continuity at  the value of $\phi$ where $U(\phi)=M$ leads to one relation between the coefficients
\be
\label{onerel}
\rho(M)+{\tilde \rho}(M)=\rho_1(M) + \rho_2(M)
\ee
The full solution is then given by summing over all solutions for different values of $M$,
\be
\label{fullsol}
\Psi(l,\phi)=\int dM \Psi_M(l,\phi)
\ee

In general, for any solution $\Psi_M(l,\phi)$,  there can be multiple disconnected regions where $U(\phi)>M$ or $U(\phi)<M$. In this case there will be  expansion coefficients,
$\rho, {\tilde \rho}$, and, $\rho_1, \rho_2$, for each such region, and they  will be related by the continuity relations analogous to eq.(\ref{onerel}) which arise at every values 
at $\phi$ where $U(\phi)=M$. 

In the rest of the paper we will work with the following wavefunction
\begin{equation}
	\hat{\Psi} = \frac{1}{l} e^{\pm il\sqrt{U(\phi)-M}} \label{psi}
\end{equation}
which is the solution to the equation,
\begin{equation}
	\left(\partial_l \partial_{\phi} + \frac{1}{2} l U'(\phi)\right) \hat{\Psi}=0           \label{WdW}
\end{equation}
It can be shown that the above equation can be recast as a massive Klein-Gordon(KG) equation of mass $m^2=1$ in a flat metric, see appendix \ref{altsol}. { The resulting solutions of this KG equation can be written in terms of Bessel functions. We can then define a Rindler basis for the general $U(\phi)$ similar to the JT dS case. More details are given in appendix \ref{altsol}}.

Before going further let us make one more comment. 
It can be shown for a general potential that the WKB approximation to eq.\eqref{WdW} becomes valid when $U(\phi)\rightarrow \infty$ and $l\rightarrow \infty$ keeping ${l\over \sqrt{U}}$ fixed.
For example for JT dS this happens in the far past or future where the dilaton becomes big and the universe grows to infinite size. 
In this limit the universe will be well described by a classical solution  (if one takes an appropriate wave packet) given by the geometry eq.\eqref{metcanquant} with an evolving dilaton, eq.\eqref{valdil}.
\subsection{Norm and Expectation Values}
\label{genponorm}
As was discussed in \cite{nanda2023jt} we will take  our definition of physical time to be the value of the dilaton.

A norm  can then be defined as 
\begin{equation}
	\langle \hat{\Psi},\hat{\Psi} \rangle = \pm i\int_{0}^{\infty} dl (\hat{\Psi}^* \partial_l \hat{\Psi}-\hat{\Psi} \partial_l \hat{\Psi}^*) \label{norma}
\end{equation}
where  the integral is to be carried out at a fixed value of $\phi$ and sign in front has to be chosen so that the RHS is positive.
Requiring the norm to be finite and to be conserved, i.e. independent of the value the dilaton,  gives important restrictions on the expansion coefficients which appear in the wave function, eq.(\ref{forma}) and eq.(\ref{formb}). These are discussed in appendix \ref{moredetquan}. In particular to try and get a normalisable wave function we will usually set 
\be
\label{rhotil}
 \rho_2(M)=0
\ee
in eq.\eqref{formb} so that there is no term in the wave function which grows exponentially at large $l$. We also require for the conservation of norm that,
\begin{equation}
	\int (\rho + \tilde{\rho}) dM=0 \label{condrho}
\end{equation}
 An inner product can also be similarly defined,
\begin{equation}
	\langle \hat{\Psi}_1, \hat{\Psi}_2 \rangle = \frac{i}{2}\int_{0}^{\infty} dl  \,(\hat{\Psi}_1^* \partial_l \hat{\Psi}_2-\hat{\Psi}_2 \partial_l \hat{\Psi}_1^*) + \frac{i}{2}\int_{0}^{\infty} dl \, (\hat{\Psi}_2^* \partial_l \hat{\Psi}_1-\hat{\Psi}_1 \partial_l \hat{\Psi}_2^*)  \label{inner}
\end{equation}

Finally we can define expectation values of operators as follows. 
We define the probability density for the universe having length $l$ at the time when the dilaton takes value $\phi$ to be 
\begin{equation}
	p(l,\phi) = \pm \frac{ i}{\mathcal{N}} (\hat{\Psi}^* \partial_l \hat{\Psi}-\hat{\Psi} \partial_l \hat{\Psi}^*) \label{probd}
\end{equation}
with the sign  being determined as discussed after eq.(\ref{norma}) so that the norm is positive.
Then the expectation value of the  moments of the length $\langle l^n \rangle $ are given by
\be
\label{expmoml}
\langle l^n \rangle  =\pm \frac{ i}{\mathcal{N}} \int dl \,  \,   (\hat{\Psi}^* \partial_l (l^n\hat{\Psi})-l^n\hat{\Psi} \partial_l \hat{\Psi}^*)
\ee
The expectation value of the operator conjugate to $l$, $\pi_l$, can be calculated using the formula,
\begin{equation}
	\langle \pi_l\rangle =\int dl {\hat \Psi}^* (-i  (\overrightarrow{\partial}-\overleftarrow{\partial})) (-i  {\partial_l} {\hat \Psi}) \label{pilexp}
\end{equation}
where the derivatives are with respect to $l$. More generally we can write,
\begin{equation}
	\langle \pi_l ^n \rangle = \int dl {\hat \Psi}^* (-i  (\overrightarrow{\partial}-\overleftarrow{\partial})) ((-i  {\partial_l})^n {\hat \Psi}) \label{pilnexp}
\end{equation}
We remind the reader that the wavefunction $\Psi$ in the above formulas is taken to be normalized to unity.
This allows the expectation values of the operator $\hat{M}$, eq.\eqref{hatm},  whose eigenvalues we denoted by $M$ above, to  be written as, \footnote{{Note that the eq.\eqref{expmoml}, eq.\eqref{pilexp}, eq.\eqref{pilnexp} and eq.\eqref{mexp} are different from those in \cite{nanda2023jt}}.}
\begin{equation}
	\label{mexp}
	\langle \hat{M}\rangle=\int dl \frac{\Psi^*}{l} (-i) \overleftrightarrow{\del_l} \left(\frac{1}{l} (\del_l^2 + \phi^2) \Psi\right) 
\end{equation}


\subsection{Examples}
\label{exam}
It is worth considering some explicit examples in more detail. 

Let us start with the dimensionally reduced theory where the function $U(r)$ takes the form eq.(\ref{pot22})

\be
\label{pot4}
U(\phi)=4  (\sqrt{\phi^3}-\sqrt{\phi})
\ee
We take $\phi$ to lie in  the range $\phi\in [0,\infty]$. 

\begin{figure}
	\begin{center}
	\tikzset{every picture/.style={line width=0.75pt}} 
	
	\begin{tikzpicture}[x=0.75pt,y=0.75pt,yscale=-1,xscale=1]
		
		\draw   (191.5,71.5) -- (331,71.5) -- (331,211) -- (191.5,211) -- cycle ;
		\draw    (190.5,71) -- (329.5,210) ;
		\draw    (330,71.5) -- (191,210.5) ;
		\draw    (330.5,71) -- (469.5,210) ;
		\draw    (470,71.5) -- (331,210.5) ;
		\draw    (470,71.5) -- (469.5,210) ;
		\draw   (260.92,141.75) -- (284.25,211.25) -- (239.25,211.25) -- cycle ;
		\draw   (260.92,141.75) -- (237.58,72.25) -- (282.58,72.25) -- cycle ;
		\draw   (400.92,140.75) -- (424.25,210.25) -- (379.25,210.25) -- cycle ;
		\draw   (400.92,140.75) -- (377.58,71.25) -- (422.58,71.25) -- cycle ;
		\draw    (331,71.5) .. controls (332.67,69.83) and (334.33,69.83) .. (336,71.5) .. controls (337.67,73.17) and (339.33,73.17) .. (341,71.5) .. controls (342.67,69.83) and (344.33,69.83) .. (346,71.5) .. controls (347.67,73.17) and (349.33,73.17) .. (351,71.5) .. controls (352.67,69.83) and (354.33,69.83) .. (356,71.5) .. controls (357.67,73.17) and (359.33,73.17) .. (361,71.5) .. controls (362.67,69.83) and (364.33,69.83) .. (366,71.5) .. controls (367.67,73.17) and (369.33,73.17) .. (371,71.5) .. controls (372.67,69.83) and (374.33,69.83) .. (376,71.5) .. controls (377.67,73.17) and (379.33,73.17) .. (381,71.5) .. controls (382.67,69.83) and (384.33,69.83) .. (386,71.5) .. controls (387.67,73.17) and (389.33,73.17) .. (391,71.5) .. controls (392.67,69.83) and (394.33,69.83) .. (396,71.5) .. controls (397.67,73.17) and (399.33,73.17) .. (401,71.5) .. controls (402.67,69.83) and (404.33,69.83) .. (406,71.5) .. controls (407.67,73.17) and (409.33,73.17) .. (411,71.5) .. controls (412.67,69.83) and (414.33,69.83) .. (416,71.5) .. controls (417.67,73.17) and (419.33,73.17) .. (421,71.5) .. controls (422.67,69.83) and (424.33,69.83) .. (426,71.5) .. controls (427.67,73.17) and (429.33,73.17) .. (431,71.5) .. controls (432.67,69.83) and (434.33,69.83) .. (436,71.5) .. controls (437.67,73.17) and (439.33,73.17) .. (441,71.5) .. controls (442.67,69.83) and (444.33,69.83) .. (446,71.5) .. controls (447.67,73.17) and (449.33,73.17) .. (451,71.5) .. controls (452.67,69.83) and (454.33,69.83) .. (456,71.5) .. controls (457.67,73.17) and (459.33,73.17) .. (461,71.5) .. controls (462.67,69.83) and (464.33,69.83) .. (466,71.5) -- (470,71.5) -- (470,71.5) ;
		\draw    (331,211.5) .. controls (332.65,209.82) and (334.32,209.8) .. (336,211.45) .. controls (337.69,213.1) and (339.35,213.08) .. (341,211.39) .. controls (342.65,209.71) and (344.32,209.69) .. (346,211.34) .. controls (347.69,212.99) and (349.35,212.97) .. (351,211.28) .. controls (352.65,209.6) and (354.32,209.58) .. (356,211.23) .. controls (357.68,212.88) and (359.35,212.86) .. (361,211.18) .. controls (362.65,209.49) and (364.31,209.47) .. (366,211.12) .. controls (367.68,212.77) and (369.35,212.75) .. (371,211.07) .. controls (372.65,209.38) and (374.31,209.36) .. (376,211.01) .. controls (377.68,212.66) and (379.35,212.64) .. (381,210.96) .. controls (382.65,209.27) and (384.31,209.25) .. (386,210.9) .. controls (387.68,212.55) and (389.35,212.53) .. (391,210.85) .. controls (392.65,209.17) and (394.32,209.15) .. (396,210.8) .. controls (397.69,212.45) and (399.35,212.43) .. (401,210.74) .. controls (402.65,209.06) and (404.32,209.04) .. (406,210.69) .. controls (407.69,212.34) and (409.35,212.32) .. (411,210.63) .. controls (412.65,208.95) and (414.32,208.93) .. (416,210.58) .. controls (417.68,212.23) and (419.34,212.21) .. (420.99,210.53) .. controls (422.64,208.84) and (424.3,208.82) .. (425.99,210.47) .. controls (427.67,212.12) and (429.34,212.1) .. (430.99,210.42) .. controls (432.64,208.73) and (434.3,208.71) .. (435.99,210.36) .. controls (437.67,212.01) and (439.34,211.99) .. (440.99,210.31) .. controls (442.64,208.62) and (444.3,208.6) .. (445.99,210.25) .. controls (447.67,211.9) and (449.34,211.88) .. (450.99,210.2) .. controls (452.64,208.52) and (454.31,208.5) .. (455.99,210.15) .. controls (457.68,211.8) and (459.34,211.78) .. (460.99,210.09) .. controls (462.64,208.41) and (464.31,208.39) .. (465.99,210.04) -- (469.5,210) -- (469.5,210) ;
		
		\draw (353,105.4) node [anchor=north west][inner sep=0.75pt]    {$r_{1}$};
		\draw (297,103.4) node [anchor=north west][inner sep=0.75pt]    {$r_{2}$};
		\draw (269,132.4) node [anchor=north west][inner sep=0.75pt]    {$P_{1}$};
		\draw (410,132.4) node [anchor=north west][inner sep=0.75pt]    {$P_{2}$};
		\draw (255,83.4) node [anchor=north west][inner sep=0.75pt]    {$F$};
		\draw (255,183.4) node [anchor=north west][inner sep=0.75pt]    {$B$};
		\draw (388,83.4) node [anchor=north west][inner sep=0.75pt]    {$BH$};
		\draw (385,189.4) node [anchor=north west][inner sep=0.75pt]    {$WH$};

	\end{tikzpicture}
	
	\end{center}
	\caption{Penrose Diagram with regions $F,B,BH, WH$ specified.}
	\label{pendia}
\end{figure}
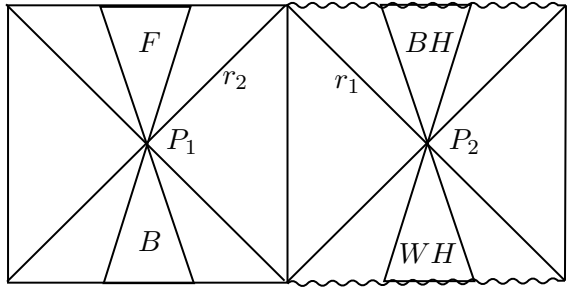

Taking the mass $M$  to be negative and in the range 
\be
\label{rangeM}
0>M>M_0
\ee
where $M_0$ is given by eq.\eqref{M0eq},  there are two zeros, for $r>0$, to the equation 
\be
\label{zeoa}
U(r)=M
\ee
which we denote as $r_2,r_1,$ with $r_2>r_1$.
Without the identification $x\simeq x+1$, these  correspond to the cosmological and black hole horizons respectively (and are of course present in the $4$ dim theory too). 
The Penrose diagram is shown in fig.\ref{pendia}. 

The region F, with, $r>r_2$,  is the future  cosmological or Milne patch; S,  with  $ r_1<r<r_2$, is the static patch; BH and WH, with $r<r_1$, are the black hole and white hole regions respectively, and B,  is the past cosmological patch. 
After the identification $x\simeq x+1$, the wedge shaped regions in F and B are retained  and similarly the wedge shaped regions  in  BH and WH  are also retained. The geometry now has an orbifold singularity at $P_1$, where the F and  B regions intersect and also at $P_2$, where the BH and WH regions intersect. 

In the quantum theory, when $\phi>r_2$ and $U(\phi)>M$, the wave function is oscillatory and of the form eq.(\ref{forma}). The $e^{-il\sqrt{U-M}}$ branch corresponds to future region (F) where the universe expands, while the $e^{il\sqrt{U-M}}$ branch corresponds to the B region where the universe contract (when one goes  forward in coordinate time as shown on the Penrose diagram). In contrast when $r_1<\phi<r_2$ the wave function is exponentially damped or growing. Setting $ \rho_2$ to vanish, see eq.(\ref{onerel}), we get, from  continuity at  $\phi=r_2$, that 
\be
\label{conda}
\rho(M)+{\tilde \rho}(M)=\rho_1(M)
\ee
Similarly  when $\phi<r_1$ we get oscillator phases in the wave function which can be expressed in this region as 
\be
\label{condca}
\Psi_M(l,\phi)={\hat \rho}(M) e^{-il\sqrt{U-M}}+{\hat {\tilde \rho}}(M) e^{i l \sqrt{U-M}}
\ee
Here the $e^{-il\sqrt{U-M}}$ term corresponds to the expanding branch in  the white hole region,   WH, while  the $e^{i l\sqrt{U-M}}$ term corresponds to the contracting branch in the BH region. Continuity at $\phi=r_1$ gives
\be
\label{contac}
{\hat \rho}(M)+{\hat {\tilde {\rho}}} (M)=\rho_1(M)= \rho(M)+{\tilde \rho}(M)
\ee
Thus the solution $\Psi_M$ has three unknown parameters, which can be taken to be $\rho(M), {\tilde \rho}(M), {\hat \rho}(M)$. If we restrict ourselves to a wave function with ${\tilde \rho}(M)=0$, so that it only has an   expanding branch in the F region and no contracting branch in the B region, there are still  two unknown coefficients which specify the state. 

As discussed after equation eq.\eqref{WdW},  the WKB approximation is  only  valid  when $U,l\rightarrow \infty$, this means for this case only in the far future/past of the  F and B regions, when $\phi, l\rightarrow \infty$.
The spacetime  in the WH/BH region is  quantum, although the wave function is oscillatory in this region. 

Loosely, one can describe the solution above as corresponding to a universe which is born at the white hole singularity, evolves forward till the orbifold point $P_2$, where 
a part of the wave function continues past the orbifold into the BH region and begins to contract, while another component, after tunneling, reappears in the F/B regions and gives rise to an expanding universe. However, we hasten to add that these words are not precise due to the quantum nature of the geometry everywhere, except, as mentioned, in the asymptotic regions of F and B. 

What about values of $M$ which do not lie in the window, eq.(\ref{rangeM})?
For $M>0$ it is easy to see that there is only one real root (with $r>0$)  for  eq.(\ref{zeoa}). This corresponds to the fact that the classical geometry has a naked singularity inside the static patch for these values of $M$. 
After the orbifold identification we are making, the static patch region is removed from the geometry,  and the wave function then has support in the classically allowed regions lying in the future and past Milne patches (analogous to the F and B regions of fig.\ref{pendia}), where it is oscillatory, and for the  classically disallowed region when  $\phi<r_2$, where we take it to be  exponentially damped . 
Similarly when $M<M_0$ there is no root with $r>0$ for eq.(\ref{zeoa})  and universe is in the classically allowed region for all values of $\phi>0$.

A similar description in fact also applies in the JT dS case if one bears in mind the conditions arising from dimensional reduction. We have not been very explicit about the range the dilaton can take in this case above. Starting from the near Nariai limit in $4$ dimensions one finds a lower limit for the dilaton with $\phi$ being bigger than a critical value, 
\be
\label{crtic}
\phi>\phi_c,
\ee
 where $\phi_c<0$. Let us consider the $2$ dim. theory also   with the dilaton meeting the constraint,  eq.(\ref{crtic}).  
If one takes a value of $M$ such that $0<M<\phi_c^2$ then the situation is analogous to the case of the dimensionally reduced system discussed above when
$M$ lies in the range eq. (\ref{rangeM}). Namely, the classical geometry is analogous to fig.\ref{pendia} and in the quantum theory a  description very analogous to the dimensionally reduced case applies, with classically allowed regions in the Milne patches, analogous to F and B, and in the black holes and white hole regions. In contrast, the case  $M>\phi_c^2$  is analogous to $M>0$ above and there is only the classically allowed regions in the F and B Milne patches. Finally when $M<0$ the wave function is oscillatory for all  values of $\phi$.

\subsection{Classical Singularities and Their Quantum Resolution}
Next we consider the example of a potential in which the second derivative of the potential diverges with a delta function singularity at $\phi_0$,
\be
\label{derdiv}
U''(\phi_0) \approx C \delta(\phi-\phi_0)
\ee
where $C$ is a constant. The classical equation of motion, eq.(\ref{Req}), leads to the Ricci scalar
\be
\label{rics}
R=U''\approx C \delta(\phi-\phi_0)
\ee
also blowing up at $\phi_0$, leading to a curvature singularity.  

The behaviour near this singularity of the potential is 
\be
\label{potbeha}
U\simeq f(\phi) + C (\phi-\phi_0) \theta(\phi-\phi_0)
\ee
where $f(\phi)$ is a smooth   function with a finite second derivative at $\phi_0$. 
We see that $U(\phi)$ is continuous function near $\phi_0$. 
As a result the wave function which follows from  eq.(\ref{psi}) is 
\be
\label{wvnsin}
\Psi=Ae^{il\sqrt{U-M}}+ B e^{-il\sqrt{U-M}}
\ee
with $A$ and $B$ being arbitrary constants, 
will also be  smoothly behaved near $\phi_0$, leading to the conclusion that  the universe evolves smoothly past the singularity. 
This will also be true for a more general wave function which is obtained by taking a linear combination of the solutions in eq.(\ref{wvnsin}) for different values of $M$. 


\section{Exponential Potentials}
\label{condefpot}
Here we consider a deformation of the JT dS theory where the first derivative of $U(\phi) $ is given by 
\be
\label{altU}
U'(\phi)=2 \phi+2 \sum_{i=1}^n\epsilon_i e^{-\alpha_i\phi}
\ee
We take $\alpha_i>0, \forall i $,  so that $U'(\phi)\rightarrow 2 \phi$ asymptotically, as $\phi\rightarrow \infty$, and  is of the deSitter form.
We also restrict ourselves to the case of ``sharp" defects, for which 
\be
\label{resta}
\pi<\alpha_i<2\pi
\ee
And  take 
\be
\label{restb}
\sum_i\epsilon_i=0
\ee
so that $U'(0)=0$. 

Starting with \cite{witten2020matrix},\cite{eberhardt20232d} and \cite{lin2023revisiting} we now have a fairly elaborate understanding of  the the theory in the AdS case, which arises after carrying an analogous deformation to the dilaton potential, see \cite{witten2020matrix}. This includes  cases where the defects are ``blunt" and do not satisfy the condition, eq.(\ref{resta}), eq.(\ref{restb}). Here, our focus is the dS case and for the sake of simplicity we will restrict ourselves as mentioned above,  to case of sharp defects, eq.(\ref{resta}), meeting condition eq.(\ref{restb}). 

For such deformations, in  the AdS case, the path integral was evaluated perturbatively in the coefficients $\epsilon_i$. An insertion of an $O(\epsilon)$ term gives rise to a puncture in the geometry and it was shown in \cite{witten2020matrix}, \cite{eberhardt20232d} and \cite{lin2023revisiting}, how the path integral can be evaluated in the presence of such a puncture.  Furthermore, it  was argued that when eq.(\ref{resta}) is met and the defects are sharp, and in addition eq.(\ref{restb}) is also true, the correction to the partition function  at order $\epsilon$, i.e. as a sum over the single puncture terms, would in fact give the full result for  the partition function, to all orders in $\epsilon$, with no further corrections. 

We are interested in the deformed JT dS theory, eq.(\ref{altU}),  and in particular in calculating the HH wave function in this theory.
One can think of evaluating this wave function asymptotically, at large $\phi,l,$ by carrying out a path integral along a contour where the metric has a mixed signature
which goes from  signature $(0,2)$, i.e. with two time like directions, to  signature $(1,1)$ at late ``times". In this late time region the geometry is approximately dS and we take the boundary of this late time region  to have length $l_B$ with  the dilaton  along it taking value $\phi_B$. 

The calculation of this HH wave function is closely analogous to the partition function in the deformed  Euclidean AdS theory. The main additional subtlety is  to be careful about  are the different signs involved and the analytic continuation required to go from signature $(0,2)$ to $(1,1)$. 

\subsection{Signs and Continuations}
\label{signcon}
It is worth discussing our sign conventions and the analytic continuations involved in more detail before proceeding. 
We will discuss the $-AdS_2$ and $dS_2$ metrics here which arise as saddle points in the JT dS calculation.
The conventions for the deformed  case which we discuss in the following subsection will be essentially the same. 

Denoting the metric of of a disk geometry  in $-AdS_2$, with signature $(0,2)$, by 
\be
\label{metmads2}
ds^2=-{dr^2\over r^2-1}-(r^2-1)dx^2
\ee
The continuation to signature $(1,1)$ dS is obtained  by taking 
\be
\label{takingd}
r\rightarrow \mp i r.
\ee
and gives rise to the metric
\be
\label{metdsa}
ds^2=-{dr^2\over r^2+1}+(r^2+1)dx^2
\ee
With our conventions the continuation $r\rightarrow -i r$ will give rise to the expanding branch wave function in dS space and the continuation $r\rightarrow +i r$ will give rise to the contracting branch.

Both the $-AdS_2$ and $dS_2$ geometries are solutions to the same action. It is worth being precise about our conventions here.
The path integral is taken to be of the form 
\be
\label{pidef}
\int \mathcal{D}g \mathcal{D}\phi \,e^{-S}
\ee
With this definition the action in $dS$ space is given (for the expanding branch) by 
\begin{equation}
\label{actas}
S_{JT , dS }=- {i\over 2} \left(\int d^2 x \sqrt{\abs{g}} \phi(R-2) - 2 \int_{\partial} \sqrt{\abs{\gamma}} \phi K\right)
\end{equation}
In $-AdS_2$ space the action is given by 
\be
\label{actmads}
S_{JT,-AdS}={1\over 2} \left(\int d^2 x \sqrt{\abs{g}} \phi(R-2) - 2 \int_{\partial} \sqrt{\abs{\gamma}} \phi K\right)
\ee
Finally for good measure the action in Euclidean $AdS_2$ is given by 
\be
\label{actads}
S_{JT,AdS}=-{1\over 2} \left(\int d^2 x \sqrt{\abs{g}} \phi(R+2) + 2 \int_{\partial} \sqrt{\abs{\gamma}} \phi (K-1)\right)
\ee
(Note the boundary term in the $AdS$ case has been defined with a counter term subtracted).

In addition to the metric, eq.(\ref{metmads2}), eq.(\ref{metdsa}) the dilaton takes the value 
\be
\label{valdila}
\phi=\pm i A r
\ee
in the $-AdS_2$ region,
where the $+$ sign is correct for the expanding branch (and therefore related to choosing the $-$ sign in eq.(\ref{takingd})) and the $-$ sign is the correct one for the contracting branch (and corresponds to taking the $+$ sign in eq.(\ref{takingd})).
We therefore see that for both branches in the dS region the dilaton is given   by 
\be
\label{valdilb}
\phi=Ar
\ee
where $A$ is real and positive so that the dilaton becomes large and positive in the asymptotic dS region. 

Henceforth we will discuss the calculations for the expanding branch, the contracting brach can be dealt with  in a similar manner. 
Summarising our discussion above the relevant continuation eq.(\ref{takingd}) is then
\be
\label{contrel}
r\rightarrow -i r
\ee
and the dilaton $\phi$ in the $-AdS_2$ region in this case is given by 
\be
\label{valdilabs}
\phi=i Ar
\ee

As mentioned above we denote the values of the dilaton and the length on the boundary in  $dS$ space by $\phi_B, l_B$.
Let ${\tilde \phi}_B, {\tilde l}_B$ denote the boundary values in $-AdS_2$.
Then 
as a result of the analytic continuations  eq.(\ref{contrel}), eq.(\ref{valdilabs}), we get that 
\be
\label{contcombi}
{\tilde l }_B{\tilde \phi}_B\rightarrow -i l_B \phi_B 
\ee
and 
\be
\label{contcc}
{{\tilde l}_B\over {\tilde \phi}_B}\rightarrow - i {l_B\over \phi_B}
\ee

A mnemonic to keep in mind is that both of these can be achieved by taking 
\be
\label{mne}
{\tilde \phi}_B\rightarrow \phi_B, {\tilde l}_B \rightarrow -i l_B.
\ee 

\subsection{Additional  Calculations}
We are now ready to turn to the case of the deformed theory. The continuation from the $(0,2)$ to the $(1,1)$ signature will be carried out in the asymptotic region and  the rules discussed above for the undeformed theory will continue to be the same as in the deformed case.

The calculation is closely parallel to the one carried out in \cite{witten2020matrix} and \cite{eberhardt20232d} for the Euclidean AdS theory with the deformation 
\be
\label{wittena}
U'(\phi)=-2\phi -2 \sum_i \epsilon_i e^{-\alpha_i\phi}
\ee
The partition function   upto $O(\epsilon)$ in that case is given as a function of $\beta$, the inverse temperature, to be,  
\be
\label{resoe}
	Z =   \left(\frac{\exp(\frac{2\pi^2}{\beta})}{\sqrt{2\pi} \beta^{\frac{3}{2}}} + \sum_i \epsilon_i \frac{\exp(\frac{(2\pi-\alpha_i)^2}{2\beta})}{\sqrt{2\pi} \beta^{\frac{1}{2}}}\right)
 \end{equation}
where the second term arises due to the $O(\epsilon)$ deformation. 
%
%
%

In the $-AdS_2$ case the action, eq.\eqref{actmads},  is now given by 
\be
\label{actbs}
S={1\over 2} \left(\int d^2 x \sqrt{\abs{g}} (\phi R-2\phi - 2 \sum_i \epsilon_i e^{-\alpha_i \phi}) - 2 \int_{\partial} \sqrt{\abs{\gamma}} \phi K\right)
\ee
The HH wave function is calculated by evaluating   the path integral 
\be
\label{Nbpi}
\Psi_{HH}(l,\phi)=\int \mathcal{D}g \mathcal{D}\phi e^{-S}
\ee
Expanding to first order in $\epsilon$ gives 
\be
\label{fohh}
\Psi_{HH}(l,\phi)=\Psi_{HH}^{0}(l,\phi) + \sum_i\epsilon_i  \int d^2y \sqrt{\abs{g(y)}} \int \mathcal{D}g \mathcal{D}\phi e^{-S_{JT}-\alpha_i \phi(y)}
\ee
where the first term $\Psi_{HH}^{0}$ is the HH wave function in the JT theory and the second term is the additional term of interest here. 
In the sum in the  second term consider the contribution  containing the exponent $e^{-\alpha_i\phi}$. Dropping the  coefficient $\epsilon_i$, and the integral 
$\int d^2y\sqrt{|g(y)|}$  in front, we get the path integral  to be 
\be
\label{pIt}
\int \mathcal{D}g \mathcal{D}\phi e^{-{1\over 2}\left[\left(\int d^2x \sqrt{\abs{g}} \phi(R-2)\right)+ 2 \alpha_i \phi(y)- 2 \int_\partial \sqrt{\abs{\gamma}}\phi K\right]}
\ee
In the saddle point approximation we get from the $\phi$ EOM
\be
\label{spa}
\sqrt{|g(x)|}(R(x)-2)+2 \alpha_i \delta(x-y)=0 
\ee
It is easy to see that the equation is solved by a  metric which is locally the same as in the $-AdS_2$ case, eq.(\ref{metmads2}), where we have taken the point $x=y$ to correspond to 
$r=1$. However now the range of the angular variable $x$ is given by 
\be
\label{rangex}
\Delta x=2\pi-\alpha_i
\ee
This can be seen easily by noting that in the $(0,2)$ signature space 
\be
\label{condeuler}
\int \sqrt{\abs{g}}R -2 \int \sqrt{\abs{\gamma}}  K=-4\pi \chi
\ee
where $\chi$ is the Euler character the manifold (here $1$). 
Substituting for the curvature $R$ from eq.(\ref{spa}) then leads to eq.(\ref{rangex}). 

This is exactly analogous to what happens in the Euclidean $AdS$ case. 
The rest of the calculation then  involves evaluating   determinants, keeping track of zero modes, etc, and these steps are  identical to the Euclidean signature case as well. As a result, the final answer can be directly obtained   from the Euclidean $AdS$ case, eq (\ref{resoe}) after the substiution $\beta\rightarrow {\tilde \beta}$ where 
${\tilde \beta}$ is the renormalised inverse temperature given by
\be
\label{deftbeta}
{\tilde \beta}={{\tilde l}_B\over {\tilde \phi}_B}
\ee
with ${\tilde \phi}_B$ and ${\tilde l}_B$ being the asymptotic values of the dilaton and the length in the $-AdS$ case. This gives the result in $-AdS$ to be   
\begin{equation}
	Z =  e^{{\tilde l}_B {\tilde \phi}_B} \left(\frac{\exp(\frac{2\pi^2}{{\tilde \beta}})}{\sqrt{2\pi} {\tilde \beta}^{\frac{3}{2}}} + \sum_i \epsilon_i \frac{\exp(\frac{(2\pi-\alpha_i)^2}{2{\tilde \beta}})}{\sqrt{2\pi} {\tilde \beta}^{\frac{1}{2}}}\right) \label{calhere}
\end{equation}
where we have inserted the prefactor in front, $e^{{\tilde l}_B {\tilde \phi}_B} $, which arises because we have not subtracted a counter term in the calculation we are currently doing.

Using the rule given in eq.(\ref{contcombi}), eq.(\ref{contcc}), eq.(\ref{mne}),  this can now be continued to give the HH wave function, 
\begin{equation} 
	\Psi^+_{HH}[l_B,\phi_B] = \frac{1}{\sqrt{2\pi}} e^{-i l_B \phi_B} \left( \left(\frac{\phi_B}{l_B}\right)^{\frac{3}{2}}e^{\frac{2i\pi^2\phi_B}{l_B}} e^{3i \frac{\pi}{4}} + \sum_i \epsilon_i \left(\frac{\phi_B}{l_B}\right)^{\frac{1}{2}} e^{\frac{i(2\pi-\alpha_i)^2 \phi_B}{2 l_B}} e^{i \frac{\pi}{4}}\right)  \label{hhdef}
\end{equation}

Note the first term is the result in the JT theory and the second arises due to the deformation in the potential. 
It follows for the same reason as in the Euclidean $AdS$ case that when eq.(\ref{restb}) is met this result is true to all orders in the $\epsilon$ expansion. 

\subsection{Further comments}
Let us end this section with some further comments. 

Note first that, as is discussed further in appendix \ref{match}, the result obtained in eq.(\ref{hhdef}) can be expressed in terms of the general solutions we found for the WDW equation eq.\eqref{WdW}. 
One finds that the HH state in the expanding branch is  given for all $\phi,l$ as a sum over various Hankel functions of the form
\begin{equation}
\label{relhh}
\Psi^+_{HH}=A l_B \left(\frac{v_1}{u_1} H_2 ^2 (\sqrt{v_1 u_1}) +\sum_i B_i \sqrt{\frac{v_2}{u_2^i}} H_1^2 (\sqrt{v_2 u^i_2}) \right)
\end{equation}
where,
\begin{align}
	& A = \frac{1}{2} e^{3 i \frac{\pi}{2}} , B_i = \epsilon_i \nonumber \\
	&	u_1= l^2 - 4 \pi^2, u^i_2 = l^2- (2\pi-\alpha_i)^2, v_1= U(\phi_B)=v_2 \label{defAetc}
\end{align}

Second, the contracting branch wave function $\Psi_{HH}^-$ is given by the complex conjugate of the expanding branch one. 
This is to be expected from the fact that the two are related by time reversal (more correctly CPT transformation), and can also be seen by a direct calculation. 

Third, it is known in the Euclidean $AdS_2$ case the partition function eq.(\ref{resoe}) can be expressed in terms of a density of states $\rho(M)$ \footnote{{ The reason why the factor of $1/2$ appears in the exponent $e^{-\beta M\over 2}$ is that in our conventions $M/2$ is  the ADM mass of the AdS black hole, see appendix \ref{therm}.}}
 \be
 \label{partaa}
 Z(\beta)=\int dM  \rho(M) e^{-\beta \frac{M}{2}}
 \ee
where 
\be
\label{defra}
\rho(M)= e^{S_0}\left(\frac{\sinh(2 \pi \sqrt{M})}{4 \pi^2} +\sum_i  \frac{\epsilon_i}{2 \pi \sqrt{M}} \cosh((2\pi-\alpha_i) \sqrt{M})\right)
\ee
As a result the HH wave function eq.(\ref{hhdef}), asymptotically, for large $\phi_B,l_B$,   can also be expressed in terms of $\rho(M)$ and is given by 
\be
\label{HHrho}
\Psi_{HH}^+(l_B,\phi_B)=e^{-i l_B \phi_B}  \int \rho(M) e^{i \beta \frac{M}{2}} dM = e^{-i l_B \phi_B} \int \rho(M) e^{ \frac{i l_B}{2 \phi_B} M} dM
\ee 
It was argued in section \ref{canquan}, where we canonically quantised the theory, that   the expanding branch solutions to the WDW equation must be of the form given in eq.(\ref{forma}), with ${\tilde \rho}$ set to vanish. 
Since $U(\phi)\rightarrow \phi^2$ asymptotically we find that the result obtained above eq.(\ref{HHrho}) is indeed of this form with the coefficient function 
being given by the density of states, $\rho(M)$ eq.(\ref{defra}). 

A related comment is the following. Precisely because the deformed potential is also asymptotically the same as in the  JT dS theory, the HH state in the deformed theory can also be thought of as a 
state in the original undeformed JT dS theory. However it will  not be the HH state of the JT dS theory, but rather a different state. This is an example of something more general, see \cite{Chakraborty:2023yed,Chakraborty:2023los}. By considering the HH state for theories with various corrections to the dilaton potential which vanish as $\phi\rightarrow \infty$, one can asymptotically, as $\phi,l,\rightarrow \infty$, construct various different states in the JT theory. These are  different from the HH state of JT theory and also in general different from each other. 
The non-gravitational analogy we find helpful to  keep in mind here is that excited states in one   theory can sometimes be obtained as ground states of another theory.
 In the current context we would expect, based on the example of the exponential terms, that the coefficient functions for all the states we can obtain in this manner, where the deformed potential $U(\phi)\rightarrow \phi^2$,  as $\phi\rightarrow \infty$, would asymptotically, as $M\rightarrow \infty$, reduce to that of the JT HH state given in eq.\eqref{dosa}. 

We end with a few more comments. Note that  in section \ref{JTMM} we mentioned how states different from the HH state could also be associated with a generalised trace in the Matrix model, eq.\eqref{gentrace}. 
For the deformed theory being considered here, eq.(\ref{altU}) meeting the condition eq.(\ref{resta}), eq.(\ref{restb}), we find, see appendix \ref{topHH}, that 
\be
\label{deffh}
f(H)=1+\sum_i 2\pi \epsilon_i \frac{\cosh({(2\pi-\alpha_i)\sqrt{H}})}{\sqrt{H}\sinh{2\pi\sqrt{H}}}
\ee
Now as per the identification made in section \ref{JTMM}, see  eq.\eqref{gentrace},\eqref{gentracec} and discussion thereafter, we can calculate the transition amplitude   to go from a contracting universe in this state in the far past to an expanding universe in this state in the far future in the SSS matrix theory. Or an amplitude to produces two universes in the expanding branch in this state from ``nothing" in this matrix model, etc. 
These would be of the form 
\be
\label{varampl}
e^{\pm i l_1 \phi_1}
e^{\pm i l_2\phi_2} \langle \Tr(e^{{\mp i l_1\over 2 \phi_1}H} f(H)) 
\Tr(e^{{\mp i l_2\over 2 \phi_2}H} f(H))\rangle
\ee
where the expectation value is being calculated in the matrix theory.
We calculate this in some  detail in appendix \ref{topHH}. 
In contrast we can calculate the same transition amplitudes in the deformed theory for its HH states. 
As discussed in the appendix \ref{topHH} the two results are different. 
This is not a contradiction. To return to the non-gravitational analogy mentioned above, one can have a theory whose excited state is the ground state in another, say  deformed theory,
but transition amplitudes involving the excited state  in the original theory cannot usually be correctly calculated in the deformed one. 

Finally, it is interesting to ask if there is an alternate dual matrix model for the deformed theory. Of course we can ask this in the Euclidean $AdS$ case itself. 
{ It has been suggested in \cite{witten2020matrix} that a matrix model dual exists for the more general potential $U(\phi)$ with an appropriate density of states. For the problem at hand where the potential is of the form eq.\eqref{altU} satisfying the conditions eq.\eqref{resta},\eqref{restb}, the corresponding double scaled matrix model dual is specified by the density of states given as in eq.\eqref{defra}. The second piece of information that is needed to fully specify the double scaled matrix is the double trumpet partition function, which remains unchanged from the pure JT case, for the potential we consider \cite{witten2020matrix}. We provide evidence for this matrix duality in appendix \ref{matmodrec}, although it is not a complete proof.}

\section{HH State in Various Theories}
\label{HHwf}
A striking feature of the HH wave function in the JT theory is that the coefficient function which appears, eq.\eqref{genex} is related to the entropy of the cosmological horizon for  black holes 
in dS space. More precisely, we can easily calculate the entropy, $S_{cos}$, of the cosmological horizon associated with a  black hole solution of mass  $M$ in dS space in the JT theory and  see that for $M\gg 1$ the coefficient function  $\rho_{HH}(M)$ one obtains describing the HH state, eq.\eqref{dosa},  is given by 
\be
\label{relas}
\rho_{HH}(M)=e^{S_{cos}(M)}
\ee. 

To compute $S_{cos}$ note that for the black hole geometry in JT dS given by eq.\eqref{classjtab},\eqref{valdilab2},
the entropy  of the cosmological horizon  is 
\be
\label{entcos}
S_{cos}=2\pi \phi_h=2\pi r_h
\ee
where $\phi_h$ is the value of the dilaton at the cosmological horizon at radial location $r_h$. 
From eq.(\ref{classjtab}) we see that the cosmological horizon lies at $r_h=\sqrt{M}$ giving 
\be
\label{cosb}
S_{cos}=2\pi \sqrt{M}
\ee
In the semiclassical regime the dilaton must be big in $G_N$ units, and $\phi\gg 1$, so that $M\gg 1$. In this limit we see from eq.\eqref{dosa} that the coefficient function  becomes 
\be
\label{coeffiab}
\rho(M)={e^{2\pi \sqrt{M}}\over 8\pi^2}
\ee
which  agrees with $e^{S_{cos}}$, upto an overall $M$ independent coefficient, which just goes into the normalisation of the wave function. 
Note also that black hole horizon is located at $r_{BH}=-\sqrt{M}$ and the entropy associated with this horizon $S_{BH}=2\pi |\phi_{BH}|$ is equal in this model with that of the cosmological horizon. Summing the two would not have given agreement with the coefficient function of the HH state. 

The purpose of this section is to explore this connection between the entropy of the cosmological horizon for black holes and the HH wave function in more general theories. 
We will find that in a large class of theories, meeting some conditions, where the HH wave function can be calculated by a saddle point geometry,
this agreement will continue to be true. Namely that the HH wave function computed in the saddle point approximation will be given in terms of a coefficient function  $\rho(M)$ which agrees in these theories, after exponentiation,  with the entropy of the cosmological horizon for black holes of mass $M$. 

{The condition that the dilaton potential needs to satisfy for the forthcoming analysis to be applicable is,
\begin{equation}
	U(\phi) = -U(i \phi) \label{condimp}
\end{equation}
where $U(\phi)$ is defined in eq.\eqref{jtacta}. It can be seen easily that the JT potential, $U(\phi)=\phi^2$, satisfies the above condition.}

In the JT theory of course we have much more control and the path integral leading to the HH state can be done exactly, this is not true more generally but it is tempting to speculate that this agreement which we find at  the semi-classical level, between the coefficient function in the HH wave function and the entropy of the cosmological horizons for  black holes,  may  continue to hold even beyond the leading semi-classical approximation. 

One way to understand the agreement  discussed above for the JT case is by relating the behavior of the JT theory in dS with its AdS counterpart. 
In the AdS case the partition function at inverse temperature $\beta$ can be understood as an integral with the density  of states  being given, after exponentiation,  by the entropy of  black hole states.
After carrying out an analytic continuation to the dS case it then follows that the HH wave function must also be expressible in terms of a coefficient function which is  equal, after exponentiation,  to the entropy of the cosmological horizon for black hole states. 
We will first sketch out this argument below in the JT case and then use a related argument  to establish the correspondence for more general potentials. 

%
\subsection{JT Theory, dS and AdS via Analytic Continuation}
{
	 
	We will try to phrase the arguments in this section as much as possible in  terms of a general   potential $U(\phi)$ rather than its explicit form given in eq.(\ref{valpot}), 
	even though this might make some of the formulae less transparent. 
	The reason for this is that we will be able to directly apply these arguments then,  for the more general case in the following subsection, in a more straightforward manner.
	The action for the JT dS theory is given in eq.\eqref{jtacta} where 
	\be
	\label{valpot}
	U(\phi)=\phi^2
	\ee
In the asymptotic limit it is known  that the HH wave function can be calculated by carrying out a path integral
over a geometry with mixed signature. Here we discuss the contour suggested by Maldacena where one segment has signature $-2$ and corresponds to a $-AdS_2$ geometry which is then analytically continued (at large values of radial coordinate $r$) to dS space, \cite{Maldacena:2019cbz, Moitra:2021uiv}.

The first step is to show that the on-shell action in $-AdS$ is related to the value of the on-shell action in $EAdS$ which we shall do now. 
The action in $-AdS$ is given by 
\be
\label{actmadsha}
S_{JT,-AdS}={1\over 2} \left(\int d^2x \sqrt{\abs{g}}(\phi R -U'(\phi)) - 2 \int_\partial  \sqrt{\abs{\gamma}}\phi K \right)
\ee
The on-shell action can be  computed in the  $-AdS_2$ region and then continued asymptotically to obtain the  saddle point approximation to $\Psi_{HH}$.
The solution  in the $-AdS_2$ region has the metric
\be
\label{metminusa}
ds^2=-\left[{dr^2\over U(r)-M}+(U(r)-M) {dx^2\over A^2}\right]
\ee
with dilaton given by 
\be
\label{valdilab}
\phi=i r
\ee
Let $\tilde{l}_B, \tilde{\phi}_B$ are the values of the length of the boundary and the dilaton at the boundary in $-AdS$ geometry. Let $\tilde{r}_B$ be the radial location of the boundary.  From the on-shell configuration eq.\eqref{metminusa},\eqref{valdilab}, for potentials, which asymptote to the value eq.\eqref{valpot} for large enough $\phi$, we have the relation
\begin{align}
	\tilde{l}_B=\frac{\tilde{r}_B}{A}\Rightarrow\frac{1}{A}=\frac{i\tilde{l}_B}{\tilde{\phi}_B}\equiv i\tilde{\beta}\label{altpht}
\end{align}
The constant $A$ can be related to the boundary value of the dilaton and the length of the universe in the asymptotic dS space by the analytic continuation eq.\eqref{mne}
\be
\label{bvals}
{1\over A}={l_B\over \phi_B}\equiv  \beta
\ee
The geometry eq.(\ref{metminusa}) smoothly closely at $r_h$ given by the condition
\be
\label{cond}
{4\pi \over U'(r_h)}=\frac{1}{A}={l_B\over \phi_B}
\ee
with $M$ in eq.(\ref{metminusa}) being given in terms of $r_h$ by 
\be
\label{valMa}
U(r_h)=M
\ee

The action in eq.\eqref{actmadsha} can be written this as follows
\be
\label{sabs}
S_{JT,-AdS}=-{\tilde \phi}_B {\tilde l}_B + \Delta S_{JT,-AdS}({\tilde \phi}_B, {\tilde l}_B)
\ee
where 
\be
\label{deltas}
\Delta S_{JT,-AdS}= {1\over 2} \left(\int d^2x \sqrt{\abs{g}}(\phi R -U'(\phi)) - 2 \int_\partial  \sqrt{\abs{\gamma}}\phi (K-1) \right)
\ee
The first term is a boundary term with ${\tilde \phi}_B, {\tilde l}_B$ being the boundary values in the $-AdS_2$ region for  the dilaton and the length of the spatial slice; these are related to 
$\phi_B, l_B$ as given in eq.(\ref{contcombi}) and eq.(\ref{contcc}).  This first term is in  fact   the ``counter-term" which is subtracted in the $AdS$ case. 
The on-shell action $\Delta S_{JT,-AdS}$ for the solution eq.(\ref{metminusa}),  eq.(\ref{valdilb}) becomes 
\be
\label{deltatwo}
\Delta S_{JT,-AdS}={1\over 2A}\left(\int_{r_h}^{{\tilde r}_B} dr  (ir U''(ir)-U'(ir)) - 2 i {\tilde r}_ B\sqrt{\abs{\gamma}}(K-1)\right)
\ee
where we have set the range of $x$ to be $ [0,1]$, and also used the fact that in this case $R=U''(\phi)$. 
Note that the parameter $A$  which appears in the eq.(\ref{deltatwo}) is given   in terms of  $l_B,\phi_B$, after analytic continuation,  by eq.(\ref{bvals}).
%
The lower limit of the integral on the RHS of eq.(\ref{deltatwo}) is the location $r_h$ where the geometry closely smoothly. The value of $r_h$ is related  to $A$ through  the condition eq.(\ref{cond}).
Now we come to an important point. Because the potential $U(\phi)$ satisfies the condition, eq.\eqref{condimp} it follows that 
\begin{align}
\label{condab}
U'(ir)= i U'(r)\\
\label{condac}
U''(ir)=U''(r)
\end{align}
where primes indicate derivatives wrt arguments. 
(this is of course easy to check from the explicit form for $U(\phi)$  in eq.\eqref{valpot}). 
Inserting eq.(\ref{condab})  and eq.(\ref{condac}) in eq.(\ref{deltatwo}) then gives 
\be
\label{conadad}
\Delta S_{JT,-AdS}={i\over 2A}\left(\int_{r_h}^{{\tilde r}_B} dr  ( rU''(r)-U'(r)) - 2 \sqrt{\abs{\gamma}}  {\tilde r}_ B(K-1)\right)
\ee
We can explicitly evaluate the value of the action above by noting that the extrinsic curvature for a boundary at constant value of $r$ is given by 
\begin{align}
	K=\frac{U'(r)}{2\sqrt{U(r)-M}}\label{kexmads}
\end{align}
which gives
\begin{align}
	\label{conad}
	S_{JT,-AdS}=-\frac{i}{A}\left(U(\tilde{r}_B)-\tilde{r}_B\sqrt{U(\tilde{r}_B)-M}+\frac{r_hU'(r_h)}{2}-U(r_h)\right)
\end{align}
the value of the on-shell action in eq.\eqref{conad}, for the potential eq.\eqref{valpot}, further simplifies to
\begin{align}
	\label{conadsim}
	S_{JT,-AdS}\simeq-\frac{i}{2A}\left({r_hU'(r_h)}-M\right)=i\left(\frac{M}{2}\frac{1}{A}-2\pi r_h\right)
\end{align}
Let us now turn to the computation in Euclidean AdS. The action is given by 
\begin{align}
	\label{exsa}
	S_{EAdS}=-{1\over 2} \left[\int d^2x \sqrt{g} (\phi R+U'(\phi))+ 2 \int_\partial\sqrt{\gamma} \phi_B (K-1)\right]
\end{align}
The black hole solution and the corresponding on-shell action is given by 
\begin{align}
\label{onsacta}
ds^2&=\left[{dr^2\over U(r)-M}+(U(r)-M) {dx^2\over A^2}\right],\quad \phi=r\\
S_{EAdS}&= {1\over A} \left(\int _{r_h}^{r_B} dr [rU''(r) -U'(r)]-2 r_B \sqrt{\abs{\gamma}}(K-1)\right)\label{eadsosact}
\end{align}
A similar computation as in the case of $-AdS$ for potential in eq.\eqref{valpot}, will lead to 
%
\begin{align}
	\label{conadesim}
	S_{JT,EAdS}\simeq-\frac{1}{2A}\left({r_hU'(r_h)}-M\right)=\frac{M}{2}\frac{1}{A}-2\pi r_h
\end{align}
We immediately see that the value of the on-shell action for the $-AdS$ and $EAdS$, eq.\eqref{conadsim} and eq.\eqref{conadesim}, are the same upto a factor of $i$. Thus, we have, 
\begin{align}
		\Delta S_{JT,-AdS}= i S_{EAdS}\label{sjtadsads}
\end{align}
where the quantities appearing above $r_h, M$ are related to the parameter $A$ through eq.\eqref{cond},\eqref{valMa}. Since, after analytic continuation, $A$ is related to $l_B,\phi_B$ as eq.\eqref{bvals}, the full value of the on-shell action of $-AdS$ after analytic continuation to dS is given by 
\be
\label{relfinal}
 S_{JT,-AdS}=i\phi_B l_B+ i S_{EADS}(\beta={l_B\over \phi_B})
\ee
The HH-wavefunction in dS in the saddle point approximation is given by 
\begin{align}
	\Psi_{HH}(\phi_B, l_B)&=e^{-S_{JT,-AdS}}\\&=e^{-i l_B\phi_B }e^{- i S_{EAdS}\left(\beta={l_B\over \phi_B}\right) }\label{psihhinseads}
\end{align}
Having related the HH wavefunction to EAdS on-shell action, we now use the relation between the on-shell value of the EAdS  and the black hole entropy to obtain the coefficient function for the HH wavefunction. For the Euclidean AdS, the partition function at inverse temperature $\beta$ is given by, in the saddle point approximation,   
\be
\label{pfsa}
Z(\beta)=e^{-S_{EAdS}(\beta)}
\ee
which can also be expressed in terms of the density of states of the black hole as
\be
\label{scons}
Z(\beta)=\int \rho_{BH}(M) e^{-{\beta M\over 2}} dM
\ee
where 
\be
\label{sconsa}
\rho_{BH}(M)=e^{S_{BH}}=e^{2\pi r_h}.
\ee
Thus, the HH wavefunction in the saddle point limit, eq.\eqref{psihhinseads} can be written as 
\be
\label{logeqb}
\log(\Psi_{HH})=-i l_B\phi_B+i \log I\left({\l_B\over \phi_B}\right)
\ee
where 
\be
\label{valI}
I(\beta)= \int \rho_{BH}(M) e^{-\frac{\beta M}{2}} dM
\ee
Next suppose we were calculating   $I$ ,eq.(\ref{valI}),
with   the inverse temperature $\beta={l_B\over \phi_B}$. The saddle point condition becomes, \footnote{To obtain eq.\eqref{spaa} one must trade $\frac{dr_h}{dM}$ for $U'(r_h)$ using eq.\eqref{id2}.}
\be
\label{spaa}
{4\pi \over U'(r_h)}=\beta={l_B\over \phi_B}
\ee
Let us denote the resulting  saddle point value of the horizon  radial coordinate by $r_{h*}$. Replacing $\beta$ on the RHS of eq.(\ref{valI}) by $\beta=-i {\l_B\over \phi_B}$  we then  get 
\be
\label{logrela}
I\left(-i {\l_B\over \phi_B}\right) =\int e^{2\pi r_h+i{l_B\over \phi_B} {M\over 2}} dM
\ee
In this case  the saddle point condition gives 
\be
\label{newsp}
{4 \pi i \over U'(r_h)}={\l_B\over \phi_B}
\ee
Let the saddle point value be denoted by $\bar{r}_h$.
Now we come to an important point. Notice that because the potential, eq.(\ref{valpot}), 
satisfies the condition eq.\eqref{condimp} the saddle point condition eq.(\ref{newsp})  can be met by taking the saddle point value in eq.(\ref{logrela}) to be  
\be
\label{spvc}
\bar{r}_h= i r_{h*}
\ee
where $r_{h*}$, we remind the reader, satisfies the condition eq.(\ref{spaa}). From the above relation between the saddle points, we can immediately infer that the corresponding saddle point value for the variable $M$ are related by 
\begin{align}
	\bar{M}=-M_*\label{mvalsrel}
\end{align}
As a result the saddle point value of eq.(\ref{logrela})  becomes 
\be
\label{spva}
\log I\left({-il_B\over \phi_B}\right)=i 
\log I\left(  {l_B\over \phi_B} \right)
\ee
Inserting this in eq.(\ref{logeqb}) gives agreement with  the HH wave function to be
\begin{align}
	\Psi_{HH}=e^{-il_B\phi_B}\int dM \rho_{HH}(M)e^{\frac{il_BM}{2\phi_B}} dM \label{fhha}
\end{align}
with 
\begin{align}
	\rho_{HH}(M)=e^{2\pi \phi_h}\label{rhohha}
\end{align}
where $r_h$ as a function of $M$ is obtained by solving eq.\eqref{valMa}. The above result shows that the coefficient function for the HH wavefunction is related to the exponential of the entropy of the cosmological horizon, as promised. 

}

\subsection{General Analysis}
\label{potcd}
We can now generalise the discussion of the previous subsection to more general potentials. 
For a class of potentials we will argue that in the semi-classical approximation the HH wave function can be expressed in terms of the entropy of the cosmological 
horizon of black holes,  as in the dS case. 
 
To begin we consider  a potential which   is of the form 
\be
\label{formab}
U(\phi)=\phi^2(1+ f(\phi^4))
\ee
And take  $f(\phi^4)$ to be a continuous function meeting the conditions that $f(\phi^4)\rightarrow 0$ as $\phi\rightarrow \infty$, so that asymptotically $U(\phi)\rightarrow \phi^2$, and
 to be well behaved at $\phi=0$ so that $U(\phi)$ vanishes like $ \phi^2$ as $\phi\rightarrow 0$. 
The action governing the theory is given in eq.(\ref{jtacta}) with $U'$ now being obtained by the $\phi$ derivative of eq.(\ref{formab}).
The solutions are given by,
\be
\label{bhdsa}
ds^2=-{dr^2\over U(r)-M}+(U(r)-M) {dx^2\over A^2}
\ee
with dilaton given by 
\be
\label{valdilac}
\phi= r
\ee
 We see that the resulting classical solutions will   asymptotically approach $dS$ space in this case.

Note  that the  potential  we are considering  is actually a function of $\phi^2$, rather than $\phi$ itself and importantly, for what follows, it also has the property that 
\be
\label{impp}
U(i\phi)=-U(\phi)
\ee

Parallel to   the discussion above we will also consider  another Euclidean theory  with an action given in eq.(\ref{exsa}). The solutions in this case are given by eq.(\ref{onsacta}). The reader will note that eq.(\ref{jtacta}) and eq.(\ref{exsa}) are generalizations of the JT dS and AdS theories and we will refer to them below as the ``dS-like'' and ``AdS-like'' theories below.

For very large mass $M$ both these theories will have black holes solutions very similar to the dS and AdS cases. As  $M$ decreases the effects of the $f(\phi^4)$ correction will begin to get important and the black hole geometries will also become different. 
The black hole solutions will survive in both cases as long  as 
\be
\label{eqha}
U(r_h)=M
\ee
 has a solution for real $r_h$.
In general there will be multiple solution to eq.(\ref{eqha}), we will be interested in the largest value of $r_h$ satisfying this equation. The dilaton will take its largest value at this horizon.
In the dS-like  case this horizon is the cosmological one and in the AdS-like case it is the black hole horizon in Euclidean theory. The entropy of the cosmological horizon in the dS-like  case will equal that of the black hole in the AdS-like case and will be given by eq.(\ref{entcos}).
The partition function in the AdS-like case will then be given as a function of $\beta$ by eq.(\ref{scons}) where $\rho_{BH}$ is given by eq.(\ref{sconsa}). 
 
 In the dS-like theory we will compute the HH wave function by considering a saddle point geometry which includes a section with $(0,2)$ signature, which we will in an analogous fashion call to be ``$-AdS_2$-like''. 
 The action of this $-AdS_2$-like system is also given by eq.(\ref{actmadsha})  with solution given by eq.(\ref{metminusa}), eq.(\ref{valdilab}). 
 This solution will be continued in the $r\rightarrow \infty$ region using the continuation eq.(\ref{contrel}) to give rise to the solution in the asymptotically dS region. 
 The constant $A$ is fixed in terms of $l_B,\phi_B$ by eq.(\ref{bvals}).
 Also we note that due to the properties of $f$ mentioned above $U'(r)\rightarrow 0$ as $r \rightarrow 0$ and $U'(r)\rightarrow r$ and goes to infinity as $r \rightarrow \infty$. Therefore a solution to eq.(\ref{cond}) always exists for $r\in[0,\infty]$, ensuring that the geometry smoothly closes for all values of $l_B,\phi_B$.  
 
 Now we come to a crucial point. Because the potential $U(\phi)$, while being different from $\phi^2$, is of the form given in eq.(\ref{formab}), the conditions   in eq.(\ref{condimp}), eq.(\ref{condab}), eq.(\ref{condac}) will  continue to be true. The reader can easily check that the arguments on the previous subsection will then also go through 
 resulting in the conclusion that the HH wave function in these cases  can be expressed in terms of the entropy of the cosmological horizon 
 and takes the form eq.(\ref{fhha}),(\ref{rhohha}) where now $\phi_h$ is given by $\phi=r_h$ and $r_h$ being the solution to eq.(\ref{eqha}) as discussed above. 
 
 As two concrete examples consider 
 \be
 \label{ex1}
 f(\phi^4)=a e^{-b \phi^4} 
 \ee
 and 
 \be
 \label{ex2}
 f(\phi^4)={c\over 1+ d \phi^4}
 \ee
 where we take $a,b,c,d>0$.
 In the first case $f$ vanishes exponentially and in the second as a power law, as $\phi\rightarrow \infty$.
 
 We see that in these examples the geometries involved will begin to depart from their counterparts in the dS/AdS/(-AdS) cases when $f\sim O(1)$.
 In the first  example this happens when $\phi_h^4\sim 1/b$ (for $a\sim O(1)$), and in the second when 
 \be
 \label{condcca}
 \phi_h^4\sim 1/d. 
 \ee
 where $\phi_h$ is the horizon value for the dilaton. The additional contribution due to the corrections are $O(\phi^2)$ when these conditions are met. 
 Loop corrections which invalidate the semi-classical approximaion would give a contribution to the action of  order $\sim \log(\phi_h)$. Thus for $b,d\ll 1$ we can ensure that the  corrections in the semi-classical calculation to the leading behavior  due to the $f(\phi^4)$ deformation are more important than the quantum loop effects. 
 
 Note also that the deformed potential in both cases is positive (for real argument) and  take all values $[0,\infty]$. Thus there should be black holes in the AdS-like 
 and dS-like cases for all values of $M$. The integral in eq.(\ref{valI}) would therefore range from $[0,\infty]$. For other functions $f$ meeting the other required properties above this may not be true and the range might be different. We have been  rather cavalier in applying the relation eq.(\ref{spva}) for the saddle point approximation to 
 $I\left({-il_B\over \phi_B}\right)$. In general one would have to ensure that the contour can be deformed to pass through the saddle point located at eq.(\ref{spvc}).
 
 One might wonder if  black holes in AdS with radius $r_h $ given in eq.(\ref{eqha}) are thermodynamically stable? This requires 
 \be
 \label{condada}
 U''(r_h)>0.
 \ee 
 Actually strictly speaking this is not needed for our arguments above, which ultimately pertain to the dS-like cosmology and only require that the partition function  in the AdS case can be obtained as a saddle point approximation of eq.(\ref{valI}) - a fact which follows  from the  Einstein equations. However it is worth noting that eq.(\ref{condada})  is in fact easy enough to ensure. For example in the two cases above, eq.(\ref{ex1}), eq.(\ref{ex2}), we can ensure this  by take $a$, $\beta$ to be $O(1)$ and  somewhat smaller than $1$, see also the discussion in appendix \ref{therm}. 
 
 Let us end with some additional comments. 
 
 Note that since we are considering cosmologies which are asymptotically dS the HH  states of the  deformed theory can also be regarded as states in the JT dS theory, which are different from its HH state. These states would have a coefficient function different from that of the JT HH state. For the JT HH  state  the coefficient function given in eq.(\ref{dosa}), which  in the semi-classical approximation takes the form, eq.(\ref{coeffiab}). 
 In contrast for the states we obtained here as HH states of  deformed theories, the coefficient functions are given by eq.(\ref{rhohha}) where $\phi_h$  is obtained  in terms of the mass $M$ by the relation 
 \be
 \label{relat}
 U(\phi_h)=M
 \ee
  The resulting coefficient functions are different  in general from eq.(\ref{coeffiab}). By changing the potential we can obtain a range of coefficient functions and therefore varying states of the JT theory. 
  
 In fact, we found  something similar in section \ref{condefpot} where we considered deformations of $U$ by adding exponential terms which also vanish as $\phi\rightarrow \infty$. 
 There too   the resulting HH state of the deformed theory could be thought of as a state in the JT theory, but with a   coefficient function eq.(\ref{defra}) different from eq.(\ref{dosa}) and which depends on the additional  exponential terms we add.  The calculations  in section \ref{condefpot} were not done in the saddle point approximation discussed here. In fact the saddle point calculation we discussed above for the HH wave function will not work in these cases because the potential does not meet the condition, eq.(\ref{impp}) \footnote{In contrast for computing the partition function of a black hole in the AdS-like theory the saddle point solution can be found. It yields a result which is closely related to eq.\eqref{resoe} but with a  different normalisation for the correction term,  as was discussed in \cite{witten2020matrix}, see also appendix \ref{semdos}.}.

 For the $dS$ case the value of the dilaton  at  the cosmological horizon, at $r=r_h=\sqrt{M}$, and the black hole horizon, at $r=-\sqrt{M}$, are both equal in magnitude and given by $2\pi \sqrt{M}$. This might suggest that more generally  the coefficient function of the HH wave function could depend on the value of the dilaton at both these horizons. In fact we have argued above that this is not true. In more general cases the dilaton takes different values in magnitude at the cosmological and black hole horizons and  it is the value at the cosmological horizon, and its associated entropy, which enters in the coefficient function determining the HH wave function.

 Finally here we discussed a potential which is asymptotic to dS space. However one could consider situations where $U(\phi)$  does not tend to $\phi^2$, at large $\phi$, and the geometry is not asymptotically dS. In some of these cases too a HH wave function can be defined and can be calculated using semi-classical methods along the lines above. This is discussed further in appendix \ref{Hhgenpot}.
 
\section{Finite Rank  Completions}
\label{fincom}
In this section,  we discuss possible non-perturbative completions of the bulk theory which could be a Matrix theory, away from the double scaled limit, or a system like the SYK model at finite $N$. The discussion  which pertains to  JT theory with  corrections, on account of the finite rank, 
is admittedly speculative. We will also discuss some steps on how to make it more concrete below. 

In section \ref{Jtdsrevsum} we discussed how the canonical quantisation procedure could be consistently carried out in the single universe sector. In section \ref{JTMM}, see \ref{hsn},
we also discussed how states obtained in canonical quantisation could be mapped to states in a typical single realisation, or in  a random   matrix theory\footnote{One could alternatively take  to states in a typical single realisation of the SYK model, or the SYK model with random couplings.}. 
Under the map one identifies the bulk wave function with a generalised trace in the dual theory, eq.(\ref{gentrace}), eq.(\ref{gentracec}).
For the single realisation case this identification follows from the map between a bulk eigenstate of ${\hat M}$ with eigenvalue $M$ and an eigenstate  state  of the Hamiltonian in the matrix theory, 
\be
\label{mapaa}
\Psi^+_M\rightarrow |E=M\rangle
\ee
In the random average case there is no direct map between eigenstates, but the bulk wave function can still be related to the expectation value of a generalised trace
in the RMT, see eq.(\ref{exmat}). 

The discussion in section \ref{hsn} was for the double scaled matrix theory, or infinite rank SYK model. Here we  consider the large but finite rank case. Suppose a gravitational dual continues to exist in this case, involving corrections to JT gravity. Below we will examine when these corrections will set in, in more detail. For now, assuming that this is true and also that the map between bulk states and the boundary, as discussed above, eq.(\ref{mapaa}), eq.(\ref{gentrace}), eq.(\ref{gentracec})   continues to hold, we see that the dimension of the bulk Hilbert space would also be finite in this case and equal to the number  of states, or the  dimension, of the Hilbert space  in the boundary theory. 

If the matrix is $L\times L$,  this dimension is its rank  $L$. For the SYK model it is the  number of states\footnote{We are considering the Majorana model with $N$ flavours.} 
\be
	\label{nusta}
	{\cal N}=2^{N/2}\sim e^{c N}
	\ee
It then follows, that this will be the number of states in the bulk theory. 

Now we come to the main point of this section. The above comments   lead to an interesting conclusion in the finite rank case. JT gravity has a topology counting parameter $S_0$, eq.\eqref{sojt}. As was mentioned in section \ref{Jtdsrevsum}, when JT theory is obtained from dimensional reduction, $S_0$ corresponds to the gravitational entropy of the higher dimensional  Nariai black hole\footnote{More correctly the entropy of the Naria black hole, obtained by summing the cosmological and black hole horizons, is $2S_0$.}.  $S_0$ is the topology counting parameter because higher genus contributions in JT theory are weighted by $(e^{S_0})^{2-2g-b}$.

In the finite rank case this topology counting parameter should  be identified, by standard large $N$ arguments, with the rank of the matrix, $L$, 
\be
	\label{rela}
	e^{S_0}=L,
	\ee
	or  similarly with ${\cal N}$, eq.(\ref{nusta}), in the  SYK theory.	
	Thus we see that the number of bulk states, which equals $L$ or ${\cal N}$,  would  agree with the gravitational entropy of the system. 
	
Note that the identification, eq.(\ref{rela}), is really justified only in the random theory, rather than a single instantiation. In that case   amplitudes which involve multi- boundary amplitudes  on the gravity side, correspond in the matrix to the expectation value of  more than one (generalised ) single trace,  which are  are suppressed by $L^{2-2g-b}$, leading to eq.(\ref{rela}). However, these amplitudes vanish for a single realisation\footnote {Higher genus contributions to single boundary amplitudes also arise, and  in the random average case are consistent with the relation eq.(\ref{rela}), however it is not clear if they can be mapped to corrections to the leading single trace expectation values in the single instantiation case.}. For these reasons,  the connection between the number of bulk states and the gravitational entropy would be justified only in the case where the dual is a random matrix of SYK theory. 

\subsection{Corrections to JT theory}
As mentioned above, the finite rank case would require corrections in the JT theory. To understand when they would have to set in, let us first pause to briefly review the double scaled limit in the SSS model. 
The density of state, which arises for a suitable potential in the matrix theory,   is  of the form \cite{Saad:2019lba},
\be
\label{formran}
\rho(E)={e^{S_{0M}}\over 4\pi^2} \sinh(2\pi\sqrt{a^2-E^2\over 2a})
\ee
with $-a<E<a$.
In the double scaled limit $a\rightarrow \infty$ with 
\be
\label{defce}
{\cal E}=E-a, 
\ee  and $e^{S_{0M}}$, both being  held fixed. Agreement with the path integral is obtained if $S_0$ in the JT theory equals the parameter $S_{0M}$ in the matrix model. 
Note that it is easy to see that in this limit  the  rank $L\rightarrow \infty$.

Now, consider the matrix theory which  is close to the double scaled limit, i.e. with a  rank that is large but finite, $L\gg 1$.  More precisely we now take  $a\gg 1$,
but not $\infty$, and $e^{S_{0M}}$  to be held fixed,  so that 
\be
\label{ranka}
L=\int dE \rho_{MM}(E)  \sim e^{S_{0M}}\sqrt{a} e^{2\pi \sqrt{a}}\gg 1
\ee

The topology counting parameter in the matrix theory is now given by $L$ instead of $e^{S_{0M}}$, with a contribution at genus $g$ with $b$ boundaries going like 
	$L^{2-2g-b}$ as mentioned above, leading to, eq.(\ref{rela}). 
	
	The deviations away from the double scaled limit  would lead to corrections on the gravitational side. 
	These corrections would appear when  ${\cal E}$, eq.(\ref{defce}), takes a value 
	\be
	\label{estaa}
	{\cal E}\sim \order(a).
	\ee
	On the gravity side it is easy to see that these will set in when 
	\be
	\label{rataaa}
	{l\over \phi}\sim {1\over \sqrt{a}}
	\ee
	with the JT approximation being valid when. 
	\be
	\label{condlaa} {l\over \phi}\gg {1\over \sqrt{a}}
	\ee
	
	
	Similar arguments also apply if we replace the matrix theory with the SYK theory. 
	Here there are two parameters $N$, the number of flavours, and $J$, the energy scale corresponding to the fermionic self coupling. 
	The number of states is given by eq.(\ref{nusta}).
	Comparing with the matrix theory, we see that the parameter $a$ maps to $N^2$ in the SYK model, so that the two exponents in eq.(\ref{ranka}) and eq.(\ref{nusta})
	agree.  Here we are only keeping track of the leading $N$ dependence and not the subleading corrections or constants which appear in the relation between the two parameters. 
	The energy ${\cal E}$ in the matrix model similarly corresponds to ${N E\over J}$ in the SYK theory- this can be seen from comparing the  density of states as a  function of energy . In the SYK model at finite $N$ corrections from the JT limit would set in when the $T/J\sim \order(1)$ so that $E\sim N {T^2\over J}\sim N J$. This agrees with 
eq.(\ref{estaa}). On the gravity side the condition $T/J\sim \order(1)$ corresponds to 
\be
\label{condpaa}
{l\over \phi}\sim \order({N^{-1}})
\ee
which agrees with eq.(\ref{rataaa}).

Noting that we are working in conventions where the prefactor ${1\over 16 \pi G_N}$, in the action, eq.\eqref{actionjt},  has been set to unity by rescaling $\phi$, and reinserting the dependence on this prefactor, and also noting that $G_N\sim \order(1/N)$, eq.(\ref{condpaa}) becomes
\be
\label{condpas}
{l\over \phi}\sim \order(1),
\ee
which is a familiar result in understanding when the SYK theory and its Schwarzian approximation begin to differ\footnote{There are additional corrections even for ${l\over \phi}\gg 1$ when the rank is finite.  But these will set in at very large values of ${l\over \phi}$,  at a scale set by the inverse spacing between the energy levels in the Matrix/SYK theory, which is very small.  E.g. in the SYK this spacings is of order $e^{-c N}$, so they would require ${l\over \phi}\sim  e^{c N}$.}.

A first step in putting some of the remarks above on firmer footing would be to see whether the resulting corrections,  to leading order in ${\phi\over l}$, which arise at finite rank,   can be understood  on the gravitational side, as arising due to suitable higher derivative corrections or loop corrections. We leave such an analysis for the future. 

Let us end with one  comment. 
Checking whether the finite rank case has a bulk interpretation is of course also of interest in the AdS case as well. 
One expects. e.g.,  based on D-brane constructions, that at least some near extremal black holes have a matrix model type description at finite rank, and in these cases it is reasonable to expect that the  finite rank  corrections have a bulk interpretation.

\section{Conclusions}
\label{conc}
Here we briefly summarise some of the main conclusions  and open questions. 

We elaborated on the canonical quantisation of JT gravity in dS space in this paper. With the dilaton being the physical clock, we  discussed in more detail how  norm preserving boundary conditions   could be imposed, and analysed the resulting space of states. We also elaborated   on some physical consequences in the resulting quantum theory. For example, starting with a state which is close to  being classical in the far future in the big bang /big crunch branch, we found,  on evolving it backwards in the quantum theory, that it passes through the oribifold singularity and emerges in the 
far past again as a smooth classical spacetime. Extending the wave function to negative values of $\phi$,    led to additional components in the wave function  in the Black hole  and White hole regions, see section \ref{Jtdsrevsum}. 
An important conclusion is that the canonical quantisation in the single universe sector led to an infinite number of states in the quantum theory. 

We then extended this discussion by canonically quantising a large class of models where the dilaton potential is changed, with our focus especially being on   models where the asymptotic form of the dilaton potential remains unchanged, eq.(\ref{asspota}). The canonical quantisation procedure we followed in the JT case can be easily extended more generally for these models, for appropriate branches of their classical solutions  containing spacetime regions which can be foliated by constant dilaton hypersurfaces. Quite generally, we found that these models also had an infinite number of physical states. We showed that when the asymptotic behaviour of $U(\phi)$ agrees with the JT case,   physical states in the deformed theory, asymptotically, can also  be regarded as being states in the JT theory.

Our discussion in this paper serves to highlight the fact that  states different from the Hartle Hawking state (or its expanding branch alone), should be given further attention, in general, in cosmology. 

We also discussed here how some of the  states, other than the HH state,  could be understood, at least to some extent,  in the path integral formulation. The idea was to deform JT theory by changing the dilaton potential, keeping its asymptotic form, for $\phi\rightarrow \infty$ the same. The HH state in the deformed theory then becomes an allowed state in the JT theory, asymptotically, but one which is different from the JT HH state, see section \ref{condefpot}. 

However, the path integral understanding we obtained was limited. 
An important open question which remains is to gain a better understanding in the path integral formulation of the additional states we have obtained in the canonically quantised theories. This will greatly help in a deeper understanding, allowing one to go beyond the single universe sector and compute  topology changing amplitudes for these states as well. 

We also showed how in the single universe sector  wave functions  for physical sates, in the big bang/big crunch branch, could be mapped to states in the double scaled matrix theory, studied in \cite{Saad:2019lba}. A path integral understanding for these states  will hopefully allow this map to be tested beyond the single universe sector, by allowing for a comparison  between the path integral and matrix theory results for topology changing  amplitudes, as has been done for the HH state, \cite{Maldacena:2019cbz}.

An important conclusion to come out of our study is that JT dS gravity can be consistently quantised, within the framework of canonically quantisation,  in the single universe sector. This quantisation, which was discussed in \cite{nanda2023jt}, and also above in section \ref{Jtdsrevsum}, uses the  Klein Gordon inner product and the Rindler basis of modes.  The resulting number of states in the Hilbert space is  infinite. 
In contrast, the path integral gives rise to a very different quantisation, which leads to  the multiverse, since single universes can evolve in the path integral formulation to
multiple  universes, and vice-versa.  Also, the inner product which  arises from the path integral  is different from the KG inner product.  Within the framework of canonical quantisation, working in the single universe sector, one can also define other inner products based on the idea of group invariants, see \cite{Held:2024rmg}, and also section \ref{hsn} above. 
 To go beyond the single universe sector, in the canonical framework  one could try to  third quantise the theory, which is something we did not attempt here. It will be  important to understand  these different ways of quantising gravitational theories,  and the resulting consequences, further.

Towards the end of the paper in section \ref{fincom} we discussed the admittedly speculative possibility that the map between the bulk and the matrix theory could continue to hold away from the double scaled limit, when the matrix theory has finite but large rank. In this case, if the identification between bulk and matrix theory  states  continues to hold, the number of states in the matrix theory would equal $e^{S_0}$- the topology counting parameter in the bulk which is related to higher dimensional deSitter entropy. 
As a first step towards making this tantalising idea more concrete, one should try and see if the modified matrix theory at finite rank, can be understood on the bulk side in terms of corrections to the JT theory, involve extra terms in the action or  higher loop effect. Or attempt to see if such a correspondence holds for the SYK model or models akin to it, at large but finite $N$.

Finally, one cannot escape the feeling that some of the key features we are finding may be  tied to the special properties of two dimensions. E.g. the HH wave function can be calculated by continuing the path integral from  $-AdS_2$  in the JT case- a fact which then got   tied to the observations above that the  double scaled matrix theory, which provides a non-perturbative definition in the AdS case,  also accounts for states  in  dS space. It would therefore be very interesting to investigate higher dimensional theories  and understand if, and to what extent, some of the features we are finding  are more general. 

We leave these interesting questions for the future. 

\section{Acknowledgements}
We thank  Sumit Das, Abhijit Gadde, Alok Laddha, Gautam Mandal, Shiraz Minwalla, Suvrat Raju, and  Onkar Parrikar for discussions. ID, AR, SKS, SPT acknowledge the organizers of ``Quantum Information, Quantum Field Theory and Gravity" held at International Center for Theoretical Sciences for their hospitality during the completion of this paper. ID, AR, SPT, acknowledge support from Government of India, Department of Atomic Energy, under Project  Identification No. RTI 4002 and from the Quantum Space-Time Endowment of the Infosys Science Foundation.  The work of SKS is supported by MEXT KAKENHI Grant-in-Aid for Transformative Research Areas A “Extreme Universe” No. 21H05184. SKS also would like to acknowledge Prof. Junggi Yoon for his hospitality at APCTP, South Korea during the course of this work.  KKN acknowledges the support to CMI from Infosys foundation. KKN and SPT acknowledge the organizers of ``National Strings Meeting 2024" at IIT Ropar for giving us the opportunity to present parts of this work.  Most of all, we are grateful to the people of India and Japan for generously supporting research in String Theory.
\newpage

\appendix

\section{Details on Rindler Basis and $M$ Basis}
\label{detrevsum}
{ In this appendix we give details on the various calculations mentioned in section \ref{Jtdsrevsum}. In subsection \ref{alg} we define an algebra of $\hat{G}_1, \hat{M},$ and $\hat{G}_3$. Then we prove that $\hat{M}$ is not hermitian in \ref{nonhermM}. We then consider the Rindler basis in subsection \ref{rinnorm} and discuss the conservation of norm in that basis. Since the Rindler basis is complete the $M$ basis can be expanded in the Rindler basis. We carry out this analysis in subsection \ref{minrind}. In the final subsection \ref{rindcont} we give details on the continuation of the wavefunction into $\phi<0$ region.}

\subsection{An Algebra}
\label{alg}
The operators we will be considering are given by,
\begin{equation}
	\hat{G}_1 =  \partial_\phi^2 +l^2, \hat{G}_3 = (-i) (l \partial_l- \phi \partial_\phi), \hat{M} = \partial_l^2 + \phi^2
\end{equation}	
 We will work with the Hamiltonian given in eq.\eqref{wda} reproduced below.
\begin{equation}
	\mathcal{H} = -\partial_l \partial_\phi + \frac{1}{l} \partial_\phi - l \phi \label{hus}
\end{equation}
The various commutators between $\mathcal{H}$ and $\hat{G}_1,\hat{M}, \hat{G}_3$ can be worked out to be,
\begin{equation}
	[\mathcal{H}, \hat{G}_1]=0, \,\,\, [\mathcal{H}, \hat{M}] =- \frac{2}{l^2} \mathcal{H}, \,\,\, [\mathcal{H}, \hat{G}_3]=0. \label{com1}
\end{equation}
The commutators between $\hat{G}_1,\hat{M}, \hat{G}_3$ are found to be,
\begin{equation}
	[\hat{G}_1, \hat{M}] = -4 i \hat{G}_3, \,\,\, [\hat{G}_1, \hat{G}_3] = 2 i \hat{G}_1, \,\,\, [\hat{M}, \hat{G}_3] = -2 i \hat{M} \label{com2}
\end{equation}
By working in the subspace where $\mathcal{H}=0$ we can set the second commutator in eq.\eqref{com1} to zero. And thus $\hat{G}_1,\hat{M}, \hat{G}_3$ form an algebra in this subspace. 

A similar algebra was considered in \cite{Held:2024rmg} with a different hamiltonian compared to eq.\eqref{hus}. Note that $\hat{M}$ equals $G_2$ in \cite{Held:2024rmg} up to an overall sign.
\subsection{Details On Non-Hermiticity of $M$}
\label{nonhermM}
{ To show $\hat{M}$ is Hermtian we must prove,
\begin{equation}
	\langle \hat{M}\hat{\Psi}_1,  \hat{\Psi}_2 \rangle- \langle \hat{\Psi}_1, \hat{M} \hat{\Psi}_2 \rangle=0 \label{defeq}
\end{equation}
Consider the LHS. Then using eq.\eqref{inner} we get,
\begin{align}
	\langle {\Psi}_1, \hat{M} {\Psi}_2 \rangle = &\frac{i}{2}\int_{0}^{\infty} dl  \left(\frac{\Psi_1^*}{l} \partial_l \left( \frac{\hat{M}\Psi_2}{l} \right)- \frac{\hat{M}\Psi_2}{l} \partial_l \left(\frac{\Psi_1^*}{l}\right)\right) \nonumber\\&+ \frac{i}{2}\int_{0}^{\infty} dl  \left( \frac{\hat{M}\Psi_2^*}{l} \partial_l \left(\frac{\Psi_1}{l}\right)-\frac{\Psi_1}{l} \partial_l \left(\frac{\hat{M}\Psi_2^*}{l}\right)\right)  \label{defeq2}
\end{align}
with,
\begin{equation}
	\hat{M}{\Psi}_i = (\partial_l^2 + \phi^2)\Psi_i
\end{equation}
where $i=1,2$. 
After a long calculation, eq.\eqref{defeq} simplifies to bulk terms and boundary terms. The bulk terms are given by,
\begin{align}
	\frac{i}{2}\int dl \,{\Psi}_1^*\left(- \frac{8\del_l^2 \Psi_2}{l^3} +\frac{24\del_l \Psi_2}{l^4} - \frac{24\Psi_2}{l^5}\right) - \frac{i}{2}\int dl \,{\Psi}_1\left(- \frac{8\del_l^2 \Psi_2}{l^3} +\frac{24\del_l \Psi_2}{l^4} - \frac{24\Psi_2}{l^5}\right)^*  \label{but}
\end{align}
while the boundary terms read,
\begin{align}
	& \frac{i}{2} \left(2\left(\frac{1}{l} \del_l {\Psi}_1^*\right) \partial_l \hat{\Psi}_2|- 2{\Psi}_1^*\del_l\left(\frac{1}{l} \partial_l \hat{\Psi}_2\right)|-\hat{\Psi}_2 \left(\frac{1}{l} \del_l^2 {\Psi}_1^*\right)| + \left(\frac{1}{l} \del_l^2 {\Psi}_2\right) \hat{\Psi}_1^*|\right)  \nonumber \\
	&-\frac{i}{2} \left(2\left(\frac{1}{l} \del_l {\Psi}_1\right) \partial_l \hat{\Psi}_2^*|- 2{\Psi}_1\del_l\left(\frac{1}{l} \partial_l \hat{\Psi}_2^*\right)|-\hat{\Psi}_2^* \left(\frac{1}{l} \del_l^2 {\Psi}_1\right)| + \left(\frac{1}{l} \del_l^2 {\Psi}_2^*\right) \hat{\Psi}_1|\right)   \label{bt}
\end{align}
Eq.\eqref{bt} must vanish at both $l=0,\infty$. The above terms will not vanish in general. Hence $\hat{M}$ is not hermitian.}

\subsection{Conservation of Norm in Rindler Basis}
\label{rinnorm}
In this subsection we review the conservation of the norm in the Rindler basis as was described in \cite{nanda2023jt}. The wavefunction in the Rindler Basis is given by,
\begin{equation}
	{\hat \Psi}=\int dk \bigl[a(k) e^{i k \theta} J_{-i |k|}( \xi)+ b(k) e^{-i k \theta}J_{i |k|}( \xi)\bigr] \label{psibessel}
\end{equation}
where,
\begin{equation}
	\xi = l \phi, e^{\theta} = \frac{\phi}{l}.
\end{equation}
We are discussing the KG norm here, given by, eq.\eqref{normaa},  
\begin{equation}
	\langle \hat{\Psi}, \hat{\Psi} \rangle = i\int dl (\hat{\Psi}^* \del_l \hat{\Psi} - \hat{\Psi} \del_l \hat{\Psi}^*) \label{normeq}
\end{equation}
The above norm is conserved if the ``flux" of probability, 
\begin{equation}
	\mathcal{C}_N = i (\hat{\Psi}^* \del_\phi \hat{\Psi} - \hat{\Psi} \del_\phi \hat{\Psi}^*) \label{cneq3}
\end{equation}
vanishes at $l=0,\infty$.

This requirement at $l=0$ is non-trivial, in particular. It can be met by taking 
\be
\label{condasa}
\partial_\phi \Psi=0
\ee
at $l=0$. 
We see below that imposing this condition will set half the  coefficients in the Rindler basis expansion above to vanish, 
 
 As $l\rightarrow 0$  
\begin{equation}
	J_{-i |k|}( \xi) \rightarrow \alpha(k) \xi^{-i \abs{k}}
\end{equation} 
where the coefficient $\alpha(k)$ is given by 
\be
\label{valalphak}
\alpha(k)=\frac{2^{i \abs{k}}}{\Gamma(1-i \abs{k})}
\ee
Hence a Rindler mode $J_{-i |k|}(\xi) e^{ik \theta}$, for $k>0$, has the limit, as $l\rightarrow 0$, 
\be
\label{limasa}
J_{-i |k|}(\xi) e^{ik \theta}\rightarrow \alpha(k) l^{-2ik}
\ee
and as a result meets the condition eq.(\ref{condasa}).
On the other hand the mode $J_{-i |k|}(\xi) e^{ik \theta}$, for $k<0$, has the limit $\alpha(k) \phi^{2ik}$ and does not meet the required condition eq.(\ref{condasa}). 
Thus we will set the coefficients $a(k)$ for $k<0$ to vanish. Similarly the coefficients $b(k), k<0$, will   be set to vanish.

This leaves the coefficients $a(k), b(k), k>0$, with the wave function taking the form
\begin{equation}
	{\hat \Psi}=\int_{k>0} dk \bigl[a(k) e^{i k \theta}J_{-i |k|}( \xi)+ b(k) e^{-i k \theta}J_{i |k|}( \xi)\bigr] \label{psibessel2}
\end{equation}

While we will not be very explicit the condition that ${\cal C}_N$, eq.(\ref{cneq3}),  vanishes at $l\rightarrow \infty$ can also  be met  by taking wave functions which decay exponentially at large $l$. 

Since we have now ensured that the norm is conserved we can  evaluate it for any value of $\phi$. In particular we can evaluate it in  the $\phi \rightarrow 0$ limit. In this limit  the norm, eq.\eqref{normeq}, reads,
\begin{equation}
	\langle \hat{\Psi}, \hat{\Psi} \rangle = \frac{2}{\pi} \int_{k>0} dk  \sinh(k\pi)(\abs{a(k)}^2 - \abs{b(k)}^2) \label{normrind}
\end{equation}

\subsection{Expansion of $M$ Basis in Rindler Basis }	
\label{minrind}
In this subsection we will derive an expression for the expansion of the wavefunction in $M$ basis in the Rindler Basis. Let us briefly recapitulate the necessary formulas. The wavefunction in the Rindler Basis is given in eq.\eqref{psibessel2}. Since,
\begin{equation}
	J_{i k} = \frac{1}{2} H_{i k}^{(1)} +  \frac{1}{2} H_{i k}^{(2)} 
\end{equation}
and,
\begin{equation}
	J_{-i k} = \frac{1}{2} H_{-i k}^{(1)} +  \frac{1}{2} H_{-i k}^{(2)}
\end{equation}
where $H_\alpha^{(1)},H_\alpha^{(2)}$ are the Hankel function of the first and second kind with index $\alpha$. 
eq.\eqref{psibessel2} can be written as,
\begin{align}
	{\hat \Psi} &=\int_{k>0} dk \frac{1}{2} \left((b(k) e^{-i k \theta}  H_{i k}^{(2)} (\xi) + a(k) e^{i k \theta}  H_{-i k}^{(2)} (\xi)) + (b(k) e^{-i k \theta}  H_{i k}^{(1)}(\xi) + a(k) e^{i k \theta}  H_{-i k}^{(1)}(\xi) ) \right) \nonumber 
\end{align}
Expanding for large $\xi$ we get,
\begin{equation}
	{\hat \Psi} =\int_{k>0} dk \frac{1}{2} \left( \frac{e^{-i \xi}}{\sqrt{\xi}}(b(k) e^{-i k \theta} c (k) + a(k) e^{i k \theta} c(-k)) + \frac{e^{i \xi}}{\sqrt{\xi}} (b(k) e^{-i k \theta}  d(k) + a(k) e^{i k \theta}  d(-k))  \right) \label{psibessel3}
\end{equation}
where,
\begin{equation}
	c(k) = \sqrt{\frac{2}{\pi}} e^{-\frac{k \pi}{2} + i \frac{\pi}{4}}, d(k) = \sqrt{\frac{2}{\pi}} e^{\frac{k \pi}{2} - i \frac{\pi}{4}} \label{valckdk}
\end{equation}
Given the above equation one can easily obtain in the case of expanding branch,
\begin{align}
	\frac{1}{2\pi}\int \hat{\Psi} e^{-i k \theta} d\theta= \frac{1}{2} \frac{e^{-i \xi}}{\sqrt{\xi}} \begin{cases}
		a(k)  c(-k) \,\,\, , k>0  \\
		b(\abs{k})  c (\abs{k}) \,\,\,, k<0  
	\end{cases} 
	\label{rel1}
\end{align}
The wavefunction for the expanding branch in the $M$ basis is given by,
\begin{equation}
	\hat{\Psi} = \frac{1}{l} \int \rho(M) e^{-i l \sqrt{\phi^2-M}} dM \label{psieq1}
\end{equation}	
The asymptotic form of eq.\eqref{psieq1} in the limit $\xi \rightarrow \infty$ is given by,
\begin{equation}
	\hat{\Psi} = \frac{e^{-i \xi}}{\sqrt{\xi}} e^{\frac{\theta}{2}} \int \rho(M) e^{i \frac{M}{2} e^{-\theta}} dM \label{psieq2}
\end{equation}
Then one can write,
\begin{equation}
	\frac{1}{2\pi}\int \hat{\Psi} e^{-i k \theta} d\theta =\frac{1}{2\pi} \frac{e^{-i \xi}}{\sqrt{\xi}} \int_{-\infty}^\infty d\theta e^{\frac{\theta}{2}} \int \rho(M) e^{i \frac{M}{2} e^{-\theta}} e^{-i k \theta} dM \label{rel}
\end{equation}	
Using the above equation along with eq.\eqref{rel1} one can obtain $a(k), b(k)$ given a coefficient function $\rho(M)$.  We now show the steps for this derivation. The integral on the RHS of eq.\eqref{rel} can be rewritten as follows,
\begin{equation}
	\frac{1}{2\pi}\int_{0}^\infty \frac{dy}{y^{\frac{3}{2}}} \int \rho(M) e^{i \frac{M}{2} y} y^{i k} dM \label{int}
\end{equation}
where,
\begin{equation}
	y = e^{-\theta}
\end{equation}	
In the limit $y \rightarrow 0$ the integral eq.\eqref{int} simplifies to,		
\begin{equation}
	\frac{1}{2\pi}\int_{0}^\infty \frac{dy}{y^{\frac{3}{2}}}  y^{i k} \int \rho(M) \left(1 + i \frac{M}{2} y \right)  dM
\end{equation}	
We see that after integrating w.r.t $y$ the first term is divergent when $y \rightarrow 0$ while the second term is convergent. To avoid the divergence we will regulate the integral as follows. Let
\begin{equation}
	\int\rho(M) dM=0 \label{cond3}
\end{equation}
Using this condition we can set the divergent term to zero. Thus the integral is convergent for $y \rightarrow 0$. In  the other limit $y \rightarrow \infty$ the integral on the rhs of eq.\eqref{rel} will be convergent as long as the integral over $M$ grows slower than $\sqrt{y}$. 

Let us now evaluate the integral over $y$ in eq.\eqref{int}. We will take the lower limit of integral to be $\epsilon$ where $\epsilon>0$. Then evaluating the $y$ integral and expanding for small $\epsilon$ we get,
\begin{equation}
	\int \rho(M) dM \left( \frac{2 i}{i + 2 k} \epsilon^{i k- \frac{1}{2}}+ \left(\frac{2}{M}\right)^{i k- \frac{1}{2}} \Gamma\left(i k -\frac{1}{2}\right) e^{- k \frac{\pi}{2} - i \frac{\pi}{4}}\right)
\end{equation}
Then using the condition eq.\eqref{cond3} we have,
\begin{equation}
	\int \rho(M) dM \left(\frac{2}{M}\right)^{i k- \frac{1}{2}} \Gamma\left(i k -\frac{1}{2}\right) e^{(i k- \frac{1}{2}) i \frac{\pi}{2}}
\end{equation}	
Using eq.\eqref{rel1} we can now determine $a(k), b(k)$. We get,
\begin{equation}
	{a} (k) =  {\frac{-ie^{-k \pi}}{\sqrt{2\pi}}} \Gamma\left(i k -\frac{1}{2}\right) \int  dM \, \rho(M)\, \left(\frac{2}{M}\right)^{i k- \frac{1}{2}} \label{aeq1}
\end{equation}	
and,
\begin{equation}
	b (k) ={\frac{-ie^{k \pi}}{\sqrt{2\pi}}}  \Gamma\left(-i k -\frac{1}{2}\right) \int  dM\, \rho(M)\, \left(\frac{2}{M}\right)^{-i k- \frac{1}{2}}\label{beq1}
\end{equation}

\subsection{Analytic Continuation to Other Branches}
\label{rindcont}
In this subsection we will show how to obtain other branches of the wavefunction given one knows the expanding branch wavefunction. In particular we will show the continuation from the expanding branch to contracting branch and the continuation of the wavefunction to $\phi<0$ region starting from $\phi>0$.

\subsubsection{From Expanding to Contracting Branch}
\label{expcon}
Consider the wavefunction eq.\eqref{psibessel3} which can be rewritten as follows.
\begin{equation}
	\hat{\Psi} (\xi,\theta) = \hat{\Psi}_+ (\xi,\theta) + \hat{\Psi}_{-} (\xi,\theta)
\end{equation}
where the subscripts ${}_{\pm}$ denote expanding and contracting branch respectively and asymptotically as $\xi \rightarrow \infty$ with $\theta$ kept fixed they read,
\begin{align}
	&{\hat \Psi}_+ (\xi,\theta) =\int_{k>0} dk  \sqrt{\frac{1}{2\pi}} e^{i \frac{\pi}{4}}  \frac{e^{-i \xi}}{\sqrt{\xi}}\left(b(k) e^{-i k \theta} e^{-k \frac{\pi}{2}} + a(k) e^{i k \theta} e^{k \frac{\pi}{2}}\right) = \sqrt{\frac{1}{2\pi}} e^{i \frac{\pi}{4}}  \frac{e^{-i \xi}}{\sqrt{\xi}} f(\theta)\nonumber \\
	&{\hat \Psi}_{-} (\xi,\theta) = \int_{k>0} dk  \sqrt{\frac{1}{2\pi}} e^{-i \frac{\pi}{4}} \frac{e^{i \xi}}{\sqrt{\xi}} \left(b(k) e^{-i k \theta} e^{k \frac{\pi}{2}}   + a(k) e^{i k \theta}  e^{-k \frac{\pi}{2}}\right)= \sqrt{\frac{1}{2\pi}} e^{-i \frac{\pi}{4}}  \frac{e^{i \xi}}{\sqrt{\xi}} g(\theta)
\end{align}
where we have put eq.\eqref{valckdk}. From the above two equations it is clear that
\begin{equation}
	f(\theta+i \pi) = i g(\theta) \label{rel2}
\end{equation}
Thus if one knows the expanding branch, the contracting branch can be derived using the above analytic continuation. Let us see how it works by applying the above relation to an expanding branch wavefunction of the form eq.\eqref{psieq1}. Using the asymptotic form given in eq.\eqref{psieq2} one has,
\begin{equation}
	f(\theta) =  e^{\frac{\theta}{2}} \int_{-M_0}^{M_0} \rho(M)  e^{i e^{-\theta} \frac{M}{2}} \label{feq}
\end{equation}
In an analogous fashion from the asymptotic form of the contracting branch one can write, 
\begin{equation}
	g(\theta) = e^{\frac{\theta}{2}} \int_{-M_0}^{M_0} \tilde{\rho}(M) e^{-i e^{-\theta} \frac{M}{2}} \label{geq}
\end{equation}
Using eq.\eqref{rel2} , eq.\eqref{feq}, and eq.\eqref{geq} we then get,
\begin{equation}
	\tilde{\rho} (M) = \rho(M) \label{rhoeq}
\end{equation}

As an example, we consider an expanding branch coefficient function which has only support for $M>0$, so it belongs to the big bang/big crunch branch. Concretely, we take the coefficient function to be 
\be
\label{exprhoaa}
\rho(M)=2 \sin[(M-M_0)x_0] e^{-{1\over 2}(M-M_0)^2}
\ee
 when $-\Delta<M-M_0<\Delta$
 and  to vanish otherwise. 
 Here $\Delta<M_0$ so that the smallest value,, $M=M_0-\Delta>0$ and the coefficient function  corresponds  entirely to  the big bang branch. In addition, we 
  take $\sigma\gg1$ to get a ``tight" Gaussian in the classical limit, and $x_0\sim O(1)$.  This example was discussed in \cite{nanda2023jt}, see the discussion in section 4.2.3 around eq.(4.87).   Note that the coefficient function meets the condition, eq.(\ref{condcoeffa}).
  
  At late times when 
  \be
  \label{condaad}
  l,\phi\gg 1, \ \ {\rm with} \  {l\over \phi} \ \  {\rm kept  \ fixed}
  \ee
  and 
  \be 
  \label{condaada}
   \phi^2-M_0\gg \sigma,   \phi^2\gg M_0, 
  \ee
  the wave function takes the form 
 \be
 \label{formalae}
 \hat{\Psi}=i {\sqrt{2\pi \sigma}\over 2l}e^{-il\phi}\{-e^{-\sigma/2(x_0+{l\over 2 \phi})^2}+e^{-\sigma/2(x_0-{l\over 2 \phi})^2}\}
 \ee
 
 In the late time  limit we are taking the first term can be dropped and 
 \be
 \label{romad}
 \hat{\Psi}=i {\sqrt{2\pi \sigma}\over 2l}e^{-il\phi}e^{-\sigma/2(x_0-{l\over 2 \phi})^2}
 \ee
 This  corresponds to a wave packet which is well described by a classical solution in the expanding branch. 
 
 Evolving back in time the classical universe will go hit an orbifold singularity, see section \ref{Jtdsrevsum}. 
 What happens in the quantum theory?
 
 From eq.(\ref{rhoeq})  the contracting branch wave coefficient function  ${\tilde \rho}$   is also given by eq.(\ref{exprhoaa}). As a result in the far past eq.(\ref{condaad}) and eq.(\ref{condaada}) are again met, and the contracting branch 
  wave function is given by  eq.(\ref{formalae})
 \be
 \label{formalad}
 \hat{\Psi}=i {\sqrt{2\pi \sigma}\over 2l}e^{il\phi}\{-e^{-\sigma/2(x_0-{l\over 2 \phi})^2}+e^{-\sigma/2(x_0+{l\over 2 \phi})^2}\}
 \ee
 Now the second term on the RHS can be dropped leading to 
 \be
 \label{romade}
\hat{\Psi}=(-i) {\sqrt{2\pi \sigma}\over 2l}e^{il\phi} e^{-\sigma/2(x_0-{l\over 2 \phi})^2}
 \ee
 
 Thus the universe goes through the orbifold singularity essentially unscathed, in accordance with our general discussion in section \ref{addcom}.  The expanding  and contracting branch wave functions   are  
  closely related,  with an overall   phase shift, since the coefficient in front changes sign, and with the factor $e^{-il\phi} $  being replaced by $e^{il\phi}$ , as expected. 

\subsubsection{Continuation to Negative values of $\phi$ }
In the section \ref{secrindlerquan} the continuation of $\Psi$ to  negative values of $\phi$ branches was discussed. Here we give some more details about how 
the continuation is carried out.  

\begin{figure}
	
	\centering
	\tikzset{every picture/.style={line width=0.75pt}} 
	
	\begin{tikzpicture}[x=0.75pt,y=0.75pt,yscale=-1,xscale=1]
		
		\draw    (313,89.6) -- (312.05,54) ;
		\draw [shift={(312,52)}, rotate = 88.48] [color={rgb, 255:red, 0; green, 0; blue, 0 }  ][line width=0.75]    (10.93,-3.29) .. controls (6.95,-1.4) and (3.31,-0.3) .. (0,0) .. controls (3.31,0.3) and (6.95,1.4) .. (10.93,3.29)   ;
		\draw    (224,152.6) -- (182,152.03) ;
		\draw [shift={(180,152)}, rotate = 0.78] [color={rgb, 255:red, 0; green, 0; blue, 0 }  ][line width=0.75]    (10.93,-3.29) .. controls (6.95,-1.4) and (3.31,-0.3) .. (0,0) .. controls (3.31,0.3) and (6.95,1.4) .. (10.93,3.29)   ;
		\draw    (301,140.6) -- (118,140.6) ;
		\draw [shift={(116,140.6)}, rotate = 360] [color={rgb, 255:red, 0; green, 0; blue, 0 }  ][line width=0.75]    (10.93,-3.29) .. controls (6.95,-1.4) and (3.31,-0.3) .. (0,0) .. controls (3.31,0.3) and (6.95,1.4) .. (10.93,3.29)   ;
		\draw    (301.5,257.4) -- (300.51,25.8) ;
		\draw [shift={(300.5,23.8)}, rotate = 89.75] [color={rgb, 255:red, 0; green, 0; blue, 0 }  ][line width=0.75]    (10.93,-3.29) .. controls (6.95,-1.4) and (3.31,-0.3) .. (0,0) .. controls (3.31,0.3) and (6.95,1.4) .. (10.93,3.29)   ;
		
		\draw (248,146) node [anchor=north west][inner sep=0.75pt]   [align=left] {$\displaystyle l$};
		\draw (309,102) node [anchor=north west][inner sep=0.75pt]   [align=left] {$\displaystyle \phi $};
		\draw (195,75) node [anchor=north west][inner sep=0.75pt]   [align=left] {$\displaystyle \xi ,\theta $};
		\draw (203,178) node [anchor=north west][inner sep=0.75pt]   [align=left] {$\displaystyle \tilde{\xi } ,\ \tilde{\theta }$};
		\draw (139,46) node [anchor=north west][inner sep=0.75pt]   [align=left] {$\displaystyle \phi  >0$};
		\draw (136,213) node [anchor=north west][inner sep=0.75pt]   [align=left] {$\displaystyle \phi < 0$};

	\end{tikzpicture}
	\caption{Extension of the wavefunction to negative $\phi$ }
	\label{negphiwf}
\end{figure}
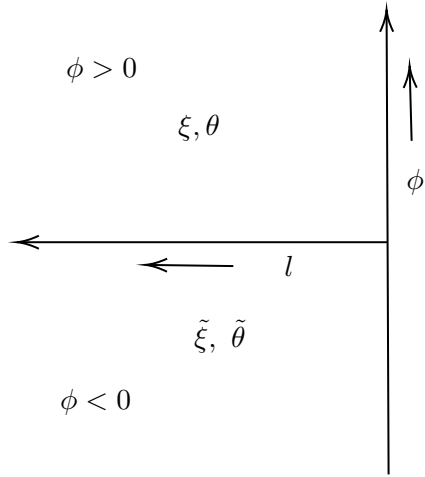

Consider a mode $J_{-i|k|}(\xi) e^{ik\theta}$, with $k>0$ present in the wave function eq.(\ref{psibessel2}). In the limit when $\phi\rightarrow 0$, with $l$ held fixed, this behaves as 
\be
\label{limasa2}
J_{-i|k|}(\xi) e^{ik\theta} \rightarrow \alpha(k) l^{-2ik},
\ee
where $\alpha(k)$ is given in eq.\eqref{valalphak}, 
and becomes independent of $\phi$. It can be continued to $\phi<0$ by simply imposing continuity at $\phi=0$. In fig.\ref{negphiwf} we show the two regions $\phi>0, \phi<0$, which meet along the $-x$ axis at $\phi=0$, with $l$ taking positive values. We impose continuity $\forall l>0$, at $\phi=0$. Similarly the mode $J_{i|k|}(\xi) e^{-ik\theta}\rightarrow \alpha(-k) l^{2ik}$ and can also be continued at $\phi=0$.

More explicitly, for $\phi<0$ the WdW equation eq.\eqref{wda} can  also be solved in terms of a similar set of Rindler basis modes. 
In terms of the variables
\begin{equation}
	\tilde{\xi} = - l \phi, e^{\tilde{\theta}} = - \frac{\phi}{l}, \,\,\, \phi<0. \label{xieq}
\end{equation}
the solutions are of the form $J_{\pm i |k|}({\tilde \xi}) e^{\pm i k \theta}$. 

The KG norm in the region $\phi<0$ continues to be of the form eq.(\ref{normeq}) with the integral now being evaluated on a $\phi<0$ surface. 
Imposing the condition eq.(\ref{condasa}) on the mode functions, to ensure no leakage of flux at $l=0$, again removes half the modes, so that the wave function  can be expanded as 
\be
\label{expndas}
\Psi=\int_{k>0} dk[{\tilde a}(k) e^{ik{\tilde \theta}}
J_{-i|k|}({\tilde \xi})
+ e^{-ik {\tilde \theta}} {\tilde b}(k) J_{i|k|}({\tilde \xi}) ]
\ee
in the region $\phi<0$. 

Now imposing continuity at $\phi=0$ gives
\begin{eqnarray}
\label{condabs}
a(k)& = & {\tilde a}(k) \\
b(k) & = & {\tilde b}(k)
\end{eqnarray}
The full solution then takes the form eq.\eqref{psibessel} in region $\phi>0$ and eq.(\ref{expndas}) for $\phi<0$, with the coefficients being related by eq.(\ref{condabs}).

Let us note before proceeding that the limiting form eq.(\ref{limasa}) as a function of $l$ has a branch cut at $l=0$. We have defined it so that  the branch cut  runs along the negative $l$ axis
(the positive $x$ direction in fig.\ref{negphiwf}). The resulting function of $l$ is analytic along the positive $l$ axis where we are imposing the continuity condition. 

We now turn to the implications of this continuation for the various types of solutions, see section \ref{addcom} above.

$\underline{\textbf{Big Bang/Crunch Sector}}$:
We can expand the wave function for $\phi<0$ in the asymptotic region where $\phi\rightarrow -\infty$, with $l/\abs{\phi}$ kept fixed and match to the coefficient functions
 ${\hat \rho}, {\tilde {\hat \rho}}$, which arise in the expansion in the M basis in this region, eq.(\ref{formabp}).
Since the coefficients meet the conditions, eq.(\ref{condabs}), it is easy to see that these coefficient functions satisfy the conditions, eq.(\ref{relcaa}), eq.(\ref{relcab}) and thus are determined by the coefficient function $\rho(M)$ which arises in the expanding branch of the big bang/big crunch sector.
 Recall that the coefficient function ${\tilde \rho}$, which determines the contracting branch wave function in the big bang/big crunch sector, also equals $\rho$,  eq.(\ref{rhoeq}). Thus all these four coefficient functions turn out to be equal. 

In this way, the wave function in the $\phi<0$ branch,  corresponding to the black hole /white hole interior, is  determined in terms of the coefficient functions in the big bang/ big crunch branch. Quite remarkably, the wave function  evolves smoothly from the big bang/ big crunch branch to the interior of the black hole/white hole region, even though after the orbifold identification is carried out the two regions are disconnected see fig \ref{dsfig}(b) where the big bang/big crunch region is in red and the black hole/white hole region is in green.

$\underline{\textbf{Bounce Sector}}$:
{Classically there are two branches in the bounce sector, an expanding branch with $\phi\rightarrow\infty$ and a contracting branch with $\phi\rightarrow-\infty$. In the wavefunction in the $M$ basis these would correspond to coefficient functions $\rho(M)$ and $\tilde{\hat{\rho}}(M)$ in eq.\eqref{genex} and eq.\eqref{formabp}. To remind the reader, the expanding branch corresponds to $\pi_{\phi}<0$ and the contracting branch has $\pi_{\phi}>0$. However in the quantum theory four different branches are possible, as is discussed in section\ref{addcom}). An expanding branch with $\phi>0$ and $\pi_{\phi}<0$ and another expanding branch with $\pi_{\phi}<0$ and $\phi<0$. Similarly in the contracting branch,  one branch with $\pi_{\phi}>0$ and $\phi<0$ and another with $\pi_{\phi}>0$ and $\phi>0$. As in the case of big bang/crunch sector all the coefficient functions turn out to be equal.}

\section{More Details on The Classical Properties}
\subsection{Variation of The Action}
\label{varprin}
In this section we give a brief derivation of equations eq.\eqref{Req} and eq.\eqref{eomJT} and also the form of the dilaton gravity action in various spacetimes. The equation of motion by varying the action with respect to the dilaton is straightforward and gives rise to eq.\eqref{Req}. Thus we will focus only on eq.\eqref{eomJT}.

Consider then the action,
\begin{equation}
	S_{U} = \frac{\eta}{2} \pqty{\int d^2 x\,\sqrt{\abs{g}}\, (\phi R- U'(\phi))+ 2 \epsilon \int_{bdy}\sqrt{\abs{\gamma}}\phi K } \label{actgenu}
\end{equation}
Here,
\begin{align}
	\eta = \begin{cases}
		-i \, , \text{Lorentzian metric} \\
		-1 \, , \text{Euclidean metric}
	\end{cases}
\end{align}
and in case of Lorentzian metric,
\begin{align}
	\epsilon = \begin{cases}
		+1 \, , \text{Timelike Hypersurface} \\
		-1 \, , \text{Spacelike Hypersurface}
	\end{cases}
\end{align}
while for Euclidean metric $\epsilon=1$ and for a metric with two timelike directions $\epsilon=-1$. It can easily be checked that the action given in eq.\eqref{actgenu} gives rise to the correct actions for simple cases like JT in Lorentzian AdS or dS or euclidean AdS using the above conventions. Choosing $\eta=1$ for a $-$AdS$_2$ metric we get from eq.\eqref{actgenu} with $U'(\phi)=2 \phi$,
\begin{equation}
	S_{-AdS} = \frac{1}{2} \pqty{\int d^2 x\,\sqrt{\abs{g}}\, (\phi R- 2 \phi)- 2 \int_{bdy}\sqrt{\abs{\gamma}}\phi K }
\end{equation}
Finally note that in AdS case, the potential $U$ is negative while for dS it is positive.

Let us now prove that the action given in eq.\eqref{actgenu} has the correct variational principle. Varying the action with respect to the metric we get,
\begin{equation}
	\delta S_{U} = \frac{\eta}{2} \pqty{\int d^2 x \,\sqrt{\abs{g}}\, (\phi \delta R_{\mu \nu} g^{\mu \nu} +U'(\phi) \frac{1}{2} g_{\mu \nu} \delta g^{\mu \nu})+ \text{Boundary terms}}
\end{equation}
Here we used the equation,
\begin{equation}
	R_{\mu \nu} = \frac{1}{2} R g_{\mu \nu}
\end{equation}
Using standard results it is easy to show that,
\begin{equation}
	\delta S_U = \frac{\eta}{2}\int \sqrt{\abs{g}} \delta g^{\mu \nu} \left( g_{\mu \nu} \Box \phi - \nabla_\mu \nabla_\nu \phi + U'(\phi) \frac{1}{2} g_{\mu \nu} \right)  + \text{Boundary terms}
\end{equation}
The terms in the parentheses is the equation of motion of the dilaton, see eq.\eqref{eomJT}. The boundary terms give,
\begin{equation}
	\delta S_B = \frac{\eta}{2} \phi \left(- \epsilon\int \sqrt{\abs{\gamma}} \gamma^{\alpha \beta} n^\mu \partial_\mu \delta g_{\alpha \beta} + 2 \epsilon \int \sqrt{\abs{\gamma}} \delta K \right)
\end{equation}
Using,
\begin{equation}
	\delta K  = \frac{1}{2} \gamma^{\alpha \beta} n^\mu \partial_\mu \delta g_{\alpha \beta}
\end{equation}
we see that,
\begin{equation}
	\delta S_B =0.
\end{equation}

\subsection{Dimensional Reduction}
\label{dimred}
In this section we will consider a potential obtained through dimensional reduction of the action describing Schwarzschild black holes in four dimensional dS spacetime assuming spherical symmetry, \cite{Nayak:2018qej, Fanaras:2021awm}.

The four dimensional black hole is described by the action,
\begin{equation}
	S= \frac{1}{16 \pi G}\int d^{4}x \sqrt{\hat{g}} (\hat{R} -  2 \hat{\Lambda}) -\frac{1}{8 \pi G} \int d^3 x \sqrt{\hat{\gamma}} \hat{K}^{(3)} \label{4Dact}
\end{equation}
Here  $\hat{}$  denotes the four dimensional quantities. We first decompose the 4D metric as follows.
\begin{equation}
	ds^2 = \bar{g}_{a b} (t,r) dx^{a} dx^{b} + \Phi(t,r)^2  d\Omega_2 ^2 \label{metric}
\end{equation}
Here $\Phi$ is the radius of the 2 sphere which is assumed to be independent of the angular coordinates $(\theta, \phi)$. $\bar{g}_{a b}$ is the two dimensional metric. The Ricci scalar $R$ in terms of the metric can be written as,
\begin{align}
	\hat{R} & = 	\hat{g}^{\mu \nu} \{-\partial_{\lambda} \hat{g}^{\lambda \rho} \partial_{\rho} \hat{g}_{\mu \nu}- \hat{g}^{\lambda \rho} \partial_{\mu} \partial_{\nu} \hat{g}_{\lambda \rho} - \frac{3}{4} \partial_{\mu} \hat{g}^{\lambda \sigma} \partial_{\nu} \hat{g}_{\lambda \sigma} - \frac{1}{2} \partial_{\mu} \hat{g}^{\lambda \alpha} \partial_{\alpha} \hat{g}_{\nu \lambda}- \frac{1}{4} \hat{g}^{\lambda \rho} \hat{g}^{\sigma \alpha} \partial_{\sigma} \hat{g}_{\rho \lambda} \partial_{\alpha} \hat{g}_{\mu \nu}\} - \partial_{\mu} \partial_{\nu} \hat{g}^{\mu \nu} \label{ricci}
\end{align}
Then using the metric eq.\eqref{metric} in eq.\eqref{ricci} we get after some simplification,
\begin{equation}
	\hat{R} = \bar{R} + \frac{2}{\Phi^2} + \frac{2}{\Phi^2} (\nabla \Phi)^2  - \frac{2}{\Phi^2 }\nabla^a \nabla_a \Phi^2
\end{equation}
Here $\bar{R}$ is the Ricci scalar corresponding to $\bar{g}_{ab}$. Using the above result in eq.\eqref{4Dact} and simplifying further we get,
\begin{equation}
	S=\frac{1}{4 G}\int d^{2}x \sqrt{{\bar{g}}} \left(2+ \Phi^2(\bar{R} -  2\hat{\Lambda}) + 2 (\nabla \Phi)^2\right)- \frac{1}{2 G} \int \sqrt{\bar{\gamma}} \Phi^2 \bar{K}\label{swaveact}
\end{equation}
Let,
\begin{equation}
	V_1 (\Phi) = \Phi^2, V_2(\Phi) = (2 \hat{\Lambda} \Phi^2-2), V_3(\Phi) = 2.
\end{equation}
Then the action eq.\eqref{swaveact} can be written as,
\begin{equation}
	S = \frac{1}{4 G}\int d^{2}x \sqrt{{\bar{g}}} (V_1(\Phi) \bar{R}- V_2(\Phi) + V_3(\Phi) (\nabla \Phi)^2)- \frac{1}{2 G} \int \sqrt{\bar{\gamma}} V_1 (\Phi) \bar{K} \label{swaveacta}
\end{equation}
We can eliminate the kinetic term of the dilaton by redefining the metric to be,
\begin{equation}
	\bar{g}_{\mu \nu} = \Omega(\Phi)^{-2} g_{\mu \nu}.
\end{equation}
Then the Ricci scalar transforms as follows,
\begin{equation}
	\bar{R} = \Omega(\Phi)^2 (R+2 \Box \log(\Omega(\Phi))).
\end{equation}
If the conformal factor $\Omega(\Phi)$ satisfies the equation
\begin{equation}
	V_3(\Phi) = 2 V_1'(\Phi) \partial_\Phi(\log(\Omega(\Phi))) \label{omegaeq}
\end{equation}
we get,
\begin{equation}
	S= \frac{1}{4 G}\int d^{2}x \sqrt{g} (V_1(\Phi) R - \Omega(\Phi)^{-2} V_2(\Phi)) - \frac{1}{2 G} \int \sqrt{\gamma} V_1 (\Phi) K\label{swaveactb}
\end{equation}
Using the values of $V_1,V_3$ one can solve eq.\eqref{omegaeq} to obtain,
\begin{equation}
	\Omega(\Phi) = \sqrt{\Phi} \label{weylf}
\end{equation}
We can rewrite the action eq.\eqref{swaveactb} in a simpler form  by taking
\begin{equation}
	V_1(\Phi) =\phi
\end{equation}
and so we finally get,
\begin{equation}
	S= \frac{1}{4 G}\int d^{2}x \sqrt{g} (\phi R - U'(\phi)) - \frac{1}{2 G} \int \sqrt{\gamma} \phi K \label{swaveactc}
\end{equation}
where,
\begin{equation}
	U'(\phi) = \Omega(\phi(\Phi))^{-2} V_2(\phi(\Phi))= 2 \left(\hat{\Lambda} \sqrt{\phi}- \sqrt{\frac{1}{\phi}}\right)  \label{pot}
\end{equation}
Given eq.\eqref{pot} we can write, with $\hat{\Lambda}=3$,
\begin{equation}
	U(\phi) = 4 \left(\sqrt{\phi^3}- \sqrt{\phi}\right) \label{pot2}
\end{equation}
The action eq.\eqref{swaveactc} is of the desired form eq.\eqref{jtacta} with the potential given in eq.\eqref{pot2}. Note that the potential in eq.\eqref{pot2} is a consequence of the 4D action eq.\eqref{4Dact} and hence is fixed. It can also be shown that the form of the potential in eq.\eqref{pot2} is the same starting from the action eq.\eqref{4Dact} in higher dimensions. It suggests that different potentials will correspond to different actions in four or higher dimensions. Also note that the form of the action eq.\eqref{swaveacta} will be the same irrespective of the higher dimensional theory assuming that the higher dimensional theory contains a maximum of two derivatives. And as a result the action eq.\eqref{swaveactc} is also the general form of a two derivative 2D dilaton gravity theory. 

\subsubsection{Classical Properties}
\label{swavered}
In this section we will consider the potential eq.\eqref{pot2} and discuss its classical properties in some detail.  
Notice that the form of the potential in eq.\eqref{pot2} has fractional powers and to avoid having to deal with branch points, we  will restrict   $\phi$ in this section   to be positive. 

The metric of this 2D system is given by eq.\eqref{solds}, and reproduced below
\begin{equation}
	dS_2 ^2 = - \frac{d \phi^2}{U(\phi)-M} + (U(\phi)-M) dx^2 \label{solds2}
\end{equation}
where, $U(\phi)$ is given in eq.\eqref{pot2}. 

It is useful to relate this metric to the familiar case  of  a 4 dimensional deSitter black hole spacetime which is given by 
\be
\label{metds4bh}
ds^2=-\left(1-r^2-{\mu\over r}\right) dt^2 + \frac{1}{\left(1-r^2-{\mu\over r}\right)} dr^2 + r^2 d\Omega_2^2
\ee
Let us define a radial coordinate $r$ by 
\be
\label{defradr}
\phi=r^2
\ee
The metric eq.(\ref{solds2}) in terms of this coordinate becomes
\be
ds^2=r \left(\frac{d r^2}{-r^2 + 1+\frac{M}{4r}} - 4\left(-r^2 +1+\frac{M}{4r}\right) dx^2\right) \label{2dmetr}
\ee
upto an overall conformal factor we see that this metric  agrees with the $r-t$ plane metric in eq.(\ref{metds4bh}).
We also learn that $M$ in eq.(\ref{metds4bh}) is related to the $4$ dim mass parameter $\mu$ by 
\begin{equation}
	M = - 4\mu\label{m4m}
\end{equation}

It is easy to see that when $M<0$  there are two horizons, a cosmological and black hole horizon, at the real roots of 
\begin{align}
	\label{rooteq}
	r^3-r-\frac{M}{4}=0
\end{align}
This requires $M>M_0$ where,
\begin{equation}
	M_0=- \frac{8}{3 \sqrt{3}} \label{M0eq}
\end{equation} 
The case $M=M_0$ corresponds to the extremal or Naria limit when the two horizons meet. 
These conclusions agree with what is known in the  $4$ dimensional case once we relate $M$ to $\mu$ using eq.\eqref{m4m}. 

Notice also that for $0<M<M_0$ we have a space-like singularity at $\phi=0$. 
And for  $M>0$, which corresponds to a negative mass black hole in $dS_4$, there is a cosmological horizon at the one real root of eq.(\ref{rooteq}), and once we cross it and enter a static patch,  a naked singularity at $\phi=0$. 

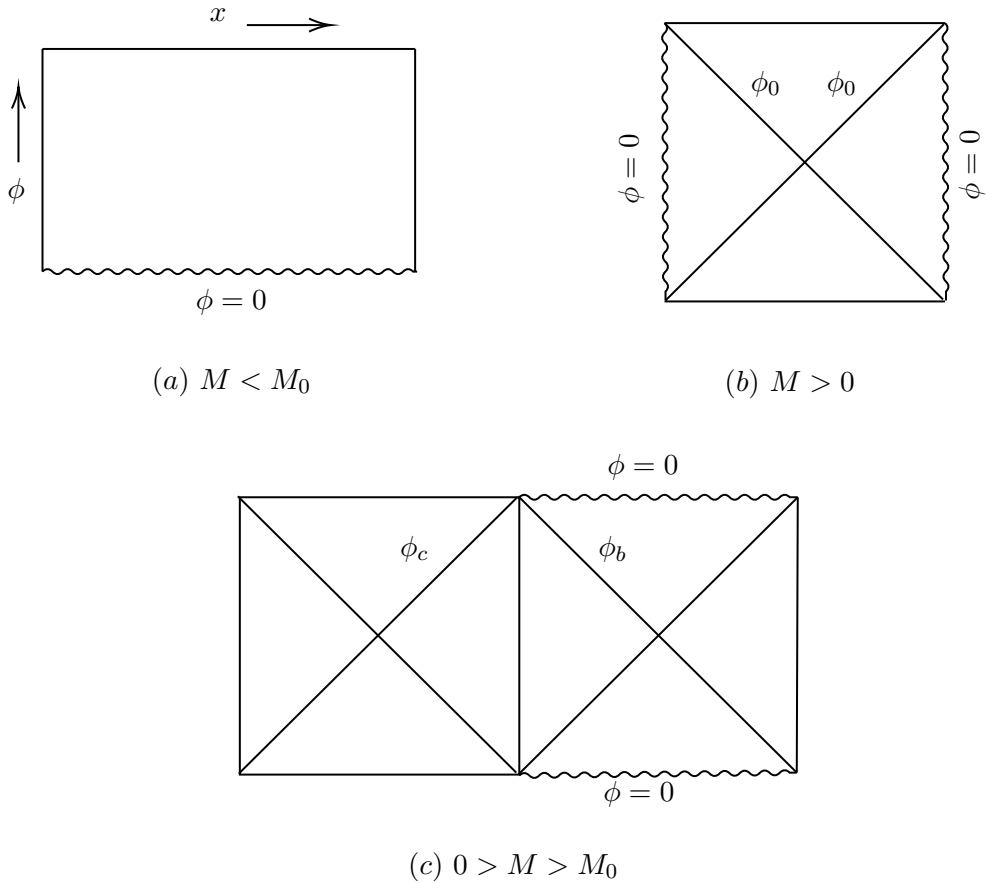
\begin{figure}

	\tikzset{every picture/.style={line width=0.75pt}} 
	
	\begin{tikzpicture}[x=0.75pt,y=0.75pt,yscale=-1,xscale=1]
		
		\draw    (104,57) -- (290,57) ;
		\draw    (104,57) -- (104,169) ;
		\draw    (290,57) -- (290,169) ;
		\draw    (104,169) .. controls (105.67,167.33) and (107.33,167.33) .. (109,169) .. controls (110.67,170.67) and (112.33,170.67) .. (114,169) .. controls (115.67,167.33) and (117.33,167.33) .. (119,169) .. controls (120.67,170.67) and (122.33,170.67) .. (124,169) .. controls (125.67,167.33) and (127.33,167.33) .. (129,169) .. controls (130.67,170.67) and (132.33,170.67) .. (134,169) .. controls (135.67,167.33) and (137.33,167.33) .. (139,169) .. controls (140.67,170.67) and (142.33,170.67) .. (144,169) .. controls (145.67,167.33) and (147.33,167.33) .. (149,169) .. controls (150.67,170.67) and (152.33,170.67) .. (154,169) .. controls (155.67,167.33) and (157.33,167.33) .. (159,169) .. controls (160.67,170.67) and (162.33,170.67) .. (164,169) .. controls (165.67,167.33) and (167.33,167.33) .. (169,169) .. controls (170.67,170.67) and (172.33,170.67) .. (174,169) .. controls (175.67,167.33) and (177.33,167.33) .. (179,169) .. controls (180.67,170.67) and (182.33,170.67) .. (184,169) .. controls (185.67,167.33) and (187.33,167.33) .. (189,169) .. controls (190.67,170.67) and (192.33,170.67) .. (194,169) .. controls (195.67,167.33) and (197.33,167.33) .. (199,169) .. controls (200.67,170.67) and (202.33,170.67) .. (204,169) .. controls (205.67,167.33) and (207.33,167.33) .. (209,169) .. controls (210.67,170.67) and (212.33,170.67) .. (214,169) .. controls (215.67,167.33) and (217.33,167.33) .. (219,169) .. controls (220.67,170.67) and (222.33,170.67) .. (224,169) .. controls (225.67,167.33) and (227.33,167.33) .. (229,169) .. controls (230.67,170.67) and (232.33,170.67) .. (234,169) .. controls (235.67,167.33) and (237.33,167.33) .. (239,169) .. controls (240.67,170.67) and (242.33,170.67) .. (244,169) .. controls (245.67,167.33) and (247.33,167.33) .. (249,169) .. controls (250.67,170.67) and (252.33,170.67) .. (254,169) .. controls (255.67,167.33) and (257.33,167.33) .. (259,169) .. controls (260.67,170.67) and (262.33,170.67) .. (264,169) .. controls (265.67,167.33) and (267.33,167.33) .. (269,169) .. controls (270.67,170.67) and (272.33,170.67) .. (274,169) .. controls (275.67,167.33) and (277.33,167.33) .. (279,169) .. controls (280.67,170.67) and (282.33,170.67) .. (284,169) .. controls (285.67,167.33) and (287.33,167.33) .. (289,169) -- (291,169) -- (291,169) ;
		\draw    (92,114) -- (92,78) ;
		\draw [shift={(92,76)}, rotate = 90] [color={rgb, 255:red, 0; green, 0; blue, 0 }  ][line width=0.75]    (10.93,-3.29) .. controls (6.95,-1.4) and (3.31,-0.3) .. (0,0) .. controls (3.31,0.3) and (6.95,1.4) .. (10.93,3.29)   ;
		\draw    (206,45) -- (246,45) ;
		\draw [shift={(248,45)}, rotate = 180] [color={rgb, 255:red, 0; green, 0; blue, 0 }  ][line width=0.75]    (10.93,-3.29) .. controls (6.95,-1.4) and (3.31,-0.3) .. (0,0) .. controls (3.31,0.3) and (6.95,1.4) .. (10.93,3.29)   ;
		\draw    (414.5,44) -- (553.5,183) ;
		\draw    (554,44.5) -- (415,183.5) ;
		\draw    (414.5,44) .. controls (416.17,45.66) and (416.18,47.33) .. (414.52,49) .. controls (412.86,50.67) and (412.87,52.34) .. (414.54,54) .. controls (416.21,55.67) and (416.21,57.33) .. (414.55,59) .. controls (412.89,60.67) and (412.9,62.34) .. (414.57,64) .. controls (416.24,65.66) and (416.25,67.33) .. (414.59,69) .. controls (412.93,70.67) and (412.94,72.34) .. (414.61,74) .. controls (416.28,75.66) and (416.29,77.33) .. (414.63,79) .. controls (412.97,80.67) and (412.97,82.33) .. (414.64,84) .. controls (416.31,85.66) and (416.32,87.33) .. (414.66,89) .. controls (413,90.67) and (413.01,92.34) .. (414.68,94) .. controls (416.35,95.66) and (416.36,97.33) .. (414.7,99) .. controls (413.04,100.67) and (413.05,102.34) .. (414.72,104) .. controls (416.39,105.67) and (416.39,107.33) .. (414.73,109) .. controls (413.07,110.67) and (413.08,112.34) .. (414.75,114) .. controls (416.42,115.66) and (416.43,117.33) .. (414.77,119) .. controls (413.11,120.67) and (413.12,122.34) .. (414.79,124) .. controls (416.46,125.67) and (416.46,127.33) .. (414.8,129) .. controls (413.14,130.67) and (413.15,132.34) .. (414.82,134) .. controls (416.49,135.66) and (416.5,137.33) .. (414.84,139) .. controls (413.18,140.67) and (413.19,142.34) .. (414.86,144) .. controls (416.53,145.66) and (416.54,147.33) .. (414.88,149) .. controls (413.22,150.67) and (413.22,152.33) .. (414.89,154) .. controls (416.56,155.66) and (416.57,157.33) .. (414.91,159) .. controls (413.25,160.67) and (413.26,162.34) .. (414.93,164) .. controls (416.6,165.66) and (416.61,167.33) .. (414.95,169) .. controls (413.29,170.67) and (413.3,172.34) .. (414.97,174) .. controls (416.64,175.67) and (416.64,177.33) .. (414.98,179) -- (415,183.5) -- (415,183.5) ;
		\draw    (554.5,44) .. controls (556.17,45.66) and (556.18,47.33) .. (554.52,49) .. controls (552.86,50.67) and (552.87,52.34) .. (554.54,54) .. controls (556.21,55.67) and (556.21,57.33) .. (554.55,59) .. controls (552.89,60.67) and (552.9,62.34) .. (554.57,64) .. controls (556.24,65.66) and (556.25,67.33) .. (554.59,69) .. controls (552.93,70.67) and (552.94,72.34) .. (554.61,74) .. controls (556.28,75.66) and (556.29,77.33) .. (554.63,79) .. controls (552.97,80.67) and (552.97,82.33) .. (554.64,84) .. controls (556.31,85.66) and (556.32,87.33) .. (554.66,89) .. controls (553,90.67) and (553.01,92.34) .. (554.68,94) .. controls (556.35,95.66) and (556.36,97.33) .. (554.7,99) .. controls (553.04,100.67) and (553.05,102.34) .. (554.72,104) .. controls (556.39,105.67) and (556.39,107.33) .. (554.73,109) .. controls (553.07,110.67) and (553.08,112.34) .. (554.75,114) .. controls (556.42,115.66) and (556.43,117.33) .. (554.77,119) .. controls (553.11,120.67) and (553.12,122.34) .. (554.79,124) .. controls (556.46,125.67) and (556.46,127.33) .. (554.8,129) .. controls (553.14,130.67) and (553.15,132.34) .. (554.82,134) .. controls (556.49,135.66) and (556.5,137.33) .. (554.84,139) .. controls (553.18,140.67) and (553.19,142.34) .. (554.86,144) .. controls (556.53,145.66) and (556.54,147.33) .. (554.88,149) .. controls (553.22,150.67) and (553.22,152.33) .. (554.89,154) .. controls (556.56,155.66) and (556.57,157.33) .. (554.91,159) .. controls (553.25,160.67) and (553.26,162.34) .. (554.93,164) .. controls (556.6,165.66) and (556.61,167.33) .. (554.95,169) .. controls (553.29,170.67) and (553.3,172.34) .. (554.97,174) .. controls (556.64,175.67) and (556.64,177.33) .. (554.98,179) -- (555,183.5) -- (555,183.5) ;
		\draw    (414.5,44) -- (554.5,44) ;
		\draw    (414.5,184) -- (554.5,184) ;
		\draw   (202.5,282.5) -- (342,282.5) -- (342,422) -- (202.5,422) -- cycle ;
		\draw    (201.5,282) -- (340.5,421) ;
		\draw    (341,282.5) -- (202,421.5) ;
		\draw    (341.5,282) -- (480.5,421) ;
		\draw    (481,282.5) -- (342,421.5) ;
		\draw    (481,282.5) -- (480.5,421) ;
		\draw    (342,282.5) .. controls (343.67,280.83) and (345.33,280.83) .. (347,282.5) .. controls (348.67,284.17) and (350.33,284.17) .. (352,282.5) .. controls (353.67,280.83) and (355.33,280.83) .. (357,282.5) .. controls (358.67,284.17) and (360.33,284.17) .. (362,282.5) .. controls (363.67,280.83) and (365.33,280.83) .. (367,282.5) .. controls (368.67,284.17) and (370.33,284.17) .. (372,282.5) .. controls (373.67,280.83) and (375.33,280.83) .. (377,282.5) .. controls (378.67,284.17) and (380.33,284.17) .. (382,282.5) .. controls (383.67,280.83) and (385.33,280.83) .. (387,282.5) .. controls (388.67,284.17) and (390.33,284.17) .. (392,282.5) .. controls (393.67,280.83) and (395.33,280.83) .. (397,282.5) .. controls (398.67,284.17) and (400.33,284.17) .. (402,282.5) .. controls (403.67,280.83) and (405.33,280.83) .. (407,282.5) .. controls (408.67,284.17) and (410.33,284.17) .. (412,282.5) .. controls (413.67,280.83) and (415.33,280.83) .. (417,282.5) .. controls (418.67,284.17) and (420.33,284.17) .. (422,282.5) .. controls (423.67,280.83) and (425.33,280.83) .. (427,282.5) .. controls (428.67,284.17) and (430.33,284.17) .. (432,282.5) .. controls (433.67,280.83) and (435.33,280.83) .. (437,282.5) .. controls (438.67,284.17) and (440.33,284.17) .. (442,282.5) .. controls (443.67,280.83) and (445.33,280.83) .. (447,282.5) .. controls (448.67,284.17) and (450.33,284.17) .. (452,282.5) .. controls (453.67,280.83) and (455.33,280.83) .. (457,282.5) .. controls (458.67,284.17) and (460.33,284.17) .. (462,282.5) .. controls (463.67,280.83) and (465.33,280.83) .. (467,282.5) .. controls (468.67,284.17) and (470.33,284.17) .. (472,282.5) .. controls (473.67,280.83) and (475.33,280.83) .. (477,282.5) -- (481,282.5) -- (481,282.5) ;
		\draw    (342,422.5) .. controls (343.65,420.82) and (345.32,420.8) .. (347,422.45) .. controls (348.69,424.1) and (350.35,424.08) .. (352,422.39) .. controls (353.65,420.71) and (355.32,420.69) .. (357,422.34) .. controls (358.69,423.99) and (360.35,423.97) .. (362,422.28) .. controls (363.65,420.6) and (365.32,420.58) .. (367,422.23) .. controls (368.68,423.88) and (370.35,423.86) .. (372,422.18) .. controls (373.65,420.49) and (375.31,420.47) .. (377,422.12) .. controls (378.68,423.77) and (380.35,423.75) .. (382,422.07) .. controls (383.65,420.38) and (385.31,420.36) .. (387,422.01) .. controls (388.68,423.66) and (390.35,423.64) .. (392,421.96) .. controls (393.65,420.27) and (395.31,420.25) .. (397,421.9) .. controls (398.68,423.55) and (400.35,423.53) .. (402,421.85) .. controls (403.65,420.17) and (405.32,420.15) .. (407,421.8) .. controls (408.69,423.45) and (410.35,423.43) .. (412,421.74) .. controls (413.65,420.06) and (415.32,420.04) .. (417,421.69) .. controls (418.69,423.34) and (420.35,423.32) .. (422,421.63) .. controls (423.65,419.95) and (425.32,419.93) .. (427,421.58) .. controls (428.68,423.23) and (430.34,423.21) .. (431.99,421.53) .. controls (433.64,419.84) and (435.3,419.82) .. (436.99,421.47) .. controls (438.67,423.12) and (440.34,423.1) .. (441.99,421.42) .. controls (443.64,419.73) and (445.3,419.71) .. (446.99,421.36) .. controls (448.67,423.01) and (450.34,422.99) .. (451.99,421.31) .. controls (453.64,419.62) and (455.3,419.6) .. (456.99,421.25) .. controls (458.67,422.9) and (460.34,422.88) .. (461.99,421.2) .. controls (463.64,419.52) and (465.31,419.5) .. (466.99,421.15) .. controls (468.68,422.8) and (470.34,422.78) .. (471.99,421.09) .. controls (473.64,419.41) and (475.31,419.39) .. (476.99,421.04) -- (480.5,421) -- (480.5,421) ;
		
		\draw (85,120.4) node [anchor=north west][inner sep=0.75pt]    {$\phi $};
		\draw (179,176.4) node [anchor=north west][inner sep=0.75pt]    {$\phi =0$};
		\draw (186,35.4) node [anchor=north west][inner sep=0.75pt]    {$x$};
		\draw (157,215.4) node [anchor=north west][inner sep=0.75pt]    {$( a) \ M< M_{0}$};
		\draw (389.9,136.5) node [anchor=north west][inner sep=0.75pt]  [rotate=-270]  {$\phi =0$};
		\draw (559.9,134.5) node [anchor=north west][inner sep=0.75pt]  [rotate=-270]  {$\phi =0$};
		\draw (456,66.4) node [anchor=north west][inner sep=0.75pt]    {$\phi _{0}$};
		\draw (494,66.4) node [anchor=north west][inner sep=0.75pt]    {$\phi _{0}$};
		\draw (443,216.4) node [anchor=north west][inner sep=0.75pt]    {$( b) \ M >0$};
		\draw (382.5,423.4) node [anchor=north west][inner sep=0.75pt]    {$\phi =0$};
		\draw (384.5,259.4) node [anchor=north west][inner sep=0.75pt]    {$\phi =0$};
		\draw (281,300.4) node [anchor=north west][inner sep=0.75pt]    {$\phi _{c}$};
		\draw (380,300.4) node [anchor=north west][inner sep=0.75pt]    {$\phi _{b}$};
		\draw (285,459.4) node [anchor=north west][inner sep=0.75pt]    {$( c) \ 0 >M >M_{0}$};

	\end{tikzpicture}

	\caption{(a): Penrose diagram for $M <M_0$.  (b): Penrose diagram for $0 >M> M_0$. $\phi_c, \phi_b$ are respectively the cosmological horizon and the black hole horizon, obtained as solutions to the equation $U(\phi)=M$. (c): Penrose diagram for $M>0$. The value of $\phi_0$ is determined by $U(\phi_0)=M$.}
	\label{pd4dds}
\end{figure}

The Penrose diagrams corresponding to various cases are shown in fig.\ref{pd4dds}.

Finally we note that taking $M=M_0$ and expanding about the double root at $\phi=\phi_c=\frac{1}{3}$ we get 
\be
\label{expv}
U(\phi)-M_0= 3\sqrt{3} (\phi-\phi_c)^2
\ee
After a further redefinition $(\phi-\phi_c)\rightarrow \phi$  we see that this gives rise to the quadratic potential of JT $dS_2$ theory. 

\subsection{ Spacetimes  where $R$ changes sign }
\label{rchanpot}
So far, we have  considered potentials which are monotonic functions of $\phi$. For such potentials, the Ricci scalar,  $R$, being proportional to the second derivative of the potential, will always be of a fixed sign. 
In this subsection we explore spacetimes where the Ricci scalar can change sign. This will lead to examples  where a dS bubble is inside AdS space,  or  an  AdS bubble is inside dS space, and so on. Such spacetimes were considered in \cite{Anninos:2022hqo}. 

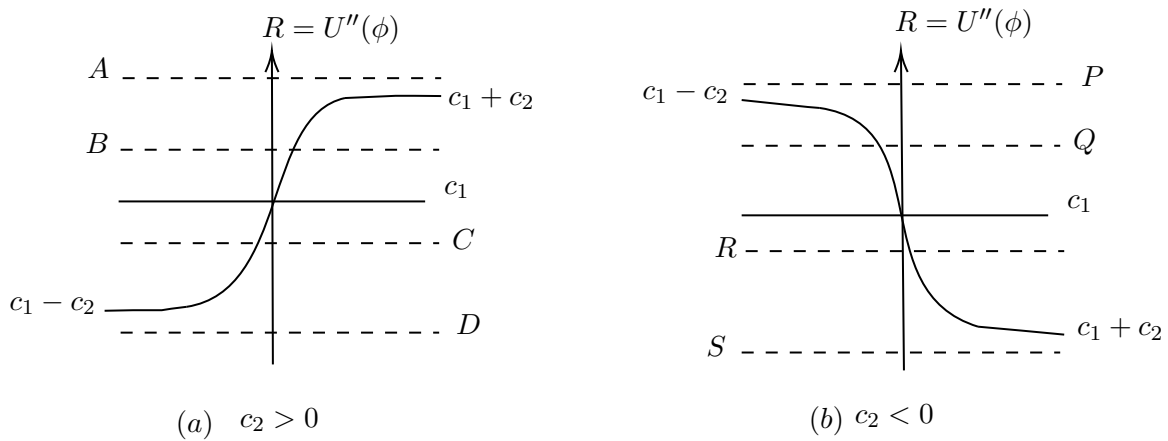
\begin{figure}[h!]

	\tikzset{every picture/.style={line width=0.75pt}} 
	
	\begin{tikzpicture}[x=0.75pt,y=0.75pt,yscale=-1,xscale=1]
		
		\draw    (75,183) .. controls (119,182) and (91,184) .. (116,181) ;
		\draw    (195,76) .. controls (240,74) and (217,75) .. (243,75) ;
		\draw    (116,181) .. controls (167,173) and (154,85) .. (195,76) ;
		\draw    (511,191) .. controls (534,193) and (532,193) .. (554,195) ;
		\draw    (393,77) .. controls (441,82) and (411,79) .. (432,81) ;
		\draw    (432,81) .. controls (492,90) and (453,173) .. (511,191) ;
		\draw    (159,210) -- (158.51,54) ;
		\draw [shift={(158.5,52)}, rotate = 89.82] [color={rgb, 255:red, 0; green, 0; blue, 0 }  ][line width=0.75]    (10.93,-3.29) .. controls (6.95,-1.4) and (3.31,-0.3) .. (0,0) .. controls (3.31,0.3) and (6.95,1.4) .. (10.93,3.29)   ;
		\draw    (82,128) -- (235,128) ;
		\draw    (393,135) -- (546,135) ;
		\draw    (474,213) -- (472.52,54) ;
		\draw [shift={(472.5,52)}, rotate = 89.47] [color={rgb, 255:red, 0; green, 0; blue, 0 }  ][line width=0.75]    (10.93,-3.29) .. controls (6.95,-1.4) and (3.31,-0.3) .. (0,0) .. controls (3.31,0.3) and (6.95,1.4) .. (10.93,3.29)   ;
		\draw  [dash pattern={on 4.5pt off 4.5pt}]  (83,102) -- (242,102) ;
		\draw  [dash pattern={on 4.5pt off 4.5pt}]  (83,149) -- (242,149) ;
		\draw  [dash pattern={on 4.5pt off 4.5pt}]  (83,194) -- (242,194) ;
		\draw  [dash pattern={on 4.5pt off 4.5pt}]  (84,66) -- (243,66) ;
		\draw  [dash pattern={on 4.5pt off 4.5pt}]  (394,69) -- (436,69) -- (553,69) ;
		\draw  [dash pattern={on 4.5pt off 4.5pt}]  (393,100) -- (552,100) ;
		\draw  [dash pattern={on 4.5pt off 4.5pt}]  (395,153) -- (554,153) ;
		\draw  [dash pattern={on 4.5pt off 4.5pt}]  (393,204) -- (552,204) ;
		
		\draw (243,116.4) node [anchor=north west][inner sep=0.75pt]    {$c_{1}$};
		\draw (554,123.4) node [anchor=north west][inner sep=0.75pt]    {$c_{1}$};
		\draw (245,69.4) node [anchor=north west][inner sep=0.75pt]    {$c_{1} +c_{2}$};
		\draw (26,172.4) node [anchor=north west][inner sep=0.75pt]    {$c_{1} -c_{2}$};
		\draw (342,66.4) node [anchor=north west][inner sep=0.75pt]    {$c_{1} -c_{2}$};
		\draw (559,185.4) node [anchor=north west][inner sep=0.75pt]    {$c_{1} +c_{2}$};
		\draw (152,32.4) node [anchor=north west][inner sep=0.75pt]    {$R=U''( \phi )$};
		\draw (65,54.4) node [anchor=north west][inner sep=0.75pt]    {$A$};
		\draw (64,92.4) node [anchor=north west][inner sep=0.75pt]    {$B$};
		\draw (247,140.4) node [anchor=north west][inner sep=0.75pt]    {$C$};
		\draw (249,183.4) node [anchor=north west][inner sep=0.75pt]    {$D$};
		\draw (561,58.4) node [anchor=north west][inner sep=0.75pt]    {$P$};
		\draw (557,90.4) node [anchor=north west][inner sep=0.75pt]    {$Q$};
		\draw (378,144.4) node [anchor=north west][inner sep=0.75pt]    {$R$};
		\draw (374,194.4) node [anchor=north west][inner sep=0.75pt]    {$S$};
		\draw (141,230.4) node [anchor=north west][inner sep=0.75pt]    {$c_{2}  >0$};
		\draw (448,228.4) node [anchor=north west][inner sep=0.75pt]    {$c_{2} < 0$};
		\draw (109,231.4) node [anchor=north west][inner sep=0.75pt]    {$( a)$};
		\draw (425,229.4) node [anchor=north west][inner sep=0.75pt]    {$( b)$};
		\draw (468,29.4) node [anchor=north west][inner sep=0.75pt]    {$R=U''( \phi )$};

	\end{tikzpicture}
	\caption{This plot shows the behaviour of Ricci Scalar $R$ qualitatively.}
	\label{rplots}
\end{figure}

As  a simple example, where the sign of the Ricci scalar can be changed by suitably controlling the parameters of the potential, take  the case where  
Ricci scalar is given by,
\begin{equation}
	R = c_1 + c_2 \tanh(\phi) \, \label{Reqds}
\end{equation}
Here $c_1$ and $c_2$ are  constants whose signs and relative magnitudes determine  Ricci scalar to be  positive, negative or zero in suitable limits. 
The limiting values of $R$ are as follows
\begin{align}
	R\simeq\begin{cases}
		c_1+c_2 \quad &\phi\gg 1\\
		c_1\quad \quad &\phi\simeq0\\
		c_1-c_2\quad &\phi\ll -1\\
	\end{cases}
	\label{Rlimits}
\end{align}
The qualitative nature of $R$ is as shown in fig \ref{rplots}.
	The potential $U(\phi)$ can be obtained by integrating eq.\eqref{Reqds}, this introduces two integration constants which correspond to  changing $U(\phi)$ by an overall constant and a linear term. The constant term does not enter in the action eq.\eqref{jtacta} and is irrelevant. 
	The asymptotic behaviour of the potential is given by 
	\begin{eqnarray}
		\label{aspot}
		U(\phi) & \rightarrow &  (c_1+c_2){\phi^2\over 2} +d \phi, \, \phi\rightarrow \infty\\
		U(\phi) & \rightarrow & (c_1-c_2) {\phi^2\over 2} + e \phi, \, \phi\rightarrow -\infty
	\end{eqnarray}
	where one combination of $d, e$ can be changed by changing the integration constant mentioned above.  
	
	We see from eq.(\ref{Rlimits}) that when $c_1+c_2>0$ and $c_1-c_2<0$ we have a dS region   at large positive values of $\phi$ which becomes AdS space at sufficiently negative values for the dilaton. Alternatively,  when $c_1+c_2<0$ and $c_1-c_2>0$, we have a dS bubble inside AdS space. 
	
	For a solution corresponding to  mass $M$, eq.\eqref{solds}, the dilaton corresponds to a time like direction when 
	$U>M$ and a space-like direction when $U<M$. This means, e.g., for  the case where $c_1+c_2<0$ and  $c_1-c_2>0$ the spacetime is AdS space, when $\phi\rightarrow \infty$,  with a time-like killing vector.  As $\phi$ becomes smaller one encounters a horizon, where $U(\phi)=M$, and thereafter for smaller $\phi$, the region of spacetime has a space-like killing vector. Eventually, for $\phi\rightarrow -\infty$, the geometry is well described by dS space. 
	In this example then the dS bubble is shielded by a horizon from the asymptotic AdS region. 
   A similar description can be obtained for the other cases \footnote{{ For a related work see \cite{Iizuka:2025vkl}.}}.

\subsection{Theories With Asymptotics Different Than dS}
\label{asymdiffds}
{As our final example let us consider now a potential of the form,
\begin{equation}
	U(\phi) = \phi^n. \label{pot3}
\end{equation}
In this section we will understand a few properties of the geodesics in the spacetime described by the above potential. To simplify the analysis let us focus on $n>2$. From eq.\eqref{solds} one gets,
\begin{equation}
	d\tau^2 = \frac{dr^2}{r^n-M} - (r^n-M) dx^2 \label{met3}
\end{equation}
where $\tau$ is the proper time. Since the metric has a spacelike killing vector $x$, the following is a conserved quantity.
\begin{equation}
	p_x = (r^n-M) \frac{d x}{d \tau}
\end{equation}
Solving for $\tau$ one gets,
\begin{align}
	\tau =\frac{2}{n-2} M^{-\frac{1}{2}+\frac{1}{n}} {}_2 F_{1} \left(\frac{1}{2}, \frac{1}{2}-\frac{1}{n}, \frac{3}{2}-\frac{1}{n}, 1- \frac{p_x^2}{M}\right) 
\end{align}
where we have integrated over $r$ from $M^{\frac{1}{n}}$ to $\infty$. Thus the proper time of a massive particle is finite. In comparison for a massless particle the affine time is infinite. This suggests that spacetime is geodesically incomplete. For $n\leq 2$, a similar analysis reveals that the proper time is infinite and thus it's geodesically complete which is expected since the $n=2$ limit is the dS$_2$ spacetime.}

\subsection{Thermodynamics }
\label{therm}
We end this section by discussing the thermodynamic properties of a black hole in an AdS spacetime described by a generic potential $U(\phi)$.  The following analysis is a review of a similar analysis carried out in \cite{witten2020deformations}. We include this here for completion. 

Consider the Euclidean AdS like metric
\begin{equation}
	ds^2 = \frac{d r^2}{U(r)-M} +  (U(r)-M) dx^2,\,\, \phi=r \label{metads}
\end{equation}
where $x$ is periodic with period $\beta$. We will require that the corresponding Ricci scalar,
\begin{equation}
	R = - U''(r) \label{Rads}
\end{equation}
is always negative. This will ensure that eq.\eqref{metads} always describes an AdS like metric.

From the usual arguments of smoothness near the horizon where 
\begin{equation}
	U(r_h)=M \label{horloc}
\end{equation}
we find
\begin{equation}
	T=\beta^{-1}= \frac{U'(r_h)}{4 \pi} \label{temp}
\end{equation}
The entropy of the black hole is defined to be,
\begin{equation}
	S = 2 \pi \phi_h (M) \label{ent}
\end{equation}
The black hole mass then can be obtained from the first law of black hole thermodynamics as,
\begin{equation}
	dM_{BH} = T dS \implies  dM_{BH} = \frac{U'(\phi_h)}{4 \pi} 2 \pi \phi_h' (M) dM =  \frac{dM}{2} \implies M_{BH} = \frac{M}{2} \label{bhm}
\end{equation}
where we used,
\begin{equation}
	U(\phi_h) = M \implies U'(\phi_h) \phi_h'(M)  =1 \label{id2}
\end{equation}
The thermodynamic stability of the black hole described by eq.\eqref{metads} requires,
\begin{equation}
	\frac{dM}{d T} = \frac{dM}{dr_h} \frac{dr_h}{d T} = \frac{16\pi^2T}{U''(r_h)} >0
\end{equation}
where we used eq.\eqref{horloc} along with eq.\eqref{temp}. Now we require $U'(r_h)>0$ and from eq.\eqref{temp} we see that $T>0$. Since eq.\eqref{Rads} is always met we see that the black holes described by eq.\eqref{metads} are thermodynamically stable.

\section{Constraint Analysis}
\label{conana}
In this section, we show that the Hamiltonian and momentum constraints, eq.\eqref{hammomcon} reproduced below are first class constraints. 
\begin{align}
	&	0=\mathcal{H}=2\pi_\phi \pi_{g_1}\sqrt{g_1}-\left(\frac{\phi'}{\sqrt{g_1}}\right)'-\frac{\sqrt{g_1} U'(\phi)}{2}\nonumber\\
	&0=\mathcal{P}=2g_1\pi_{g_1}'+\pi_{g_1}g_1'-\pi_\phi \phi'
\end{align}

To show this, we need to evaluate the Poisson bracket of the two constraints. For this, it will be useful to consider the integrated version of the constraints as
\begin{align}
	Q_H&=\int \sqrt{g_1(x)}\epsilon_H(x)\mathcal{H}(x)dx\nonumber\\
	Q_P&=\int \epsilon_P(y)\mathcal{P}(y)dy\label{qpqh}
\end{align}
The pairs $(\phi, \pi_\phi)$, $(g_1,\pi_{g_1})$ satisfy the canonical Poisson bracket relations
\begin{align}
	\{\phi(x),\pi_\phi(y)\}=\delta(x-y),\,\{g_1(x),\pi_{g_1}(y)\}=\delta(x-y)\label{pbrel}
\end{align}

\begin{enumerate}
	\item We will first consider
	\begin{align}
		\{Q_P, Q_P\} &= \int dx dy \epsilon_{P_1} (x) \epsilon_{P_2} (y) \{\mathcal{P}(x), \mathcal{P}(y)\} \nonumber \\
		&= \int dx dy\epsilon_{P_1} (x) \epsilon_{P_2} (y) \{2g_1\pi_{g_1}'+\pi_{g_1}g_1'-\pi_\phi \phi' (x), 2g_1\pi_{g_1}'+\pi_{g_1}g_1'-\pi_\phi \phi'(y)\}
	\end{align}
	Some of the relevant Poisson brackets are 
	\begin{align}
		& \int dx dy\epsilon_{P_1} (x) \epsilon_{P_2} (y) 4 \{g_1\pi_{g_1}'(x), g_1\pi_{g_1}'(y)\} = -4 \int dx \pi_{g_1}' g_1(\epsilon_{P_1} \epsilon_{P_2}' - \epsilon_{P_2} \epsilon_{P_1}') \nonumber \\
		& \int dx dy\epsilon_{P_1} (x) \epsilon_{P_2} (y) \{\pi_{g_1}g_1'(x), \pi_{g_1}g_1'(y)\} =\int dx g_1'\pi_{g_1}(\epsilon_{P_1} \epsilon_{P_2}' - \epsilon_{P_2} \epsilon_{P_1}') \nonumber \\
		& \int dx dy\epsilon_{P_1} (x) \epsilon_{P_2} (y) \{\pi_\phi \phi'(x), \pi_\phi \phi'(y)\} 
		= \int dx \phi' \pi_{\phi} (\epsilon_{P_1} \epsilon_{P_2}' - \epsilon_{P_2} \epsilon_{P_1}') \nonumber 
	\end{align}
	Adding everything we get,
	\begin{equation}
		\{Q_P, Q_P\} = -\int dx \mathcal{P} (\epsilon_{P_1} \epsilon_{P_2}' - \epsilon_{P_2} \epsilon_{P_1}') = \int dx \epsilon_{P_3}\mathcal{P} = Q_P \label{QpQPpb}
	\end{equation}
	Thus,
	\begin{equation}
		\label{PPpb}
		\{\mathcal{P}(x),\mathcal{P}(y)\} = \mathcal{P}(x) \partial_y \delta(x-y) - \mathcal{P}(y) \partial_x \delta(x-y) 
	\end{equation}
	\item Next we consider,
	\begin{align}
		\{Q_P, Q_H\} &= \int dy \epsilon_H (y)  \{Q_P, \sqrt{g_1(y)} \mathcal{H}(y)\} \nonumber \\
		&= \int dx dy\epsilon_P (x) \epsilon_H (y) \sqrt{g_1(y)} \{\mathcal{P}(x), \mathcal{H}(y)\} +  \int dy \epsilon_H (y) \mathcal{H}(y) \{Q_P, \sqrt{g_1(y)}\} \label{QpQhpb}
	\end{align}

	First, we analyze the first term in eq.\eqref{QpQhpb}. We will not present the analysis in full. Rather we give only few of the Poisson brackets.
	
	\begin{align}
		&4 \int dx dy \epsilon_P (x) \epsilon_H (y) \sqrt{g_1(y)} \{g_1\pi_{g_1}' (x), \pi_\phi \pi_{g_1}\sqrt{g_1} (y)\} = 2\int\epsilon_H \left(g_1' \epsilon_P \pi_\phi \pi_{g_1} + \epsilon_P' \pi_\phi \pi_{g_1} g_1 +2\pi_\phi g_1 \pi_{g_1}' \epsilon_P\right)\nonumber  \\
		& -\int dx dy \epsilon_P (x) \epsilon_H (y)	\sqrt{g_1(y)} \{2g_1\pi_{g_1}' (x), \left(\frac{\phi'}{\sqrt{g_1}}\right)' (y)\} = -\int dx (g_1 \epsilon_P)' (\epsilon_H \sqrt{g_1})' \phi' \frac{1}{g_1^{\frac{3}{2}}} \nonumber \\
		& -\int dx dy \epsilon_P (x) \epsilon_H (y) \sqrt{g_1(y)}\{2g_1\pi_{g_1}' (x), \frac{\sqrt{g_1} U'(\phi)}{2} (y)\}  =- \frac{1}{2}\int dx \epsilon_H U'(\phi) (\epsilon_P g_1)'\nonumber
	\end{align}
	Similarly the second term in eq.\eqref{QpQhpb}becomes
	\begin{align}
		\{Q_P, \sqrt{g_1} (x)\} & =  \left\{\int \epsilon_P(y)\mathcal{P}(y)dy, \sqrt{g_1} (x) \right\} = (\epsilon_P \sqrt{g_1})'\nonumber 
	\end{align}
	Combining everything 
	and simplifying further we finally get,
	\begin{equation}
		\{Q_P, Q_H\} = - \int \epsilon_P (\epsilon_H \sqrt{g_1})'\mathcal{H} + \int \epsilon_H  \mathcal{H} (\epsilon_P \sqrt{g_1})' \label{QpQhpb2}
	\end{equation}
	This can be simplified to,
	\begin{equation}
		\{Q_P, Q_H\} = \int (\epsilon_P' \epsilon_H - \epsilon_P \epsilon_H') \sqrt{g_1} \mathcal{H} = \int \epsilon_{\tilde{H}} \sqrt{g_1} \mathcal{H} = Q_H
	\end{equation}
	From the first term in eq.\eqref{QpQhpb2} one concludes that,
	\begin{equation}
		\label{PHpb}
		\{\mathcal{P} (x), \mathcal{H} (y)\} = \mathcal{H} (x) \partial_y \delta(x-y)
	\end{equation}
	\item Finally we consider
	\begin{align}
		\{Q_H, Q_H\} &= \int dx dy\epsilon_{H_1} (x) \epsilon_{H_2} (y) \sqrt{g_1(x)} \sqrt{g_1(y)} \{\mathcal{H}(x), \mathcal{H}(y)\}  \nonumber \\
		&+ \int dx dy\epsilon_{H_1} (x) \epsilon_{H_2} (y) \left(\sqrt{g_1(y)} \mathcal{H}(x)  \{\sqrt{g_1(x)}, \mathcal{H}(y)\} + \sqrt{g_1(x)} \mathcal{H}(y)  \{\mathcal{H}(x),\sqrt{g_1(y)}\}\right) \label{QhQhpb}
	\end{align}
	Computing the relevant Poisson brackets we find that
	\begin{align}
		\{Q_H,Q_H\} = \int dx (\epsilon_{H_1}' \epsilon_{H_2}-\epsilon_{H_1} \epsilon_{H_2}') \mathcal{P} = \int dx \epsilon_P \mathcal{P} = Q_P
	\end{align}
	One can verify,
	\begin{equation}
		\label{HHpb}
		\{\mathcal{H}(x) , \mathcal{H}(y)\} = \mathcal{P} (x) \frac{1}{g_1(x)} \partial_y \delta (x-y) -  \mathcal{P} (y) \frac{1}{g_1(y)} \partial_x \delta (x-y) 
	\end{equation}
	Results eq.\eqref{PPpb},eq.\eqref{PHpb} and eq.\eqref{HHpb} match with the corresponding results in \cite{isham1985representations}.
	
\end{enumerate}

\section{More details on Quantization}
\label{moredetquan}
In this appendix we will provide more details about the canonical quantization described in section \ref{canquan}.
\subsection{Rindler Basis for $U(\phi)$}
\label{altsol}
In this subsection we derive an alternate representation of the wavefunction which we will be useful for our analysis of potentials with conical deficit.  Consider the Wheeler de Witt equation, eq.\eqref{WdW}. It can be rewritten as,
\begin{equation}
	\partial_u \partial_v \hat{\Psi} + \frac{1}{4} \hat{\Psi} =0 \label{WdW3}
\end{equation}
where,
\begin{equation}
	u=l^2, v= U(\phi). \label{uveq}
\end{equation}
We will assume that $U(\phi)$ is everywhere positive. This equation describes a free scalar of mass $m^2=1$ in a flat metric,
\begin{equation}
	ds^2 = -du \, dv.
\end{equation}
Defining coordinates $\xi, \theta$ through,
\begin{equation}
	u = \xi e^{-\theta}, v=\xi e^{\theta}
\end{equation}
we can rewrite eq.\eqref{WdW3} as follows,
\begin{equation}
	\label{WdW4}
	\xi^2 \partial_\xi^2 \hat{\Psi} + \xi \partial_\xi \hat{\Psi}  - \partial_\theta^2 \hat{\Psi}  + \xi^2 \hat{\Psi}  =0
\end{equation}
Considering an ansatz of the form,
\begin{equation}
	\hat{\Psi}  = e^{m \theta} f(\xi)
\end{equation}
eq.\eqref{WdW4} takes the form,
\begin{equation}
	\xi^2 \partial_\xi^2 f(\xi) + \xi \partial_\xi f(\xi) + (\xi^2-m^2) f(\xi)=0
\end{equation}
The solution to this equations are Hankel functions. And hence the wavefunction takes the form,
\begin{equation}
	\hat{\Psi}_m = A_m e^{m \theta} H_m ^{(1)} (\xi) + B_m e^{m \theta} H_m ^{(2)} (\xi) 
\end{equation}
where $A_m,B_m$ are complex coefficients. More generally one can consider a linear combination of different values of $m$.
\begin{equation}
	\label{solpsi}
	\hat{\Psi} = \int dm \, e^{m \theta} (A_m  H_m ^{(1)} (\xi) + B_m H_m ^{(2)} (\xi))
\end{equation}
Notice that in terms of variables $l$ and $\phi$, $\xi$ and $\theta$ can be written as,
\begin{equation}
	\xi = \sqrt{uv}=l \sqrt{U(\phi)}, e^{\theta} =\sqrt{\frac{v}{u}}= \frac{\sqrt{U(\phi)}}{l}.
\end{equation}
One can further generalize eq.\eqref{solpsi} by considering shifted $u,v$.
\begin{equation}
	u = l^2 + c_1, v= U(\phi)+c_2. \label{uveq2}
\end{equation}
where $c_1, c_2$ are arbitrary constants. 
We can also write eq.\eqref{solpsi} as,
\begin{equation}
	\hat{\Psi} = \int dk \, (a(k)  J_{-i \abs{k}}  (\xi) e^{ik \theta}+ b(k) J_{i \abs{k}} (\xi)  e^{-i k \theta}) \label{solpsi2}
\end{equation}
The above equation is identical in form to eq.\eqref{psibessel}. And as such the arguments for conservation of norm presented in appendix \ref{rinnorm} applies in this case as well. In particular we can again set $k<0$ modes to vanish for the norm to be conserved. Then the conserved norm takes the same form as eq.\eqref{normrind}.

Before moving further let us note that if $U(\phi)$ is not everywhere positive then too we can define a Rindler basis by appropriately defining $v$, see eq.\eqref{uveq}.

\subsection{Conservation of Inner Product}
\label{consinnprod}
In this subsection we show that for normalizable wavefunction the inner product defined in eq.\eqref{inner} is conserved which means it must be independent of the value of the dilaton at which the inner product is computed. Thus,
\begin{equation}
	\partial_\phi \langle \hat{\Psi}_1, \hat{\Psi}_2 \rangle =0.
\end{equation}
Simplifying the above equation we get,
\begin{equation}
	I_N = \partial_\phi \langle \hat{\Psi}_1, \hat{\Psi}_2 \rangle = \frac{i}{2} (\partial_\phi \hat{\Psi}_1^* \hat{\Psi}_2-\partial_\phi \hat{\Psi}_2 \hat{\Psi}_1^*)|^{l=\infty} _{l=0} + \frac{i}{2} (\partial_\phi \hat{\Psi}_2^* \hat{\Psi}_1-\partial_\phi \hat{\Psi}_1 \hat{\Psi}_2^*)|^{l=\infty} _{l=0}  =0 \label{innercons}
\end{equation}
To simplify the calculation we will consider the states that have conserved norm. In the Rindler basis a wavefunction with conserved norm satisfies the following condition
\begin{equation}
	\del_\phi \hat{\Psi}_i|_{l=0} =0 
\end{equation}
Thus at $l=0$, $I_N=0$.

Similarly since norm is conserved, as $l \rightarrow \infty$ we can assume that the wavefunction dies sufficiently fast enough and hence $I_N$ vanishes as $l \rightarrow \infty$. Thus the inner product as defined in eq.\eqref{inner} is conserved for wavefunctions that have conserved norm. 

\subsection{Conservation of Probability in the $M$ basis}
\label{consprob}
{
	 Before ending this section let us briefly discuss the conservation of norm in the $M$ basis for a general potential $U(\phi)$.}

A general solution for the wavefunction satisfying WDW equation eq.\eqref{hcongenph} in case of a general potential is given by,
\begin{align}
	\hat{\Psi} &= \frac{1}{l}\left(\int_{-\infty}^{U(\phi)} dM \rho(M) e^{-il\sqrt{U(\phi)-M}} + \int_{-\infty}^{U(\phi)} dM \tilde{\rho} (M) e^{ il\sqrt{U(\phi)-M}} \right)  \nonumber \\
	&+ \frac{1}{l}\left( \int_{U(\phi)}^{\infty} dM \rho_1 (M) e^{-l\sqrt{M-U(\phi)}} + \int_{U(\phi)}^{\infty} dM \rho_2 (M) e^{l\sqrt{M-U(\phi)}}\right) \label{psigen}
\end{align}
The norm defined in eq.\eqref{norma} is computed at a particular value of the physical clock, the dilaton $\phi$. However the norm should not depend on the value of $\phi$ chosen. Thus,
\begin{equation}
	\partial_\phi \langle \hat{\Psi},\hat{\Psi} \rangle =0 \label{normcons}
\end{equation}
Then using the definition of norm given in eq.\eqref{normaa} the above condition translates to,
\begin{equation}
	\mathcal{C}_N = i (\hat{\Psi}^* \del_\phi \hat{\Psi}-\hat{\Psi} \del_\phi \hat{\Psi}^*)|_{l=0}^{l=\infty} \label{cneq}
\end{equation}
The analysis is similar to the one performed in \cite{nanda2023jt}. In particular, the $l=0$ limit of eq.\eqref{cneq}, see eq.(3.16) of \cite{nanda2023jt} becomes for a general potential,
\begin{align}
	\label{condva2}
	&\bigg(i\int_{-M_0}^{U(\phi)}(\rho-{\tilde \rho})^*\sqrt{U(\phi)-M}\int_{U(\phi)}^{M_0}{\rho_1-\rho_2\over\sqrt{M-U(\phi)}}
	+i\int_{U(\phi)}^{M_0}(\rho_1-\rho_2)^*\sqrt{M-U(\phi)} 
	\int_{-M_0}^{U(\phi)}{\rho-{\tilde \rho}\over \sqrt{U(\phi)-M}}\nonumber\\
	&+\int_{-M_0}^{U(\phi)}(\rho-{\tilde \rho})^{*}\sqrt{U(\phi)-M} \int_{-M_0}^{U(\phi)}{\rho-{\tilde \rho}\over \sqrt{U(\phi)-M}}
	- \int_{U(\phi)}^{M_0}(\rho_1-\rho_2)^*\sqrt{M-U(\phi)}\int_{U(\phi)}^{M_0}{\rho_1-\rho_2\over \sqrt{M-U(\phi)}}\bigg) \nonumber \\
	& - c.c =0
\end{align}
Let us assume that $U(\phi)$ is positive everywhere. Then following the earlier result in \cite{nanda2023jt} we can satisfy this condition by working in a compact range of negative values of $M$. This allows us to set all the coefficients to zero except $\rho(M)$ using eq.\eqref{onerel}. In this case norm is then conserved for all values of $\phi$. There are also other possible choices of the coefficients and range of $M$ that allows us to satisfy the condition eq.\eqref{condva2}. Interested reader can refer to \cite{nanda2023jt} for a more detailed discussion of these possibilities. 

Now we move on to our discussion of the limit $l \rightarrow \infty$ of eq.\eqref{cneq}. Let us assume that $U(\phi)$ is positive everywhere. Then we can satisfy this quite easily by working in a compact range of negative values of $M$. At large values of $l$, $\mathcal{C_N} \rightarrow \frac{1}{l}$ and hence vanishes as $l \rightarrow \infty$. In cases where range of $M$ is compact but includes positive values we can still achieve $\mathcal{C_N} \rightarrow 0$ as $l \rightarrow \infty$ by setting $\rho_2 (M)$ to be zero. The cases where $\rho_2 (M)$ is not zero, the leading term can be made to vanish by making $\rho_2 (M)$ a real function or having an $M-$independent phase. But one then must take into account the behavior of sub-leading terms.

In conclusion we saw that the analysis done in the case of JT gravity extends in a straightforward manner to any arbitrary positive potential.

\section{More Details on Potential with Exponential Corrections}
\label{moredetpotcd}
Here we give more details on the potential with exponential corrections considered in section \ref{condefpot}. 

\subsection{Matching With The Path Integral Result}
\label{match}
In this subsection we will show how the wavefunction in the presence of defects can be matched to the exact solutions of WDW equation in terms of  Hankel functions in eq.\eqref{solpsi}. In Euclidean AdS, the corresponding partition function in the presence of defects as obtained in \cite{witten2020matrix} and \cite{eberhardt20232d} to $\order{(\epsilon^2)}$ is given by,
\begin{align}
	\label{Pfresults}
	Z= 	\begin{cases}
		e^{S_0}\left(\frac{\exp(\frac{2\pi^2}{\tilde{\beta}})}{\sqrt{2\pi} \tilde{\beta}^{\frac{3}{2}}} + \epsilon \frac{\exp(\frac{(2\pi-\alpha)^2}{2\tilde{\beta}})}{\sqrt{2\pi} \tilde{\beta}^{\frac{1}{2}}} + \frac{\epsilon^2}{2} \sqrt{\frac{\tilde{\beta}}{2\pi}} \right), \,& \pi< \alpha < 2\pi \\
		e^{S_0}\left(\frac{\exp(\frac{2\pi^2}{\tilde{\beta}})}{\sqrt{2\pi} \tilde{\beta}^{\frac{3}{2}}} + \epsilon \frac{\exp(\frac{(2\pi-\alpha)^2}{2\tilde{\beta}})}{\sqrt{2\pi} \tilde{\beta}^{\frac{1}{2}}} + \frac{\epsilon^2}{2} \sqrt{\frac{\tilde{\beta}}{2\pi}} e^{\frac{2(\pi-\alpha)^2}{\tilde{\beta}}}\right), \, &0< \alpha < \pi 
	\end{cases}
\end{align}
The corresponding wavefunctions in dS can be obtained by doing the analytic continuation,
\begin{equation}
	\beta = i \tilde{\beta} \label{anarel}
\end{equation}
One gets for the sharp defects $\pi < \alpha < 2\pi$,
\begin{equation}
	\Psi^{+}_{sd}= e^{-i l \sqrt{U(\phi)}} \left(e^{3 i \frac{\pi}{4}}\frac{\exp(\frac{2i\pi^2}{{\beta}})}{\sqrt{2\pi} {\beta}^{\frac{3}{2}}} + \epsilon \frac{\exp(i\frac{(2\pi-\alpha)^2}{2{\beta}})}{\sqrt{2\pi} \beta^{\frac{1}{2}}} e^{i \frac{\pi}{4}} + \frac{\epsilon^2}{2} \sqrt{\frac{\beta}{2\pi}} e^{-i \frac{\pi}{4}}\right) \label{sdwf}
\end{equation}
where the factor in front of parentheses follows because we are considering expanding branch wavefunction. Similarly in the case of blunt defects $0<\alpha<\pi$ one gets,
\begin{equation}
	\Psi^{+}_{bd}= e^{-i l \sqrt{U(\phi)}} \left(e^{3 i \frac{\pi}{4}}\frac{\exp(\frac{2i\pi^2}{{\beta}})}{\sqrt{2\pi} {\beta}^{\frac{3}{2}}} + \epsilon \frac{\exp(i\frac{(2\pi-\alpha)^2}{2{\beta}})}{\sqrt{2\pi} \beta^{\frac{1}{2}}} e^{i \frac{\pi}{4}} + \frac{\epsilon^2}{2} \sqrt{\frac{\beta}{2\pi}}  e^{i\frac{2(\pi-\alpha)^2}{\beta}} e^{-i \frac{\pi}{4}}\right) \label{bdwf}
\end{equation}
The subscript $sd$, $bd$ in eq.\eqref{sdwf},\eqref{bdwf} is to denote that the wavefunctions correspond to the case of sharp and blunt defect respectively. 

Consider a wavefunction of the form,
\begin{equation}
	\Psi = \frac{-il}{2}   \left(\frac{v_1}{u_1} H_2 ^{(2)} (\sqrt{v_1 u_1}) + \epsilon \sqrt{\frac{v_2}{u_2}} H_1^{(2)} (\sqrt{v_2 u_2})+{\epsilon^2} H_0^{(2)} (\sqrt{v_3 u_3}) \right) \label{pert1}
\end{equation}
where 
\begin{align}
	&u_1= l^2 -4\pi^2, \, \,\,\,\qquad v_1= U(\phi)\nonumber\\
	&u_2 = l^2-(2\pi-\alpha)^2, v_2= U(\phi)\nonumber\\
	&u_3 = l^2+q_3, \,\,\quad\qquad v_3= U(\phi)
	\label{usfir}
\end{align}
where the constant $q_3$ takes the value
\begin{align}
	q_3 = \begin{cases}
		- 4(\pi-\alpha)^2, \quad \, \alpha \leq \pi \\
		0, \quad \,\qquad\qquad\,\,\, \alpha \geq \pi
	\end{cases}  \label{q3eq}
\end{align}
Noting the large argument expansion of the the Hankel function,
\begin{equation}
	H_{\nu} ^{(2)} (\xi) = (-1)^\nu \frac{e^{-i\xi}}{\sqrt{\xi}} \sqrt{\frac{2}{\pi}} e^{-i (2\nu-1) \frac{\pi}{4}} \label{asymhan}
\end{equation}
 the wavefunction in eq.\eqref{pert1}, at large $U(\phi), l$  can be written as,
\begin{equation}
	\Psi =e^{-i l \sqrt{U(\phi)}}  \sqrt{\frac{1}{2\pi}}\left(\frac{U(\phi) ^{\frac{3}{4}}}{l^{\frac{3}{2}}} e^{3i \frac{\pi}{4}}  e^{2 i \pi^2\frac{\sqrt{U(\phi)}}{l}}  + \epsilon \frac{U(\phi) ^{\frac{1}{4}}}{l^{\frac{1}{2}}} e^{i \frac{\pi}{4}}  e^{i (2\pi-\alpha)^2\frac{\sqrt{U(\phi)}}{2l}} +\frac{\epsilon^2}{2} \frac{l^{\frac{1}{2}}  e^{-i \frac{\pi}{4}}}{U(\phi) ^{\frac{1}{4}}} e^{-\frac{q_3 \sqrt{U(\phi)}}{2 l}} \right) \label{pert1i}
\end{equation}
Expressing it in terms of the variable
\begin{equation}
	\label{id}
{\beta} = \lim_{l,\phi \rightarrow \infty} \frac{l}{\sqrt{U(\phi)}} 
\end{equation}
we find that the wavefunction eq.\eqref{pert1i} matches with eq.\eqref{sdwf}, \eqref{bdwf} for the appropriate values of $q_3$. 

\subsubsection{General order analysis}
It is clear from the above analysis that this can be extended to any order. The general strategy for any arbitrary order will be as follows. At order $m$, the partition function $Z_m$ will be a sum of different powers of $\beta$. The highest power of $\beta$ will be $\beta^{m-\frac{3}{2}}$. To match with the wavefunction given in eq.\eqref{solpsi} we must choose a suitable Hankel. It can be done as follows. Noting that,
\begin{equation}
	e^{n\theta} \propto \frac{1}{\beta ^n}
\end{equation}
and we get $\sqrt{\beta}$ from large argument of the Hankel function, for large $l,\phi$ we can write in general,
\begin{equation}
	l	e^{n \theta}  H_m ^{(2)} (\xi) \propto \beta^{\frac{1}{2} -n}
\end{equation}
To match with $Z_m$ we then need $n= 2-m$. To summarize if,
\begin{equation}
	Z_m = e^{S_0} \epsilon^m \sum_i c_m ^i \tilde{\beta}^{m-\frac{3}{2}-i}
\end{equation}
the corresponding wavefunction one should consider is,
\begin{equation}
	\Psi_m = A l \epsilon^m \sum_i e^{(2+i-m) \theta_i} d_m ^i H_{2+i-m}^{(2)} (\xi_i)
\end{equation}
where,
\begin{equation}
	e^{\theta_i} = \sqrt{\frac{v_i}{u_i}} = \sqrt{\frac{U(\phi)+p_i}{l^2+q_i}} , \xi=\sqrt{u_i v_i} = \sqrt{(l^2+q_i)(U(\phi)+p_i)}
\end{equation}
Note that both sets of constants $c_m ^i$ and $d_m ^i$ can depend on $\alpha$, the defect angle. They will be continuous functions of $\alpha$ and will take specific forms based on the range of $\alpha$. Similarly, $p_i,q_i$ will also depend on $\alpha$ as in eq.\eqref{q3eq}.

\subsection{Semi-classical Density Of States}
\label{semdos}
In section \ref{condefpot} we found the Hartle Hawking wavefunction for the potential with exponential correction and the corresponding coefficient function, eq.\eqref{defra}. In section \ref{HHwf} we saw that the Hartle-Hawking wavefunction can be calculated in a semi-classical expression by considering a coefficient function which is given by the exponential of the cosmological entropy, eq.\eqref{relas}. In this subsection we will examine the relation between the correct density of states for the defect case given in eq.\eqref{defra} and the approximate one given in eq.\eqref{relas}.

To do so let us first find cosmological horizon $r_h$ which is given by,
\begin{equation}
	U(r_h) = M \label{rheq}
\end{equation}
where,
\begin{equation}
	U(r) = r^2 -  \sum_{i=1}^{2}  \frac{2}{\alpha_i}\epsilon_i e^{-\alpha_i \phi} \label{potcd3}
\end{equation}
with the condition,
\begin{equation}
\sum_{i=1}^{2}	\epsilon_i=0
\end{equation}
Let the horizon correct up to $\order{(\epsilon)}$ be given by,
\begin{equation}
	r_h = \sqrt{M} + \epsilon \delta r \label{rheq2}
\end{equation}
Using eq.\eqref{rheq2} and eq.\eqref{potcd3} in eq.\eqref{rheq} we get,
\begin{equation}
	\delta r = \frac{1}{\alpha_1 \sqrt{M}} e^{-\alpha_1 \sqrt{M}} - \frac{1}{\alpha_2 \sqrt{M}} e^{-\alpha_2 \sqrt{M}}
\end{equation}
Then from eq.\eqref{relas} we get,
\begin{equation}
	\rho = e^{2 \pi r_h} = e^{2 \pi \sqrt{M}} + \frac{2\pi \epsilon}{\sqrt{M}} \left(\frac{e^{( 2\pi-\alpha_1) \sqrt{M}}}{\alpha_1}-\frac{e^{( 2\pi-\alpha_2) \sqrt{M}}}{\alpha_2}\right) \label{defra2}
\end{equation}
Matching with eq.\eqref{defra} we see that the formula eq.\eqref{relas} does capture the correct behavior of the density of states at large values of energy. However there is a mismatch of an $\alpha$ factor in the denominator. This implies that the correction term in the wavefunction will also differ from the actual wavefunction by an $\alpha$ factor, see also section 2 of \cite{witten2020matrix}.

\section{Computation of Correlators for a different state Using Matrix Theory}
\label{topHH}
In this appendix, we will elaborate more on the deformed theory with the exponential deformation mentioned in eq.\eqref{altU}. Following the discussion above eq.\eqref{deffh} we can consider the HH state in this deformed theory as a state in the original undeformed JT dS theory. 
The resulting coefficient function for this state allows us to map this state into the Matrix theory of the undeformed theory in terms of a generalized trace, see eq.\eqref{exmat},\eqref{deffm}. Using this map to the Matrix theory, we compute one point and two point functions for this state. For example, the two point function corresponds to producing two universes out of nothing or a transition amplitude for going  from one universe to the other. We will find that our answer for the two point function in the matrix theory is different from the two point function for the HH state computed in the deformed theory directly and will discuss the reason for this mismatch.
%

\subsection{ One Point Function}


The HH wavefunction for the exponential deformation eq.\eqref{altU}, satisfying eq.\eqref{restb},  computed in a peturbation theory in $\epsilon_i$ reads
\begin{align}
	{\Psi=	e^{-i l \phi} \left(e^{3 i \frac{\pi}{4}}\frac{\exp(\frac{2i\pi^2}{{\beta}})}{\sqrt{2\pi} {\beta}^{\frac{3}{2}}} + \sum_{i} \epsilon_i \frac{\exp(i\frac{(2\pi-\alpha_i)^2}{2{\beta}})}{\sqrt{2\pi} \beta^{\frac{1}{2}}} e^{i \frac{\pi}{4}}\right)}\label{psicondf}
\end{align}
The above result is the exact result with no further $\order{(\epsilon^2)}$ for potential satisfying eq.\eqref{restb}.
The path integral is done with the deformed dilaton potential involving exponential correction, as discussed in section \ref{condefpot}, see eq.\eqref{hhdef}.
The corresponding density of states is given by eq.\eqref{defra} reproduced below,
\begin{align}
	\label{conrho}
	\rho(M)&=e^{S_0}\left[\frac{\sinh{2\pi\sqrt{M}}}{4\pi^2}+\sum_i\epsilon_i\frac{ \cosh{(2\pi-\alpha_i)\sqrt{M}}}{2\pi\sqrt{M}}\right]
\end{align}
Considering the wavefunction in eq.\eqref{psicondf} as that corresponding to a state in the undeformed JT theory,  we can extract $f(M)$ as defined in eq.\eqref{deffm}, to be
\begin{equation}
	\label{falpe}
	f(M)=1+\sum_i 2\pi \epsilon_i \frac{\cosh{(2\pi-\alpha_i)\sqrt{M}}}{\sqrt{M}\sinh{2\pi\sqrt{M}}}
\end{equation}
We now perform a simple consistency check by evaluating the generalized trace using the map eq.\eqref{gentrace} for the function $f(M)$ given by eq.\eqref{falpe} and verify that we indeed get the expected result for the deformed potential involving exponential correction, eq.\eqref{psicondf}.

For the following analysis it will be convenient to define,
\begin{equation}
	\hat{\beta} = \frac{\beta}{2} \label{hat}
\end{equation}
The wave function expressed in terms of the generalized trace in the matrix theory as in eq.\eqref{exmat} with the new definition in eq.\eqref{hat}, then takes the form 
\begin{equation}
	\label{zalp}
	\langle \Tr(e^{i\hat{\beta} H}f(H))\rangle=\int dM~ \rho_{MM}(M) f(M)e^{i\hat{\beta} M}
\end{equation}
where $\rho_{JT}$ is given by eq.\eqref{dosa}.
We compute the above integral in a way that can be adapted for the two-point function that we compute next. $f(M)$ in eq.\eqref{falpe} possesses a series expansion around $M=0$ as schematically given by 
\begin{align}
	\lim_{M\to0}f(M)&=1+\frac{\sum_i\epsilon_i}{M}+\sum_{n=0}^{\infty}\left(\sum_i\epsilon_iC_n(\alpha_i)\right) M^n\nonumber\\
	\label{serfalp}
	&=1+\sum_{n=0}^{\infty}\left(\sum_i\epsilon_iC_n(\alpha_i)\right) M^n
\end{align}
where in order to obtain the last equality in the above equation, we used eq.\eqref{restb}. First two $C_n(\alpha_i)$ are given by 
\begin{align}
	C_0(\alpha_i)&=\frac{4\pi^2}{3}-2\pi\alpha_i+\frac{\alpha_i^2}{2}\nonumber\\
	C_1(\alpha_i)&=-\frac{16\pi^4}{45}+\frac{2\pi^2\alpha_i^{2}}{3}-\frac{\pi\alpha_i^3}{3}+\frac{\alpha_i^4}{24}\quad \quad \nonumber
\end{align}
Now, using  eq. \eqref{serfalp} in eq.\eqref{zalp}, we can write the one point function as 
\begin{equation}
	\langle \Tr(e^{i\hat{\beta} H}f(H))\rangle=\langle \Tr(e^{i\hat{\beta} H})\rangle+\sum_{n=0}^{\infty}\left(\sum_i\epsilon_iC_n(\alpha_i)\right)\left(\frac{1}{i}\frac{\partial}{\partial\hat{ \beta}}\right)^n\langle \Tr(e^{i\hat{\beta} H})\rangle
\end{equation}
Evaluating the above equation at large $\beta$ and using eq.\eqref{restb}, one can easily find 
\begin{align}
	\label{sss1pt}
	\langle \Tr(e^{i\hat{\beta} H}f(H))\rangle
	&=e^{S_0}\frac{e^{\frac{i\pi^2}{\hat{\beta}}}}{4\sqrt{\pi}(-i\hat{\beta})^{3/2}}
	\left[1-2i\hat{\beta}\sum_j\epsilon_j\left(e^{\frac{\alpha_j(\alpha_j-4\pi)}{4(-i\hat{\beta})}}-1\right)\right]\\
	&=e^{S_0}\left(\frac{e^{\frac{i\pi^2}{\hat{\beta}}}}{4\sqrt{\pi}(-i\hat{\beta})^{3/2}}
	+\sum_j\epsilon_j\frac{e^{\frac{i(2\pi-\alpha_j)^2}{4(\hat{\beta})}}}{2\sqrt{\pi}(-i\hat{\beta})^{1/2}}\right)
\end{align}
Using eq.\eqref{hat} the above relation matches with eq.\eqref{psicondf}.


\subsection{Two Point Function}

As discussed in section \ref{JTMM}, in a state other than HH state, the amplitude to produce two expanding or contracting universes, asymptotically having length
$l_1$ and $l_2$  at $\phi_1$ and $\phi_2$ respectively, from ``nothing", in the matrix model is proportional to 
\begin{multline}
	\langle \Tr(e^{\pm i\hat{\beta}_1 H}f(H))\Tr(e^{\pm i\hat{\beta}_2 H}f(H))\rangle\\
	=\int dM_1dM_2~\langle\rho_{MM}(M_1)f(M_1)e^{\pm i\hat{\beta}_1M_1}\rho_{MM}(M_2)f(M_2)e^{\pm i\hat{\beta}_2 M_2}\rangle\label{twpfff}
\end{multline}
where in the above equation, $+$ sign corresponds to expanding branch and $-$ sign corresponds to contracting branch with $\beta_1$ and $\beta_2$ are defined by 
\begin{equation}
	\beta_1=\frac{l_1}{\phi_1}\quad \beta_2=\frac{l_2}{\phi_2}
\end{equation}
For simplicity, lets work with two expanding branches. Again considering $f(M)$  given by eq.\eqref{falpe} which possesses a series expansion around $M=0$ as in eq.\eqref{serfalp}, the two point function in eq.\eqref{twpfff} becomes
\begin{multline}
	\langle \Tr(e^{i\hat{\beta}_1 H}f(H))\Tr(e^{i\hat{\beta}_2 H}f(H))\rangle=\left[1+\sum_{n=0}^{\infty}\left(\sum_i\epsilon_iC_n(\alpha_i)\right)\left(\frac{1}{i}\frac{\partial}{\partial \hat{\beta}_1}\right)^n\right]\\
	\label{sss2pt}
	\left[1+\sum_{m=0}^{\infty}\left(\sum_i\epsilon_jC_m(\alpha_j)\right)\left(\frac{1}{i}\frac{\partial}{\partial \hat{\beta}_2}\right)^m\right] \langle \Tr(e^{ i\hat{\beta}_1 H})\Tr(e^{i\hat{\beta}_2 H})\rangle
\end{multline}  
Now, $\langle \Tr(e^{i\beta_1 H})\Tr(e^{i\beta_2 H})\rangle$ is given by the double trumpet partition function as follows
\begin{equation}
	\langle \Tr(e^{ i\hat{\beta}_1 H})\Tr(e^{i\hat{\beta}_2 H})\rangle= \frac{\sqrt{\hat{\beta}_1\hat{\beta}_2}}{2\pi(\hat{\beta}_1+\hat{\beta}_2)} \label{2ptcor}
\end{equation}
Therefore, we can further rewrite
\begin{multline}
	\langle \Tr(e^{i\hat{\beta}_1 H}f(H))\Tr(e^{i\hat{\beta}_2 H}f(H))\rangle=\frac{\sqrt{\hat{\beta}_1\hat{\beta}_2}}{2\pi(\hat{\beta}_1+\hat{\beta}_2)}\\
	+\left[\sum_{n=0}^{\infty}\left(\sum_i\epsilon_iC_n(\alpha_i)\right)\left(\frac{1}{i}\frac{\partial}{\partial \hat{\beta}_1}\right)^n
	+\sum_{m=0}^{\infty}\left(\sum_i\epsilon_jC_m(\alpha_j)\right)\left(\frac{1}{i}\frac{\partial}{\partial \hat{\beta}_2}\right)^m\right]\frac{\sqrt{\hat{\beta}_1\hat{\beta}_2}}{2\pi(\hat{\beta}_1+\hat{\beta}_2)}+\mathcal{O}(\epsilon_i\epsilon_j)
\end{multline}  
For convenience of computing derivatives, we substitute
\begin{equation}
	\frac{\sqrt{\hat{\beta}_1\hat{\beta}_2}}{2\pi(\hat{\beta}_1+\hat{\beta}_2)}=\int_{0}^{\infty} \frac{d\alpha}{2\pi} \sqrt{\hat{\beta}_1\hat{\beta}_2}e^{-\alpha(\hat{\beta}_1+\hat{\beta}_2)}
\end{equation}
We finally find that 
\begin{multline}
	\langle \Tr(e^{i\hat{\beta}_1 H}f(H))\Tr(e^{i\hat{\beta}_2 H}f(H))\rangle=\frac{\sqrt{\hat{\beta}_1\hat{\beta}_2}}{2\pi(\hat{\beta}_1+\hat{\beta}_2)}\biggl[1+\sum_i\epsilon_i\Bigl\{2 C_0(\alpha_i)\\
	-i C_1(\alpha_i)\left(\frac{2}{\hat{\beta}_1+\hat{\beta}_2}-\frac{1}{2\hat{\beta}_1}-\frac{1}{2\hat{\beta}_2}\right)+C_2(\alpha_i)\left(\frac{1}{4\hat{\beta}_1^2}+\frac{1}{4\hat{\beta}_2^2}+\frac{1}{\hat{\beta}_1\hat{\beta}_2}-\frac{4}{(\hat{\beta}_1+\hat{\beta}_2)^{2}}\right)
	+...\Bigr\}\\+\mathcal{O}(\epsilon_i\epsilon_j)\Bigl]
	\label{2ptdefrm}
\end{multline}
Note that the above series is convergent in the limit $\hat{\beta}_1,\hat{\beta_2}\to \infty$ limit which is the correct limit while considering the expansion of $f(M)$ around $M=0$. Note that  in \cite{witten2020matrix}, it was argued that the two point correlator will continue to be the one in eq.\eqref{2ptcor} in the presence of the exponential deformation of the dilaton potential and therefore will correspond to only the first term within square bracket in eq.\eqref{2ptdefrm}. Here, we find that there are additional corrections. As mentioned towards the end of section \ref{condefpot} this is not a contradiction. While the deformed theory, obtained by adding   the exponential term in the dilaton potential, asymptotically, at late time, does  give rise to a physical state in the undeformed JT theory, the two theories are in fact different and the transition amplitudes computed in the two theories need not agree. 

%

\section{Hartle Hawking State for Generic Potentials}
\label{Hhgenpot}
In this section we will generalize the discussion in section \ref{HHwf} for potentials that do not asymptote to dS. In particular we will consider potentials that satisfy eq.\eqref{condimp}, reproduced below,
\begin{equation}
	U(\phi) = - U(i \phi) \label{condimp2}
\end{equation}
but do not satisfy eq.\eqref{formab}. A simple example of this potential will be,
\begin{equation}
	U(\phi) = \phi^{4m+2} \label{pot5}
\end{equation}
where $m$ is a positive integer. This clearly satisfies eq.\eqref{condimp2}. More generally one can consider any odd function of $\phi^2$.  We will show that even for theories that do not asymptote to dS space, the no-boundary wavefunction can still be obtained by a coefficient function which is the exponential of cosmological entropy.

The argument is very similar to the one presented in section \ref{HHwf}. Hence we will be brief in our presentation and only highlight the differences. Consider then the $-AdS_2$ like metric given by,
\begin{equation}
	ds^2 = - \frac{dr^2}{U(r)-M} - (U(r)-M) \frac{dx^2}{A^2}, \phi = i r \label{metads2}
\end{equation}
The horizon $r_h$ is located at,
\begin{equation}
	U(r_h)=M \label{horeq}
\end{equation}
and the usual arguments regarding the smoothness near the horizon leads to,
\begin{equation}
	\beta = \frac{4 \pi}{U'(r_h)} = \frac{1}{A}  \label{betaeq}
\end{equation}

The path integral in the $-$AdS space takes the form
\begin{equation}
	Z = \int \mathcal{D}g_{\mu \nu} \mathcal{D}\phi e^{-S_{-AdS}}  \label{Zcd}
\end{equation}
We now proceed to compute the above path integral in the saddle point method by calculating the on-shell action given by,
\begin{equation}
	S_{-AdS} = \frac{1}{2} \int d^2x \sqrt{g} (\phi R - U'(\phi)) -  \int dx \sqrt{\gamma} \phi K \label{adsact4}
\end{equation}
We take our boundary to be at $r=r_B$. Then using the metric eq.\eqref{metads2},a short computation gives,
\begin{equation}
	K = \frac{U'(r_B)}{2 \sqrt{U(r_B)-M}} \label{Keq}
\end{equation}
Using eq.\eqref{Keq}, eq.\eqref{Req} and eq.\eqref{horeq} we can now evaluate eq.\eqref{adsact4}. In the first term of eq.\eqref{adsact4} the radial coordinate ranges from the horizon $r_h$ to the boundary $r_B$. Putting it all together we get,
\begin{align}
	S_{-AdS} &=   \frac{i}{2 A} \left(-2 (U(r_B)-M)- r_h U'(r_h)  \right)
\end{align}
In terms of $\phi$ we have,
\begin{equation}
	-S_{-AdS} = -\frac{i}{A}  (U(\phi_B)+M) + \frac{\phi_h}{2 A} U'(r_h)
\end{equation}
This can now be analytically continued and one gets,
\begin{equation}
	iS_{dS} = -\frac{i}{A}   (U(\phi_B)+M) +\frac{\phi_h}{2 A} U'(r_h) \label{sds}
\end{equation}
The dS like metric is given by,
\begin{equation}
	ds^2 = - \frac{d\rho^2}{U(\rho)+M} + (U(\rho)+M) \frac{dx^2}{A^2}, \phi = \rho \label{metds}
\end{equation}
From eq.\eqref{metds} we have,
\begin{equation}
	\frac{1}{A} = \frac{l_B}{\sqrt{U(\rho_B)+M}} 
\end{equation}
Employing the above relation along with eq.\eqref{betaeq} we get by expanding eq.\eqref{sds} to $\order{(M)}$,
\begin{equation}
		iS_{dS}  = - i l_B \sqrt{U(\phi_B)} -  \frac{il_B M}{2 \sqrt{U(\phi_B)}} + 2 \pi \phi_h
\end{equation}
Thus the wavefunction is given by,
\begin{equation}
	\Psi_{HH} (l_B, \phi_B) = \exp\left( - i l_B \sqrt{U(\phi_B)} -  \frac{il_B M}{2 \sqrt{U(\phi_B)}} + 2 \pi \phi_h\right) \label{psihh}
\end{equation}
This has the same form as eq.\eqref{logeqb} with $\sqrt{U(\phi_B)} = \phi_B$ and 
\begin{equation}
	\log I\left( \frac{-il_B}{\sqrt{U(\phi_B)}}\right) = \frac{-il_B M}{2 \sqrt{U(\phi_B)}} + 2 \pi \phi_h \label{iueq}
\end{equation}
The above equation readily follows from eq.\eqref{valI} reproduced below,
\begin{equation}
	I(\beta) = \int \rho(\tilde{M}) e^{- \frac{\beta\tilde{M}}{2}} d\tilde{M}
\end{equation}
by noting that, 
\begin{equation}
	\rho = e^{2 \pi \phi_h}  \label{rhoxx}
\end{equation}
and evaluating the integral in saddle point approximation. As promised the coefficient function that leads to the no boundary wavefunction for theories that do not asymptote to dS but still satisfies eq.\eqref{condimp2}, is of the form eq.\eqref{rhoxx} .  Note that in obtaining the above saddle point relation we have glossed over many caveats such as having multiple solutions to eq.\eqref{horeq}. These are in fact the same caveats that are present in the potentials that asymptote to dS, eq.\eqref{formab}, see the discussion towards the end of the section \ref{HHwf}. 

\subsection{Finiteness of the Wavefunction}
Before ending this section let us discuss the finiteness of the no-boundary wavefunction obtained for the general theories discussed above and in section \ref{HHwf}.

First let us recapitulate a few facts about the divergence of the norm of Hartle Hawking wavefunction in the case of JT gravity. The norm diverges chiefly because the wavefunction has a divergence near $l=2\pi$. To verify this consider the wavefunction,
\begin{equation}
	\Psi = \int_{0}^{\phi^2} \rho(M) e^{-i l \sqrt{\phi^2-M}} dM +\int_{\phi^2}^{\infty}  \rho(M) e^{-l \sqrt{M-\phi^2}}  dM\label{psihh2}
\end{equation}
For,
\begin{equation}
	\rho(M) = \frac{\sinh(2\pi\sqrt{M})}{4 \pi^2}
\end{equation}
eq.\eqref{psihh2} gives the Hartle Hawking wavefunction. For values of $M$ such that $M \gg \phi^2$ one gets from eq.\eqref{psihh2},
\begin{equation}
	\Psi \sim \int^{\infty} e^{2 \pi \sqrt{M}} e^{-l \sqrt{M}} dM \label{psihhapp}
\end{equation}
which is divergent for $l \leq 2 \pi$. Coincidentally $l=2\pi$ is also the minimum value of $l$ in a global dS spacetime. The metric is given by,
\begin{equation}
	ds^2 = - \frac{dr^2}{r^2+1} + (r^2+1) dx^2, \phi=A r
\end{equation}
where range of $x$, $\Delta x$ is such that the analytically continued $-$AdS metric has no conical deficit. Thus $\Delta x = 2 \pi$. So we get,
\begin{equation}
	l =2 \pi \sqrt{r^2+1} 
\end{equation}
Since the minimum value of $r$, $r_{min}=0$ we get, $l_{min}=2\pi$. 

Now consider the Hartle Hawking wavefunction in the potentials that asymptote to dS such as eq.\eqref{formab}. In this case, working in the semi-classical regime for large enough values of $M$ one gets the density of state to be,
\begin{equation}
	\rho(M) \sim e^{2\pi \sqrt{M}} \label{rhoexp}
\end{equation}
Thus for $M \gg \phi^2$, eq.\eqref{psihhapp} is still valid. Hence for $l \leq 2\pi$ the wavefunction has a divergence. However in this case $l=2\pi$ is not the minimum value. To see this note that in this case the global dS like metric is given by, using eq.\eqref{metds} with $M=1$,
\begin{equation}
	ds^2 = - \frac{d\rho^2}{U(\rho)+M} + (U(\rho)+M) dx^2, \phi=\rho
\end{equation}
Noting that $\Delta x$ is given by eq.\eqref{betaeq} we have,
\begin{equation}
	l_{min} = \frac{4\pi}{U'(r_h)}\sqrt{U(\rho_{min})+U(r_h)}  
\end{equation}
where we used $M= U(r_h)$. This in general will not be $2 \pi$. 

Finally for the potentials of the form eq.\eqref{pot5}, the density of states in the semi-classical regime is given by, with $p=4m+2$, 
\begin{equation}
	\rho(M) = e^{2 \pi M^{\frac{1}{p}}}
\end{equation}
The wavefunction for values of $M$ such that $M \gg \phi^p$  then reads,
\begin{equation}
	\Psi \sim \int^{\infty}e^{2 \pi M^{\frac{1}{p}}} e^{-l \sqrt{M}} dM
\end{equation}
Then for any value of $l>0$ this integral is convergent. However for $l=0$ this integral diverges. 

Note that while the above semi-classical analysis is valid for large values of $M$, the analysis of the divergence of the norm in its entirety will require the one-loop determinant which will then enable us to calculate the wavefunction and its norm exactly.

\section{Recursion Relations in The Case of Conical Deficit Potential}
\label{matmodrec}
In this section we shall understand the matrix model dual for the theory of JT gravity in dS with a defect. The analysis follows closely that of \cite{witten2020matrix}. 

The conical deficit potential we will consider is given by,
\begin{equation}
	U'(\phi) = 2\phi+ 2 \sum_{i} \epsilon_i e^{-\alpha_i \phi}\label{defg1}
\end{equation}
We shall consider the special case of the potential which does not shift the lowest energy level. As discussed in \cite{witten2020matrix}, this translates to the property on the potential as
\begin{equation}
	U'(0) =0 \implies \sum_{i} \epsilon_i=0.\label{updefg2}
\end{equation}
For the potential satisfying the above condition, the density of states given by eq.\eqref{defra}, reproduced below,
\begin{equation}
		\rho(M) = \frac{\sinh(2 \pi \sqrt{M})}{4 \pi^2} +\sum_i \epsilon_i \frac{1}{2 \pi \sqrt{M}} \cosh((2\pi-\alpha_i) \sqrt{M})\label{rhg3}
\end{equation}
is exact. 

 The matrix model, in the double scaling limit, is completely characterized by two pieces of information. The first is the leading density of states, which in the gravity side is computed by the disk partition function. In our case corresponding to the potential eq.\eqref{defg1}, this is given by eq.\eqref{rhg3}. The second piece of information is the connected component of the two point function of the density of states of the matrix theory. On the gravity side, this is obtained by the double trumpet partition function. As argued in \cite{witten2020matrix}, for the potential satisfying eq.\eqref{updefg2}, the double trumpet partition function is unchanged from the pure JT theory. For a theory with the two pieces of information mentioned above, the recursion relations of the double scaled matrix model then fix the higher point connected correlators of the density of states. 
 
 The strategy we follow to establish the matrix dual of the theory with potential eq.\eqref{defg1} is as follows. We shall compute certain integral transforms of the resolvents on the matrix model side. We then compare these with the computation of the same quantity from the gravity side. 
 
 Let us now be more explicit about the quantities we evaluate on either side. 
 
 Let us first begin with the matrix theory side. On the matrix theory side, the information of the matrix model is encapsulated in the matrix resolvent defined by 
 \begin{align}
 	R(E)=\Tr\frac{1}{E-H}=\sum_i\frac{1}{E-\lambda_i}\label{retr}
 \end{align}
where $H$ is the random matrix and $\lambda_i$ are its eigenvalues. For a double scaled matrix model, the correlation function of resolvents have an expansion of the form
\begin{align}
	\langle R(E_1)\dots R(E_n)\rangle_{\text{conn}} =\sum_{g=0}^{\infty}\frac{R_{g,n}(E_1\dots E_n)}{(e^{S_0})^{2g+n-2}}\label{re1ren}
\end{align}
where $e^{S_0}$ is a large parameter. For making comparison to gravity calculations easier, it is convenient to define the quantity $W_{g,n}$ as
\begin{align}
	W_{g,n}(z_1,\dots z_n)=(-1)^n2^nz_1\dots z_n R_{g,n}(-z_1^2,\dots,-z_n^2)\label{wgndef}
\end{align}
For the double scale matrix models, the above defined quantities $W_{g,n}$ satisfies the recursion relations, see \cite{Saad:2019lba},
\begin{align}
	W_{g,n}(z_1,\overbrace{z_2,\dots,z_n}^{J}) &=\text{Res}_{z\rightarrow 0}  \left[\frac{1}{z_1^2-z^2} \frac{1}{4 y(z)} W_{g-1,n+1} (z,-z,J) \right] \nonumber \\ 
	&+ \text{Res}_{z\rightarrow 0}  \left[\sum'_{I\bigcup I'=J, h\bigcup h'=g} \left(W_{h,1+\abs{I}} (z,I)  W_{h',1+\abs{I'}} (-z,I')\right)\right]  \label{wgn}
\end{align}
$\sum'$ means two cases are excluded, $(I=J, h=g)$ and $(I'=J, h'=g)$. The input to the above recursion relation are 
\begin{align}
	&W_{0,1}(z)=2z y(z),\qquad z^2=-E,\,\,  y(z) = - i \pi \rho(z) \nonumber\\
	&W_{0,2}=\frac{1}{(z_1-z_2)^2}\label{wo1wo2recur}
\end{align}
So, in effect, all the quantities in the double scaled matrix model for potentials satisfying the condition eq.\eqref{updefg2} can be computed by just knowing the information about the leading density of states. 

To establish the duality of the matrix model with the density of states in eq.\eqref{rhg3} to the corresponding gravity with the potential \ref{defg1}, we shall evaluate the first few non-trivial $W_{g,n}$ using the matrix model recursion relations. We then evaluate the same quantities from the gravity theory in a perturbation theory in $\epsilon_{i}$ and show that they match with the corresponding perturbative expansion of the matrix theory result.

 Before we start computing various quantities, let us mention that the  duality for the theory with  defects does not trivially follow from the one without defects. In the absence of defects, any given path integral with a fixed number of boundaries (n) and handles (g) is related to the appropriate $W_{g,n}$. For the case with defects, a given path integral with a particular configuration of defects, boundaries and handles, is not directly related to the above quantities $W_{g,n}$. Rather, one has to take add the  contributions obtained by doing the path integral   with all possible number of defects of various  types and compare the resulting sum with   the corresponding quantities $W_{g,n}$, obtained in the matrix model. Computing the path integral for all possible number of defects is a daunting task. So we shall only obtain the results for a maximum of two defects and correspondingly expand the matrix theory results to the corresponding order to show the match between the two sides.

Let us first evaluate the quantities on the matrix theory side. For sake of simplicity, we will consider minimum possible number of defects with which eq.\eqref{updefg2} can be met. The minimum number of defects needed is two labelled with $i=1,2$ satisfying the condition
\begin{align}
	\epsilon_{1}+\epsilon_{2}=0\label{espcon2}
\end{align}
From the density of states in eq.\eqref{rhg3}, we have that
\begin{equation}
	y(z) = \frac{\sin(2 \pi z)}{4\pi} - \frac{1}{2 \pi z } \sum_i \epsilon_i \cosh((2\pi-\alpha_i) z) \label{yeq}
\end{equation}

The first few non-trivial $W_{g,n}$ that we verify are $W_{0,3}$ and $W_{1,1}$. Using the recursion relations given in eq.\eqref{wgn} we can find $W_{0,3}$. We get,
\begin{equation}
	 W_{0,3} (z_1,z_2,z_3) = \text{Res}_{z \rightarrow 0} \{\frac{1}{z_1^2-z^2} \frac{1}{4 y(z)} \left( W_{0,2} (z,z_2)  W_{0,2} (-z,z_3) +  W_{0,2} (z,z_3)  W_{0,2} (-z,z_2)\right) \}\label{wo3rec}
\end{equation}
Using eq.\eqref{yeq} and eq.\eqref{wo1wo2recur} we get,
\begin{equation}
	W_{0,3} (z_1,z_2,z_3) = \frac{2}{z_1^2 z_2^2 z_3^2} \frac{1}{2- \epsilon_1 (4\pi-\alpha_1-\alpha_2) (\alpha_1-\alpha_2)} \label{w03eqcd}
\end{equation}
where we used $\epsilon_2 =- \epsilon_1$. Similarly, from the recursion relations,  we can write for $W_{1,1}$,
\begin{equation}
	W_{1,1} (z_1) = \text{Res}_{z \rightarrow 0} \{\frac{1}{z_1^2-z^2} \frac{1}{4 y(z)}  W_{0,2} (z,-z)\} \label{w11rec}
\end{equation}
which evaluates to,
\begin{equation}
		W_{1,1} (z_1) = \frac{24 + 16 \pi^2 z_1^2 + (-4\pi+\alpha_1+\alpha_2) (\alpha_1-\alpha_2) (12+z_1^2 (8 \pi^2 + \alpha_1^2 + \alpha_2^2 - 4 \pi(\alpha_1+\alpha_2) ) ) \epsilon_1}{48 z_1^4 (2 + (\alpha_1-\alpha_2) (-4\pi+\alpha_1+\alpha_2) \epsilon_1)^2} \label{w11eqcd}
\end{equation}
The results above in eq.\eqref{w03eqcd} and eq.\eqref{w11eqcd}, which are exact in $\epsilon_{1}$ need to be expanded in a perturbation theory around $\epsilon_{1}=0$ to compare with gravity answers since we will evaluate the path integrals perturbatively in $\epsilon_{i}$. We will do so upto quadratic order in $\epsilon_{1}$. Let us now turn to the computations on the gravity side. 

In the gravity theory, we shall evaluate these quantities perturbatively in $\epsilon_i$ by expanding the path integral as
\begin{align}
	&\int \mathcal{D}g\mathcal{D}\phi e^{\frac{1}{2} \int d^2x \sqrt{g}(\phi R + 2 \phi + 2 \sum_i \epsilon_i e^{-\alpha_i \phi})} = \int \mathcal{D}g\mathcal{D}\phi e^{\frac{1}{2} \int d^2x \sqrt{g} (\phi R + 2 \phi)} \left(1+ \sum_i \epsilon_i \int d^2y \sqrt{g(y)} e^{-\alpha_i  \phi(y)} \right) \nonumber \\
&+ \frac{1}{2}	\mathcal{D}g\mathcal{D}\phi e^{\frac{1}{2} \int d^2x \sqrt{g} (\phi R + 2 \phi)} \left(\sum_{i j} \epsilon_i \epsilon_j \int d^2y \sqrt{g(y)} \int d^2z \sqrt{g(z)} e^{-\alpha_i  \phi(y)}e^{-\alpha_i  \phi(z)} \right) + \cdots
\end{align}
 Since $\epsilon_{1},\epsilon_{2}$ satisfy eq.\eqref{espcon2}, we can consider just $\epsilon_{1}$ as the perturbation parameter. Let us denote the perturbation theory as
\begin{align}
	W_{g,n}=\sum_i \epsilon_1^m W_{g,n}^{(m)}\label{wgm}
\end{align}
In the gravity side, the quantities $g,n$ indicate the appropriate topology for the path integral to be carried out. In particular, $g$ specifies the genus of the surface and $n$ specifies the number of geodesic boundaries on the surface.   Let us quickly remind the reader about the  analogous calculation in the pure JT theory. For pure JT theory the partition function at any genus with specified number of asymptotic boundaries is obtained by summing over the corresponding Riemann surface with appropriate genus and geodesic boundaries with half trumpets attached to these geodesic boundaries.
\begin{align}
	Z_{g,n}(\beta_1,\dots,\beta_n)=\int \left(\prod_{i=1}^{n}b_idb_iZ_{\text{Sch}}^{\text{trumpet}}(\beta_i,b_i)\right) V_{g,n}^{(0)}(b_1,\dots,b_n) \label{zgn}
\end{align}
where $V_{g,n}^{(0)}$ is the volume of the bordered Riemann surface with genus $g$ and $n$ geodesic boundaries of lengths $b_1,\dots b_n$  with the superscript $^{(0)}$ indicating that it corresponds the case of no defects and $Z_{\text{Sch}^{\text{trumpet}}}(\beta_i,b_i)$  is the half trumpet partition function given by 
\begin{align}
	Z_{\text{Sch}}^{\text{trumpet}}(\beta,b)=\frac{1}{\sqrt{4\pi\beta}}e^{-\frac{b^2}{4\beta}}\label{halftrpjt}
\end{align}
In the pure JT theory corresponding to $m=0$ in eq.\eqref{wgm}, the quantities $W_{g,n}^{(0)}$ are obtained as the Laplace transform of the corresponding volume of bordered Riemann surfaces. 
\begin{align}
	W_{g,n}^{(0)}(z_1,\dots,z_n)=\int_0^\infty b_1 db_1 e^{-b_1z_1}\dots\int_0^\infty b_n db_n e^{-b_nz_n}V_{g,n}^{(0)}(b_1,\dots,b_n)\label{wgnvbn}
\end{align}
The above equation follows by noting that,
\begin{equation}
	R_{g,n} (E_1, E_2, \cdots E_n)  = (-1)^n \int \prod_{i=1}^{n} d \beta_i e^{\beta_i E_i} Z_{g,n}(\beta_1,\dots,\beta_n) \label{rgn}
\end{equation}
where each integral ranges from $0$ to $\infty$. Using the above equation in eq.\eqref{wgndef} along with eq.\eqref{zgn} and eq.\eqref{halftrpjt} and then integrating over $\beta_i$ leads to eq.\eqref{wgnvbn}.

At $\order{(\epsilon)^0}$, the results are that of pure JT gravity without defects. So, we immediately have
\begin{align}
	W_{0,3}^{(0)}(z_1,z_2,z_3)=\frac{1}{z_1^2z_2^2z_3^2},\nonumber\\
	W_{1,1}^{(0)}(z_1)=\frac{3+2\pi^2 z_1^2}{24 z_1^4}
	\label{purjtws}
\end{align}
For the case with defects the formula in eq.\eqref{wgnvbn} is extended straightforwardly as
\begin{align}
	W_{g,n}^{(m)}(z_1,\dots,z_n)= \frac{1}{n!}\sum_{i_1,i_2,\cdots i_m=1} ^{r} \epsilon_{i_1} \epsilon_{i_2}\cdots \epsilon_{i_m}\prod_{i=1}^{n}\int_0^\infty b_i db_i e^{-b_iz_i}V_{g,n}^{(m)}(b_1,\dots,b_n,\alpha_{i_1},\dots,\alpha_{i_m})\label{wgnvbnm}
\end{align}
In the above formula the quantity $V_{g,n}^{(m)}(b_1,\dots,b_n,\alpha_{i_1},\dots,\alpha_{i_m})$ stands for the volume of bordered Riemann surfaces with $n$ boundaries and $m$ defects corresponding to the deficit angles $\alpha_{i_1},\dots,\alpha_{i_m}$ and $r$ is the total number of defects. So, in effect, the quantity $W_{g,n}^{(m)},m>0$ is the Laplace transform of $V_{g,n}^{(m)}(b_1,\dots,b_n,\alpha_{i_1},\dots,\alpha_{i_m})$, the volumes with $m$ defects inserted summed over all possible combinations of $m$ defects. 

The crucial observation that facilitates the computation of $W_{g,n}^{(m)}$ is that volumes with defects $V_{g,n}^{(m)}$ can be obtained from the volumes $V_{g,n+m}^{(0)}$ without defects by an appropriate analytic continuation. Explicitly, this is given by 
\begin{align}
	V_{g,n}^{(m)}(b_1,\dots,b_n,\alpha_{i_1},\dots,\alpha_{i_m})=V_{g,n+m}^{(0)}(b_1,\dots,b_n,i(2\pi - \alpha_{i_1}),\dots,i(2\pi - \alpha_{i_m}))\label{vgma}
\end{align}
Let us now evaluate them explicitly for the case of $V_{0,3}^{(1)},V_{0,3}^{(2)},V_{1,1}^{(1)},V_{1,1}^{(2)}$. These can be obtained from the pure JT theory by the analytic continuation of $V_{0,4}^{(0)},V_{0,5}^{(0)},V_{1,2}^{(0)},V_{1,3}^{(0)}$ respectively. These are explicitly given by 
\begin{align}
	V_{0,4}^{(0)}(b_1,\dots,b_4)&=2 \pi^2 + \frac{1}{2} \sum_{i=1}^4 b_i^2\nonumber\\
	V_{0,5}^{(0)}(b_1,\dots,b_5)&=10\pi^4+3\pi^2\sum_{i=1}^5 b_i^2+\half \sum_{i<j}b_i^2b_j^2+\frac{1}{8}\sum_{i}^5 b_i^4  \nonumber\\
	V_{1,2}^{(0)}(b_1,b_2)&= \frac{1}{192} (4 \pi^2 + b_1^2 + b_2^2) (12 \pi^2 + b_1^2 + b_2^2 )\nonumber\\
	V_{1,3}^{(0)}(b_1,b_2,b_3)&=\frac{14 \pi ^6}{9}+\frac{13 \pi ^4 }{24}\sum_{i=1}^3 b_i^2+ \frac{\pi ^2 }{24}\sum_{i=1}^3 b_i^4+\frac{\pi ^2}{8}  \sum_{i<j}b_i^2 b_j^2+\frac{1}{192}  \sum_{i\neq j}b_i^2 b_j^4\nonumber\\
	&\,\,\,+\frac{1 }{1152}\sum_{i=1}^3 b_i^6+\frac{1}{96} b_1^2b_2^2b_3^2
	\label{vgnpujt}
\end{align}
	Let us begin with the computation of $W_{1,3}$. The volumes we need in this case are $V_{1,3}^{(1)}$ and $V_{1,3}^{(2)}$ from which we can compute $W_{1,3}^{(1)}$ and $W_{1,3}^{(2)}$ respectively. From eq.\eqref{wgnvbnm} for $g=1,n=3,r=2$, we have
\begin{align}
	W^{(1)} _{0,3} &= \epsilon_{1} \int \prod_{i=1}^{3} db_i b_i e^{-b_i z_i} (V^{(1)}_{0,3} (b_1,b_2, b_3,\alpha_1) - V^{(1)}_{0,3} (b_1,b_2, b_3,\alpha_2)) \nonumber\\
	&= \frac{\epsilon_{1}}{2z_1^2 z_2^2 z_3^2}  (4\pi-\alpha_1-\alpha_2) (\alpha_1-\alpha_2)\nonumber\\
W^{(2)} _{0,3} &= \half \epsilon_{1}^2 \int \prod_{i=1}^{3} db_i b_i e^{-b_i z_i} (V^{(2)}_{0,3} (b_1,b_2, b_3,\alpha_1,\alpha_1)+V^{(2)}_{0,3} (b_1,b_2, b_3,\alpha_2,\alpha_2) \nonumber\\
&\qquad  -V^{(2)}_{0,3} (b_1,b_2, b_3,\alpha_1,\alpha_2)-V^{(2)}_{0,3} (b_1,b_2, b_3,\alpha_2,\alpha_1)) \nonumber\\
&=\frac{\epsilon_1^2 (\alpha_1-\alpha_2)^2 (\alpha_1+\alpha_2-4 \pi )^2}{4 z_1^2 z_2^2 z_3^2}
\end{align}
The full $W_{0,3}$ to quadratic order in $\epsilon_{1}$ reads
\begin{align}
	W_{0,3}=\frac{1}{z_1^2z_2^2z_3^2}\left(1+ \frac{\epsilon_{1}}{2}  (4\pi-\alpha_1-\alpha_2) (\alpha_1-\alpha_2)+\frac{\epsilon_1^2 (\alpha_1-\alpha_2)^2 (\alpha_1+\alpha_2-4 \pi )^2}{4 }\right)\label{w03grav2}
\end{align}
It can be easily seen that the above results the corresponding expression from the matrix model side in eq.\eqref{w03eqcd} expanded to quadratic order in $\epsilon_{1}$.

A similar analysis for $W_{1,1}^{(1)},W_{1,1}^{(2)}$ gives
\begin{align}
	W_{1,1}^{(1)}(z)&= \epsilon_{1} \int b_1  db_1  e^{-b_1 z_i} (V^{(1)}_{1,1} (b_1,\alpha_1) -  V^{(1)}_{1,1} (b_1,\alpha_2))\nonumber\\
	&=\frac{\epsilon_{1}}{192 z^4} (4\pi-\alpha_1-\alpha_2) (\alpha_1-\alpha_2) (12 + z^2 (8 \pi^2 + \alpha_1^2 + \alpha_2^2 - 4 \pi(\alpha_1+\alpha_2) ))\nonumber\\
	W_{1,1}^{(2)}(z)&= \epsilon_{1} \int  b_1 db_1 e^{-b_1 z_i} (V^{(2)}_{1,1} (b_1,\alpha_1,\alpha_1) +V^{(2)}_{1,1} (b_1,\alpha_2,\alpha_2)-  V^{(2)}_{1,1} (b_1,\alpha_1,\alpha_2)-V^{(2)}_{1,1} (b_1,\alpha_2,\alpha_1))\nonumber\\
	&=\frac{\epsilon_1^2 (\alpha_1-\alpha_2)^2 (\alpha_1+\alpha_2-4 \pi )^2 \left(z^2 \left(-\alpha_1^2+4 \pi  (\alpha_1+\alpha_2)-\alpha_2^2+4 \pi ^2\right)+6\right)}{192 z^4}\label{w11eq}	
\end{align}
Again, the full $W_{1,1}$ to quadratic order in $\epsilon_{1}$ is given by 
\begin{align}
	W_{1,1}=&\frac{3+2\pi^2 z_1^2}{24 z_1^4}+\frac{\epsilon_{1}}{192 z^4} (4\pi-\alpha_1-\alpha_2) (\alpha_1-\alpha_2) (12 + z^2 (8 \pi^2 + \alpha_1^2 + \alpha_2^2 - 4 \pi(\alpha_1+\alpha_2) ))\nonumber\\
	&+\frac{\epsilon_1^2 (\alpha_1-\alpha_2)^2 (\alpha_1+\alpha_2-4 \pi )^2 \left(z^2 \left(-\alpha_1^2+4 \pi  (\alpha_1+\alpha_2)-\alpha_2^2+4 \pi ^2\right)+6\right)}{192 z^4}\label{w11qdgr}
\end{align}
It can be checked as before that this matches with the perturbative expansion of the corresponding matrix theory result in eq.\eqref{w11eqcd}.

\section{Propagator Calculation in JT Gravity}

In this section we review  amplitudes  in the JT theory which  evolves a single universe  state at some value of $(l_1,\phi_1)$ to a different value of $(l_2,\phi_2)$ and discuss some of their properties.  
We look at  two such amplitudes. One, which takes a state in the asymptotically far past,   in the contracting branch, to a state which is asymptotically, in the far future, in the expanding branch. This should be viewed as the amplitude for a  tunnelling  process, and will be referred to as the tunneling branch below. The  second, which takes a state in the expanding branch, asymptotically in the far future, to another state at a later time in the far future (i.e. at larger value of $\phi$), also in the expanding branch. This will be referred to at the expanding branch propagator below\footnote{A similar calculation can be done in the contracting branch between two states.}. 

Neither of the calculations for these two amplitudes new. The first amplitude has been discussed in \cite{Saad:2019lba, Maldacena:2019cbz} and calculated in the second order formalism in \cite{Moitra:2021uiv}. The second amplitude was calculated in \cite{Cotler:2024xzz} in the first order formalism and also discussed in canonical quantisation in \cite{Maxfield:2024rmg}.
Here we briefly review the calculation in 
\cite{Moitra:2021uiv}, for completeness, and then discuss the calculation of the expanding branch propagator   in the second order formalism.


\subsection{Tunneling Branch}
As discussed in \cite{Saad:2019lba, Maldacena:2019cbz} and \cite{Moitra:2021uiv}, the amplitude for an initially contracting  universe with length and dilaton taking values $(l_1,\phi_1)$ to tunnel to a finally expanding universe with the length and dilaton taking values, $(l_2,\phi_2)$, $\mathcal{K} (l_2,\phi_2;l_1,\phi_1)$ is obtained by summing over contributions  ${\cal K}(l_2,\phi_2;l_1,\phi_1;b)$, where $b$ is a modulus corresponding to the length of the minimum size geodesic in the tunneling geometry. 
${\cal K}(l_2,\phi_2;l_1,\phi_1;b)$ is given by 
\begin{equation}
	\mathcal{K} (l_2,\phi_2;l_1,\phi_1;b) =\exp{\frac{ib^2}{2} \left(\frac{\phi_2}{l_2} - \frac{\phi_1}{l_1}\right) - i(\phi_2 l_2 - \phi_1 l_1)}  \left(\frac{\phi_1 \phi_2}{l_1 l_2}\right)^{\frac{1}{2}} \frac{1}{8\pi} \label{propeq1}
\end{equation}
This leads to 
\begin{align}
	\mathcal{K} (l_2,\phi_2;l_1,\phi_1) &= 4\int b db 	\mathcal{K} (l_2,\phi_2;l_1,\phi_1;b) = 4 \int_{0}^{\infty} d\left(\frac{b^2}{2}\right)	\mathcal{K} (l_2,\phi_2;l_1,\phi_1;b) \nonumber \\
	&= \frac{1}{2\pi}\exp{- i(\phi_2 l_2 - \phi_1 l_1)}  \left(\frac{\phi_1 \phi_2}{l_1 l_2}\right)^{\frac{1}{2}} \frac{i}{\left(\frac{\phi_2}{l_2} - \frac{\phi_1}{l_1}\right)+i \epsilon} \label{propeq2}
\end{align}

Note that since $\mathcal{K} (l_2,\phi_2;l_1,\phi_1;b)$ must satisfy the WdW equation eq.\eqref{wda} for both $l_1,\phi_1$ and $l_2,\phi_2$ and the fact that WdW equation admits Hankel function as solutions we can write the exact form of eq.\eqref{propeq1} as follows,
\begin{equation}
	\mathcal{K} (l_2,\phi_2;l_1,\phi_1:b) = \frac{1}{16} l_1 l_2 \phi_1 \phi_2 \sqrt{\frac{1}{l_2^2-b^2}} H^{(2)}_1\left(\phi_2 \sqrt{l_2^2-b^2}\right)   \sqrt{\frac{1}{l_1^2-b^2}} \left(H^{(2)}_1\left(\phi_1 \sqrt{l_1^2-b^2}\right) \right)^*
\end{equation}
which for large $l_1,l_2,\phi_1,\phi_2$ reduces to eq.\eqref{propeq1}.

\subsection{Expanding Branch Propagator}
Next we consider the expanding branch propagator. We work in the second order formalism. The calculation is  similar to that of the tunelling branch and we will only highlight some key points below. 
We take the initial and final lengths $l_2,l_1$, and the initial and final values of the dilaton $\phi_1,\phi_2$  to satisfy the condition $l_2 > l_1 \gg 1$, $\phi_2>\phi_1 \gg 1$.

%
We will take  the metric for dS$_2$ in global coordinates given by, 
\be
\label{metads3}
ds^2=-{dr^2 \over (r^2+1)} +(r^2+1) d\theta^2 , \phi= A r
\ee
The $\theta$ coordinate is periodic with period $b$,  $\theta\in [0,b]$.


We can then follow the same argument that leads to eq.(2.35) of \cite{Moitra:2021uiv} to obtain the inner product
\begin{align}
	\langle PV_{L,m_1} PV_{L,m_2}\rangle & =  \delta_{m_1, -m_2} \frac{4 i b  \hat{c}_{m_1} \hat{c}_{-m_1} \abs{m_1}\left(m_1^2-1\right)  \left(  \left(-2 m_1^2+r_1^2+1\right)+2 i \abs{m_1} r_1\right)}{r_1^2+1}\left(\frac{r_1-i}{r_1+i}\right)^{\abs{ m_1} } \nonumber \\
	&	-\delta_{m_1, -m_2} \frac{4 i b  \hat{c}_{m_1} \hat{c}_{-m_1} \abs{m_1}\left(m_1^2-1\right)  \left(  \left(-2 m_1^2+r_2^2+1\right)+2 i \abs{m_1} r_2\right)}{r_2^2+1}\left(\frac{r_2-i}{r_2+i}\right)^{\abs{ m_1} } \nonumber \\
	&	- \delta_{m_1, -m_2} \frac{4 i b  \hat{d}_{m_1} \hat{d}_{-m_1} \abs{m_1}\left(m_1^2-1\right)  \left(  \left(-2 m_1^2+r_1^2+1\right)-2 i \abs{m_1} r_1\right)}{r_1^2+1}\left(\frac{r_1+i}{r_1-i}\right)^{\abs{ m_1} }\nonumber \\
	&	+ \delta_{m_1, -m_2} \frac{4 i b  \hat{d}_{m_1} \hat{d}_{-m_1} \abs{m_1}\left(m_1^2-1\right)  \left(  \left(-2 m_1^2+r_2^2+1\right)-2 i \abs{m_1} r_2\right)}{r_2^2+1}\left(\frac{r_2+i}{r_2-i}\right)^{\abs{ m_1} } \label{innprod}
\end{align}
where $m =\frac{2\pi \tilde{m}}{b}, \tilde{m} \in \mathcal{Z}$. Using the above equation, in the limit $r_1 \rightarrow \infty, r_2 \rightarrow \infty$ we get to leading order,
\begin{equation}
	D[PV_L] = \prod_{\tilde{m}\geq 1} d{\hat c}_m d{\hat c}_{m}^* d{\hat d}_m d{\hat d}_{m}^* \left(\frac{32 b}{3}  m^2 (m^2-1)^2 \left(\frac{1}{r_1^3}- \frac{1}{r_2^3}\right)\right)^2 \label{measure}
\end{equation}

Notice that there is a moduli for $\tilde{m}=0$. This will also contribute to the measure. See section 4 of \cite{Moitra:2021uiv}, eq.(4.17) onwards. The inner product of the metric perturbations corresponding to the moduli reads,
\begin{align}
	\langle PV_{tw}, PV_{tw}  \rangle =-\frac{2 t^2 b}{3\pi^2} \left(\frac{1}{r_1^3}- \frac{1}{r_2^3}\right) ;\quad \langle  PV_{b}, PV_{b} \rangle = {8(\delta b)^2\over 3b} \left(\frac{1}{r_1^3}- \frac{1}{r_2^3}\right) \label{moduliinp}
\end{align}
and the measure for moduli after integrating over the twist modulus turns out to be 
\be
\label{mm}
\int D[PV_{tw}]D[PV_{b}]= \frac{4 i}{3 \pi}  \int dt \, db \left(\frac{1}{r_1^3}- \frac{1}{r_2^3}\right) = 4 i \int d \tilde{t} db =  4 i \int b db
\ee
where we used the fact that $\tilde{t} = \frac{t}{3\pi} \left(\frac{1}{r_1^3}- \frac{1}{r_2^3}\right)$ has range $[0,b]$.

With the measure determined we now need the on-shell action to second order in fluctuation which is given by,
\begin{equation}
	i S =\frac{ib^2}{2} \left(\frac{\phi_2}{l_2} - \frac{\phi_1}{l_1}\right) - i(\phi_2 l_2 - \phi_1 l_1)  - \frac{1}{2}\sum_{m>1} {C*}^T (R_2-R_1) C \label{sact3}
\end{equation}
where,
\begin{equation}
	C = \begin{pmatrix}
		\hat{c}_m \\ \hat{d}_m
	\end{pmatrix}               , C* = \begin{pmatrix}
		\hat{c}_m^* \\ \hat{d}_m^*
	\end{pmatrix}
\end{equation}
and,
\begin{equation}
	R_1 = 2i b^2\frac{\phi_1}{l_1} m^2(m^2-1)\begin{pmatrix}
		\psi'_{+} (r_1) ^2 &  \psi'_{+} (r_1)  \psi'_{-} (r_1) \\
		\psi'_{+} (r_1)  \psi'_{-} (r_1) & \psi'_{-} (r_1) ^2
	\end{pmatrix}
\end{equation}
\begin{equation}
	R_2 =2i b^2\frac{\phi_2}{l_2} m^2(m^2-1)\begin{pmatrix}
		\psi'_{+} (r_2) ^2 &  \psi'_{+} (r_2)  \psi'_{-} (r_2) \\
		\psi'_{+} (r_2)  \psi'_{-} (r_2) & \psi'_{-} (r_2) ^2
	\end{pmatrix}
\end{equation}
Determinant of $(R_2-R_1)$ expanded in the limit where $r_1,r_2$ are large gives
\begin{equation}
	\det(R_2-R_1) = -\frac{64}{9} m^6 (m^2-1)^4 b^4 \left(\frac{1}{r_1^3}- \frac{1}{r_2^3}\right)^2 \frac{\phi_1 \phi_2}{l_1 l_2}
\end{equation}

Putting everything together and simplifying we obtain
\begin{equation}
	\mathcal{K} (l_2,\phi_2;l_1,\phi_1;b) =\exp{\frac{ib^2}{2} \left(\frac{\phi_2}{l_2} - \frac{\phi_1}{l_1}\right) - i(\phi_2 l_2 - \phi_1 l_1)}  \left(\frac{\phi_1 \phi_2}{l_1 l_2}\right)^{\frac{1}{2}} \frac{1}{8\pi} e^{-i \frac{\pi}{2}} \label{propfinal}
\end{equation}
where we have used the zeta function regularization for simplification, see appendix C of \cite{Moitra:2021uiv}. Integrating over the moduli
\begin{equation}
	\mathcal{K} (l_2,\phi_2;l_1,\phi_1) =  4 i \int b db 	\mathcal{K} (l_2,\phi_2;l_1,\phi_1;b) =  4 i \int_{-\infty} ^{\infty} d\left(\frac{b^2}{2}\right)	\mathcal{K} (l_2,\phi_2;l_1,\phi_1;b) \label{stsa}
\end{equation}
we obtain 
\begin{equation}
	\mathcal{K} (l_2,\phi_2;l_1,\phi_1) = \exp{- i(\phi_2 l_2 - \phi_1 l_1)}  \left(\frac{\phi_1 \phi_2}{l_1 l_2}\right)^{\frac{1}{2}} \textbf{$\delta$} \left(\frac{\phi_2}{l_2} - \frac{\phi_1}{l_1}\right) \label{propeq}
\end{equation}
The reason for taking the limit of $b^2 \in (-\infty,\infty)$ was explained in \cite{Cotler:2024xzz} and can be understood as follows. Consider the metric
\begin{equation}
	ds^2=-{dr^2 \over (r^2+1)} +(r^2+1) d\theta^2 
\end{equation}
Let us define a new angle $\xi = \frac{\theta}{b}$ and change $r$ to $r = \sinh(\zeta)$. Then we get in the large $\zeta$ limit after some simplification,
\begin{equation*}
	ds^2=-d\tilde{\zeta}^2 +\frac{1}{4} (e^{2 \tilde{\zeta}} +2 b^2) d\xi^2
\end{equation*}
where $\tilde{\zeta} = \zeta + \log(b)$. Since we want the second term to have positive sign throughout and $e^{2 \tilde{\zeta}}$ is always a large positive value we can take $b^2 \in (-\infty,\infty)$. 

Finally the exact form of eq.\eqref{propfinal} is given by,
\begin{equation}
	\mathcal{K} (l_2,\phi_2;l_1,\phi_1:b) = \frac{1}{16} e^{-i \frac{\pi}{2}} l_1 l_2 \phi_1 \phi_2 \sqrt{\frac{1}{l_2^2-b^2}} H^{(2)}_1\left(\phi_2 \sqrt{l_2^2-b^2}\right)   \sqrt{\frac{1}{l_1^2-b^2}} \left(H^{(2)}_1\left(\phi_1 \sqrt{l_1^2-b^2}\right) \right)^*\label{exfker}
\end{equation}


 Note that our results eq.(\ref{propeq2}) and eq.\eqref{propeq}  agree \cite{Cotler:2024xzz} and \cite{Maxfield:2024rmg}, upto an overall phase  factor,  which can be absorbed into the definition of the initial and final states. Also, we note that the integral eq.(\ref{stsa}) is to be carried out by first taking the asymptotic limit $l_1,\phi_1, \rightarrow \infty, l_2,\phi_2 \rightarrow \infty,$ keeping $b^2$ fixed, in the kernel $\mathcal{K} (l_2,\phi_2;l_1,\phi_1:b)$, eq.(\ref{exfker}), which then yields eq.(\ref{propeq}). 
 


\subsubsection{Additional Comments}
{\bf Inner Product:}

The expanding branch propagator allows us to calculate the inner product between two states $|l\rangle, |l'\rangle$, \cite{Cotler:2024xzz}.
Setting $\phi_2=\phi_1$ in eq.\eqref{propeq} gives, 
\begin{equation}
	\mathcal{K} (l,\phi_1;l',\phi_1)= \frac{1}{l'} \textbf{$\delta$} \left(\frac{1}{l} - \frac{1}{l'}\right)=l\delta(l-l')
\end{equation}
Identifying the above propagator with the inner product $\langle l|l'\rangle$ we get 
\begin{equation}
	\langle l | l' \rangle = {\mathcal{K}} (l,\phi;l',\phi) =  l \textbf{$\delta$} \left(l-l'\right)
\end{equation}

This inner product leads to the following  expression for the  resolution of identity in terms of the length basis states\footnote{Here we are restricting the space of states in the single universe sector.}:
\begin{equation}
\label{resid}
	\int_{0} ^{\infty} \frac{dl}{l} |l\rangle \langle l| = I
\end{equation}

And leads to a norm for the state $|\Psi\rangle$ with wave function  $\Psi(l,\phi)$ given by 
\be
\label{normsapp}
\langle \Psi|\Psi\rangle=\int{ dl\over l}\langle \Psi|l\rangle \langle l|\Psi \rangle=\int {dl\over l}\abs{\Psi|(l,\phi)}^2
\ee

This is clearly different from the Klein Gordon norm given in eq.(\ref{normaa}) (note that $\Psi$ and ${\hat \Psi}$ are related by eq.(\ref{foob})).

Asymptotically, for $\phi\rightarrow \infty$,  it follows from eq.(\ref{genex}) for a state in which $\rho(M)$ has compact support,  and is non-vanishing only for $|M|<M_0$,  that the wave function in the expanding branch  takes the form 
\be
\label{epaa}
\Psi(l,\phi)=e^{-il\phi}f\left({l\over \phi}\right)
\ee
where 
\be
\label{deff}
f\left({l\over \phi}\right)=\int dM \rho(M) e^{i l M\over 2 \phi}
\ee
The norm we have arrived at from the path integral calculation for such a state is 
\be
\label{normassa}
\langle \Psi|\Psi\rangle= \int {dl \over l} \abs{f\left({l\over \phi}\right)}^2
\ee
While the KG norm asmptotically gives 
\be
\label{norm2}
\langle \Psi|\Psi\rangle=\int {dl \over l} {\phi\over l} \abs{f\left({l\over \phi}\right)}^2
\ee
These are clearly different. However, it is worth noting that both of these  expressions  are conserved, i.e., independent of the dilaton (this follows from redefining the integration variable to be $l/\phi$ and noting that the integration limits are $[0,\infty]$).

{\bf Properties of The Propagator:}

Finally,  we note that that the expanding branch propagator meets the follows two checks. 
First, it propagates the Hartle Hawking state in a consistent manner. 
The HH state at $(l_2,\phi_2)$ is given in terms of the value it takes at an earlier time, $(l_1,\phi_1)$, by,
\begin{equation}
	\Psi_{HH} (l_2,\phi_2) =  \int_{0} ^{\infty} \frac{dl_1}{l_1} \mathcal{K} (l_2,\phi_2;l_1,\phi_1) \Psi_{HH} (l_1,\phi_1)
\end{equation}
Inserting from eq.\eqref{hhlt} for the HH wave function and evaluating the RHS, it is easy to see that the equality is met. 
%
This can also be  seen to be true  for a more general state where the late time wave function takes the form given in eq.(\ref{epaa}). 

In addition, it is easy to check that the propagator meets the following composition law, consistent with the resolution of identity, eq.(\ref{resid}):
%
\begin{equation}
	\mathcal{K} (l_2,\phi_2;l_1,\phi_1) = \int_{0} ^{\infty} \frac{dl}{l} \mathcal{K} (l_2,\phi_1;l,\phi) \mathcal{K} (l,\phi;l_1,\phi_1).
\end{equation}
%

		\vspace{-0.6cm}
		
		\newpage
		\bibliographystyle{JHEP}
		\bibliography{refs}

	\end{document}